\documentclass[longauth]{aaEC}

\usepackage{graphicx}
\usepackage{natbib}
\usepackage{scalerel}
\usepackage{multirow}
\usepackage{amssymb}

\usepackage{amsmath}
\DeclareMathOperator{\arcsinh}{arcsinh}

\newcommand{\zb}{\texttt{Zoobot}}

\usepackage[table]{xcolor}

\usepackage{txfonts}
\usepackage[pdfencoding=auto,psdextra]{hyperref}
\hypersetup{
    colorlinks=true,
    linkcolor=blue,
    filecolor=magenta,      
    urlcolor=blue,
    citecolor=blue
}
\usepackage{orcidlink}

\makeatletter
\renewcommand*\aa@pageof{, page \thepage{} of \pageref*{LastPage}}

\makeatother

\usepackage[utf8]{inputenc}

\usepackage[switch, modulo]{lineno}

\usepackage{Euclid}

\begin{document}

\title{Euclid Quick Data Release (Q1)} \subtitle{The Strong Lensing Discovery Engine C: Finding lenses with machine learning}


\newcommand{\orcid}[1]{\href{https://orcid.org/#1}{\orcidlink{#1}}}
		   
\author{Euclid Collaboration: N.~E.~P.~Lines\orcid{0009-0004-7751-1914}\thanks{\email{natalie.lines@port.ac.uk}}\inst{\ref{aff1}}
\and T.~E.~Collett\orcid{0000-0001-5564-3140}\inst{\ref{aff1}}
\and M.~Walmsley\orcid{0000-0002-6408-4181}\inst{\ref{aff2},\ref{aff3}}
\and K.~Rojas\orcid{0000-0003-1391-6854}\inst{\ref{aff4},\ref{aff1}}
\and T.~Li\orcid{0009-0005-5008-0381}\inst{\ref{aff1}}
\and L.~Leuzzi\orcid{0009-0006-4479-7017}\inst{\ref{aff5},\ref{aff6}}
\and A.~Manj\'on-Garc\'ia\orcid{0000-0002-7413-8825}\inst{\ref{aff7}}
\and S.~H.~Vincken\inst{\ref{aff4}}
\and J.~Wilde\orcid{0000-0002-4460-7379}\inst{\ref{aff8}}
\and P.~Holloway\orcid{0009-0002-8896-6100}\inst{\ref{aff9}}
\and A.~Verma\orcid{0000-0002-0730-0781}\inst{\ref{aff9}}
\and R.~B.~Metcalf\orcid{0000-0003-3167-2574}\inst{\ref{aff5},\ref{aff6}}
\and I.~T.~Andika\orcid{0000-0001-6102-9526}\inst{\ref{aff10},\ref{aff11}}
\and A.~Melo\orcid{0000-0002-6449-3970}\inst{\ref{aff11},\ref{aff10}}
\and M.~Melchior\inst{\ref{aff4}}
\and H.~Dom\'inguez~S\'anchez\orcid{0000-0002-9013-1316}\inst{\ref{aff12}}
\and A.~D\'iaz-S\'anchez\orcid{0000-0003-0748-4768}\inst{\ref{aff7}}
\and J.~A.~Acevedo~Barroso\orcid{0000-0002-9654-1711}\inst{\ref{aff13}}
\and B.~Cl\'ement\orcid{0000-0002-7966-3661}\inst{\ref{aff13},\ref{aff14}}
\and C.~Krawczyk\orcid{0000-0001-9233-2341}\inst{\ref{aff1}}
\and R.~Pearce-Casey\inst{\ref{aff15}}
\and S.~Serjeant\orcid{0000-0002-0517-7943}\inst{\ref{aff15}}
\and F.~Courbin\orcid{0000-0003-0758-6510}\inst{\ref{aff8},\ref{aff16}}
\and G.~Despali\orcid{0000-0001-6150-4112}\inst{\ref{aff5},\ref{aff6},\ref{aff17}}
\and R.~Gavazzi\orcid{0000-0002-5540-6935}\inst{\ref{aff18},\ref{aff19}}
\and S.~Schuldt\orcid{0000-0003-2497-6334}\inst{\ref{aff20},\ref{aff21}}
\and H.~Degaudenzi\orcid{0000-0002-5887-6799}\inst{\ref{aff22}}
\and L.~R.~Ecker\orcid{0009-0005-3508-2469}\inst{\ref{aff23},\ref{aff24}}
\and W.~J.~R.~Enzi\orcid{0009-0004-2992-3148}\inst{\ref{aff1}}
\and K.~Finner\orcid{0000-0002-4462-0709}\inst{\ref{aff25}}
\and A.~Galan\orcid{0000-0003-2547-9815}\inst{\ref{aff10},\ref{aff11}}
\and C.~Giocoli\orcid{0000-0002-9590-7961}\inst{\ref{aff6},\ref{aff17}}
\and N.~B.~Hogg\orcid{0000-0001-9346-4477}\inst{\ref{aff26}}
\and K.~Jahnke\orcid{0000-0003-3804-2137}\inst{\ref{aff27}}
\and S.~Kruk\orcid{0000-0001-8010-8879}\inst{\ref{aff28}}
\and G.~Mahler\orcid{0000-0003-3266-2001}\inst{\ref{aff29},\ref{aff30},\ref{aff31}}
\and A.~More\inst{\ref{aff32},\ref{aff33}}
\and B.~C.~Nagam\orcid{0000-0002-3724-7694}\inst{\ref{aff34},\ref{aff35}}
\and J.~Pearson\orcid{0000-0001-8555-8561}\inst{\ref{aff15}}
\and A.~Sainz~de~Murieta\inst{\ref{aff1}}
\and C.~Scarlata\orcid{0000-0002-9136-8876}\inst{\ref{aff34}}
\and D.~Sluse\orcid{0000-0001-6116-2095}\inst{\ref{aff29}}
\and A.~Sonnenfeld\orcid{0000-0002-6061-5977}\inst{\ref{aff36}}
\and C.~Spiniello\orcid{0000-0002-3909-6359}\inst{\ref{aff9}}
\and T.~T.~Thai\orcid{0000-0002-8408-4816}\inst{\ref{aff18},\ref{aff37}}
\and C.~Tortora\orcid{0000-0001-7958-6531}\inst{\ref{aff38}}
\and L.~Ulivi\orcid{0009-0001-3291-5382}\inst{\ref{aff39},\ref{aff40},\ref{aff41}}
\and L.~Weisenbach\orcid{0000-0003-1175-8004}\inst{\ref{aff1}}
\and M.~Zumalacarregui\orcid{0000-0002-9943-6490}\inst{\ref{aff42}}
\and N.~Aghanim\orcid{0000-0002-6688-8992}\inst{\ref{aff43}}
\and B.~Altieri\orcid{0000-0003-3936-0284}\inst{\ref{aff28}}
\and A.~Amara\inst{\ref{aff44}}
\and S.~Andreon\orcid{0000-0002-2041-8784}\inst{\ref{aff45}}
\and N.~Auricchio\orcid{0000-0003-4444-8651}\inst{\ref{aff6}}
\and H.~Aussel\orcid{0000-0002-1371-5705}\inst{\ref{aff46}}
\and C.~Baccigalupi\orcid{0000-0002-8211-1630}\inst{\ref{aff47},\ref{aff48},\ref{aff49},\ref{aff50}}
\and M.~Baldi\orcid{0000-0003-4145-1943}\inst{\ref{aff51},\ref{aff6},\ref{aff17}}
\and A.~Balestra\orcid{0000-0002-6967-261X}\inst{\ref{aff52}}
\and S.~Bardelli\orcid{0000-0002-8900-0298}\inst{\ref{aff6}}
\and P.~Battaglia\orcid{0000-0002-7337-5909}\inst{\ref{aff6}}
\and R.~Bender\orcid{0000-0001-7179-0626}\inst{\ref{aff24},\ref{aff23}}
\and F.~Bernardeau\inst{\ref{aff53},\ref{aff19}}
\and A.~Biviano\orcid{0000-0002-0857-0732}\inst{\ref{aff48},\ref{aff47}}
\and A.~Bonchi\orcid{0000-0002-2667-5482}\inst{\ref{aff54}}
\and D.~Bonino\orcid{0000-0002-3336-9977}\inst{\ref{aff55}}
\and E.~Branchini\orcid{0000-0002-0808-6908}\inst{\ref{aff56},\ref{aff57},\ref{aff45}}
\and M.~Brescia\orcid{0000-0001-9506-5680}\inst{\ref{aff58},\ref{aff38}}
\and J.~Brinchmann\orcid{0000-0003-4359-8797}\inst{\ref{aff59},\ref{aff60}}
\and S.~Camera\orcid{0000-0003-3399-3574}\inst{\ref{aff61},\ref{aff62},\ref{aff55}}
\and G.~Ca\~nas-Herrera\orcid{0000-0003-2796-2149}\inst{\ref{aff63},\ref{aff64},\ref{aff65}}
\and V.~Capobianco\orcid{0000-0002-3309-7692}\inst{\ref{aff55}}
\and C.~Carbone\orcid{0000-0003-0125-3563}\inst{\ref{aff21}}
\and V.~F.~Cardone\inst{\ref{aff66},\ref{aff67}}
\and J.~Carretero\orcid{0000-0002-3130-0204}\inst{\ref{aff68},\ref{aff69}}
\and S.~Casas\orcid{0000-0002-4751-5138}\inst{\ref{aff70}}
\and M.~Castellano\orcid{0000-0001-9875-8263}\inst{\ref{aff66}}
\and G.~Castignani\orcid{0000-0001-6831-0687}\inst{\ref{aff6}}
\and S.~Cavuoti\orcid{0000-0002-3787-4196}\inst{\ref{aff38},\ref{aff71}}
\and K.~C.~Chambers\orcid{0000-0001-6965-7789}\inst{\ref{aff72}}
\and A.~Cimatti\inst{\ref{aff73}}
\and C.~Colodro-Conde\inst{\ref{aff74}}
\and G.~Congedo\orcid{0000-0003-2508-0046}\inst{\ref{aff75}}
\and C.~J.~Conselice\orcid{0000-0003-1949-7638}\inst{\ref{aff3}}
\and L.~Conversi\orcid{0000-0002-6710-8476}\inst{\ref{aff76},\ref{aff28}}
\and Y.~Copin\orcid{0000-0002-5317-7518}\inst{\ref{aff77}}
\and A.~Costille\inst{\ref{aff18}}
\and H.~M.~Courtois\orcid{0000-0003-0509-1776}\inst{\ref{aff78}}
\and M.~Cropper\orcid{0000-0003-4571-9468}\inst{\ref{aff79}}
\and A.~Da~Silva\orcid{0000-0002-6385-1609}\inst{\ref{aff80},\ref{aff81}}
\and G.~De~Lucia\orcid{0000-0002-6220-9104}\inst{\ref{aff48}}
\and A.~M.~Di~Giorgio\orcid{0000-0002-4767-2360}\inst{\ref{aff82}}
\and C.~Dolding\orcid{0009-0003-7199-6108}\inst{\ref{aff79}}
\and H.~Dole\orcid{0000-0002-9767-3839}\inst{\ref{aff43}}
\and F.~Dubath\orcid{0000-0002-6533-2810}\inst{\ref{aff22}}
\and C.~A.~J.~Duncan\orcid{0009-0003-3573-0791}\inst{\ref{aff3}}
\and X.~Dupac\inst{\ref{aff28}}
\and S.~Escoffier\orcid{0000-0002-2847-7498}\inst{\ref{aff83}}
\and M.~Fabricius\orcid{0000-0002-7025-6058}\inst{\ref{aff24},\ref{aff23}}
\and M.~Farina\orcid{0000-0002-3089-7846}\inst{\ref{aff82}}
\and R.~Farinelli\inst{\ref{aff6}}
\and F.~Faustini\orcid{0000-0001-6274-5145}\inst{\ref{aff54},\ref{aff66}}
\and S.~Ferriol\inst{\ref{aff77}}
\and F.~Finelli\orcid{0000-0002-6694-3269}\inst{\ref{aff6},\ref{aff84}}
\and S.~Fotopoulou\orcid{0000-0002-9686-254X}\inst{\ref{aff85}}
\and M.~Frailis\orcid{0000-0002-7400-2135}\inst{\ref{aff48}}
\and E.~Franceschi\orcid{0000-0002-0585-6591}\inst{\ref{aff6}}
\and M.~Fumana\orcid{0000-0001-6787-5950}\inst{\ref{aff21}}
\and S.~Galeotta\orcid{0000-0002-3748-5115}\inst{\ref{aff48}}
\and K.~George\orcid{0000-0002-1734-8455}\inst{\ref{aff23}}
\and W.~Gillard\orcid{0000-0003-4744-9748}\inst{\ref{aff83}}
\and B.~Gillis\orcid{0000-0002-4478-1270}\inst{\ref{aff75}}
\and P.~G\'omez-Alvarez\orcid{0000-0002-8594-5358}\inst{\ref{aff86},\ref{aff28}}
\and J.~Gracia-Carpio\inst{\ref{aff24}}
\and B.~R.~Granett\orcid{0000-0003-2694-9284}\inst{\ref{aff45}}
\and A.~Grazian\orcid{0000-0002-5688-0663}\inst{\ref{aff52}}
\and F.~Grupp\inst{\ref{aff24},\ref{aff23}}
\and L.~Guzzo\orcid{0000-0001-8264-5192}\inst{\ref{aff20},\ref{aff45},\ref{aff87}}
\and S.~Gwyn\orcid{0000-0001-8221-8406}\inst{\ref{aff88}}
\and S.~V.~H.~Haugan\orcid{0000-0001-9648-7260}\inst{\ref{aff89}}
\and W.~Holmes\inst{\ref{aff90}}
\and I.~M.~Hook\orcid{0000-0002-2960-978X}\inst{\ref{aff91}}
\and F.~Hormuth\inst{\ref{aff92}}
\and A.~Hornstrup\orcid{0000-0002-3363-0936}\inst{\ref{aff93},\ref{aff94}}
\and P.~Hudelot\inst{\ref{aff19}}
\and M.~Jhabvala\inst{\ref{aff95}}
\and E.~Keih\"anen\orcid{0000-0003-1804-7715}\inst{\ref{aff96}}
\and S.~Kermiche\orcid{0000-0002-0302-5735}\inst{\ref{aff83}}
\and A.~Kiessling\orcid{0000-0002-2590-1273}\inst{\ref{aff90}}
\and B.~Kubik\orcid{0009-0006-5823-4880}\inst{\ref{aff77}}
\and M.~K\"ummel\orcid{0000-0003-2791-2117}\inst{\ref{aff23}}
\and M.~Kunz\orcid{0000-0002-3052-7394}\inst{\ref{aff97}}
\and H.~Kurki-Suonio\orcid{0000-0002-4618-3063}\inst{\ref{aff98},\ref{aff99}}
\and Q.~Le~Boulc'h\inst{\ref{aff100}}
\and A.~M.~C.~Le~Brun\orcid{0000-0002-0936-4594}\inst{\ref{aff101}}
\and D.~Le~Mignant\orcid{0000-0002-5339-5515}\inst{\ref{aff18}}
\and S.~Ligori\orcid{0000-0003-4172-4606}\inst{\ref{aff55}}
\and P.~B.~Lilje\orcid{0000-0003-4324-7794}\inst{\ref{aff89}}
\and V.~Lindholm\orcid{0000-0003-2317-5471}\inst{\ref{aff98},\ref{aff99}}
\and I.~Lloro\orcid{0000-0001-5966-1434}\inst{\ref{aff102}}
\and G.~Mainetti\orcid{0000-0003-2384-2377}\inst{\ref{aff100}}
\and D.~Maino\inst{\ref{aff20},\ref{aff21},\ref{aff87}}
\and E.~Maiorano\orcid{0000-0003-2593-4355}\inst{\ref{aff6}}
\and O.~Mansutti\orcid{0000-0001-5758-4658}\inst{\ref{aff48}}
\and S.~Marcin\inst{\ref{aff103}}
\and O.~Marggraf\orcid{0000-0001-7242-3852}\inst{\ref{aff104}}
\and M.~Martinelli\orcid{0000-0002-6943-7732}\inst{\ref{aff66},\ref{aff67}}
\and N.~Martinet\orcid{0000-0003-2786-7790}\inst{\ref{aff18}}
\and F.~Marulli\orcid{0000-0002-8850-0303}\inst{\ref{aff5},\ref{aff6},\ref{aff17}}
\and R.~Massey\orcid{0000-0002-6085-3780}\inst{\ref{aff31}}
\and S.~Maurogordato\inst{\ref{aff105}}
\and E.~Medinaceli\orcid{0000-0002-4040-7783}\inst{\ref{aff6}}
\and S.~Mei\orcid{0000-0002-2849-559X}\inst{\ref{aff106},\ref{aff107}}
\and Y.~Mellier\inst{\ref{aff108},\ref{aff19}}
\and M.~Meneghetti\orcid{0000-0003-1225-7084}\inst{\ref{aff6},\ref{aff17}}
\and E.~Merlin\orcid{0000-0001-6870-8900}\inst{\ref{aff66}}
\and G.~Meylan\inst{\ref{aff13}}
\and A.~Mora\orcid{0000-0002-1922-8529}\inst{\ref{aff109}}
\and M.~Moresco\orcid{0000-0002-7616-7136}\inst{\ref{aff5},\ref{aff6}}
\and L.~Moscardini\orcid{0000-0002-3473-6716}\inst{\ref{aff5},\ref{aff6},\ref{aff17}}
\and R.~Nakajima\orcid{0009-0009-1213-7040}\inst{\ref{aff104}}
\and C.~Neissner\orcid{0000-0001-8524-4968}\inst{\ref{aff110},\ref{aff69}}
\and R.~C.~Nichol\orcid{0000-0003-0939-6518}\inst{\ref{aff44}}
\and S.-M.~Niemi\inst{\ref{aff63}}
\and J.~W.~Nightingale\orcid{0000-0002-8987-7401}\inst{\ref{aff111}}
\and C.~Padilla\orcid{0000-0001-7951-0166}\inst{\ref{aff110}}
\and S.~Paltani\orcid{0000-0002-8108-9179}\inst{\ref{aff22}}
\and F.~Pasian\orcid{0000-0002-4869-3227}\inst{\ref{aff48}}
\and K.~Pedersen\inst{\ref{aff112}}
\and W.~J.~Percival\orcid{0000-0002-0644-5727}\inst{\ref{aff113},\ref{aff114},\ref{aff115}}
\and V.~Pettorino\inst{\ref{aff63}}
\and S.~Pires\orcid{0000-0002-0249-2104}\inst{\ref{aff46}}
\and G.~Polenta\orcid{0000-0003-4067-9196}\inst{\ref{aff54}}
\and M.~Poncet\inst{\ref{aff116}}
\and L.~A.~Popa\inst{\ref{aff117}}
\and L.~Pozzetti\orcid{0000-0001-7085-0412}\inst{\ref{aff6}}
\and F.~Raison\orcid{0000-0002-7819-6918}\inst{\ref{aff24}}
\and R.~Rebolo\orcid{0000-0003-3767-7085}\inst{\ref{aff74},\ref{aff118},\ref{aff119}}
\and A.~Renzi\orcid{0000-0001-9856-1970}\inst{\ref{aff120},\ref{aff121}}
\and J.~Rhodes\orcid{0000-0002-4485-8549}\inst{\ref{aff90}}
\and G.~Riccio\inst{\ref{aff38}}
\and E.~Romelli\orcid{0000-0003-3069-9222}\inst{\ref{aff48}}
\and M.~Roncarelli\orcid{0000-0001-9587-7822}\inst{\ref{aff6}}
\and R.~Saglia\orcid{0000-0003-0378-7032}\inst{\ref{aff23},\ref{aff24}}
\and Z.~Sakr\orcid{0000-0002-4823-3757}\inst{\ref{aff122},\ref{aff123},\ref{aff124}}
\and A.~G.~S\'anchez\orcid{0000-0003-1198-831X}\inst{\ref{aff24}}
\and D.~Sapone\orcid{0000-0001-7089-4503}\inst{\ref{aff125}}
\and B.~Sartoris\orcid{0000-0003-1337-5269}\inst{\ref{aff23},\ref{aff48}}
\and J.~A.~Schewtschenko\orcid{0000-0002-4913-6393}\inst{\ref{aff75}}
\and M.~Schirmer\orcid{0000-0003-2568-9994}\inst{\ref{aff27}}
\and P.~Schneider\orcid{0000-0001-8561-2679}\inst{\ref{aff104}}
\and T.~Schrabback\orcid{0000-0002-6987-7834}\inst{\ref{aff126}}
\and A.~Secroun\orcid{0000-0003-0505-3710}\inst{\ref{aff83}}
\and G.~Seidel\orcid{0000-0003-2907-353X}\inst{\ref{aff27}}
\and M.~Seiffert\orcid{0000-0002-7536-9393}\inst{\ref{aff90}}
\and S.~Serrano\orcid{0000-0002-0211-2861}\inst{\ref{aff127},\ref{aff128},\ref{aff129}}
\and P.~Simon\inst{\ref{aff104}}
\and C.~Sirignano\orcid{0000-0002-0995-7146}\inst{\ref{aff120},\ref{aff121}}
\and G.~Sirri\orcid{0000-0003-2626-2853}\inst{\ref{aff17}}
\and A.~Spurio~Mancini\orcid{0000-0001-5698-0990}\inst{\ref{aff130}}
\and L.~Stanco\orcid{0000-0002-9706-5104}\inst{\ref{aff121}}
\and J.~Steinwagner\orcid{0000-0001-7443-1047}\inst{\ref{aff24}}
\and P.~Tallada-Cresp\'{i}\orcid{0000-0002-1336-8328}\inst{\ref{aff68},\ref{aff69}}
\and A.~N.~Taylor\inst{\ref{aff75}}
\and I.~Tereno\inst{\ref{aff80},\ref{aff131}}
\and S.~Toft\orcid{0000-0003-3631-7176}\inst{\ref{aff132},\ref{aff133}}
\and R.~Toledo-Moreo\orcid{0000-0002-2997-4859}\inst{\ref{aff134}}
\and F.~Torradeflot\orcid{0000-0003-1160-1517}\inst{\ref{aff69},\ref{aff68}}
\and I.~Tutusaus\orcid{0000-0002-3199-0399}\inst{\ref{aff123}}
\and E.~A.~Valentijn\inst{\ref{aff35}}
\and L.~Valenziano\orcid{0000-0002-1170-0104}\inst{\ref{aff6},\ref{aff84}}
\and J.~Valiviita\orcid{0000-0001-6225-3693}\inst{\ref{aff98},\ref{aff99}}
\and T.~Vassallo\orcid{0000-0001-6512-6358}\inst{\ref{aff23},\ref{aff48}}
\and G.~Verdoes~Kleijn\orcid{0000-0001-5803-2580}\inst{\ref{aff35}}
\and A.~Veropalumbo\orcid{0000-0003-2387-1194}\inst{\ref{aff45},\ref{aff57},\ref{aff56}}
\and Y.~Wang\orcid{0000-0002-4749-2984}\inst{\ref{aff135}}
\and J.~Weller\orcid{0000-0002-8282-2010}\inst{\ref{aff23},\ref{aff24}}
\and A.~Zacchei\orcid{0000-0003-0396-1192}\inst{\ref{aff48},\ref{aff47}}
\and G.~Zamorani\orcid{0000-0002-2318-301X}\inst{\ref{aff6}}
\and F.~M.~Zerbi\inst{\ref{aff45}}
\and E.~Zucca\orcid{0000-0002-5845-8132}\inst{\ref{aff6}}
\and V.~Allevato\orcid{0000-0001-7232-5152}\inst{\ref{aff38}}
\and M.~Ballardini\orcid{0000-0003-4481-3559}\inst{\ref{aff136},\ref{aff137},\ref{aff6}}
\and M.~Bolzonella\orcid{0000-0003-3278-4607}\inst{\ref{aff6}}
\and E.~Bozzo\orcid{0000-0002-8201-1525}\inst{\ref{aff22}}
\and C.~Burigana\orcid{0000-0002-3005-5796}\inst{\ref{aff138},\ref{aff84}}
\and R.~Cabanac\orcid{0000-0001-6679-2600}\inst{\ref{aff123}}
\and A.~Cappi\inst{\ref{aff6},\ref{aff105}}
\and D.~Di~Ferdinando\inst{\ref{aff17}}
\and J.~A.~Escartin~Vigo\inst{\ref{aff24}}
\and L.~Gabarra\orcid{0000-0002-8486-8856}\inst{\ref{aff9}}
\and J.~Mart\'{i}n-Fleitas\orcid{0000-0002-8594-569X}\inst{\ref{aff109}}
\and S.~Matthew\orcid{0000-0001-8448-1697}\inst{\ref{aff75}}
\and N.~Mauri\orcid{0000-0001-8196-1548}\inst{\ref{aff73},\ref{aff17}}
\and A.~Pezzotta\orcid{0000-0003-0726-2268}\inst{\ref{aff139},\ref{aff24}}
\and M.~P\"ontinen\orcid{0000-0001-5442-2530}\inst{\ref{aff98}}
\and C.~Porciani\orcid{0000-0002-7797-2508}\inst{\ref{aff104}}
\and I.~Risso\orcid{0000-0003-2525-7761}\inst{\ref{aff140}}
\and V.~Scottez\inst{\ref{aff108},\ref{aff141}}
\and M.~Sereno\orcid{0000-0003-0302-0325}\inst{\ref{aff6},\ref{aff17}}
\and M.~Tenti\orcid{0000-0002-4254-5901}\inst{\ref{aff17}}
\and M.~Viel\orcid{0000-0002-2642-5707}\inst{\ref{aff47},\ref{aff48},\ref{aff50},\ref{aff49},\ref{aff142}}
\and M.~Wiesmann\orcid{0009-0000-8199-5860}\inst{\ref{aff89}}
\and Y.~Akrami\orcid{0000-0002-2407-7956}\inst{\ref{aff143},\ref{aff144}}
\and S.~Anselmi\orcid{0000-0002-3579-9583}\inst{\ref{aff121},\ref{aff120},\ref{aff145}}
\and M.~Archidiacono\orcid{0000-0003-4952-9012}\inst{\ref{aff20},\ref{aff87}}
\and F.~Atrio-Barandela\orcid{0000-0002-2130-2513}\inst{\ref{aff146}}
\and C.~Benoist\inst{\ref{aff105}}
\and K.~Benson\inst{\ref{aff79}}
\and P.~Bergamini\orcid{0000-0003-1383-9414}\inst{\ref{aff20},\ref{aff6}}
\and D.~Bertacca\orcid{0000-0002-2490-7139}\inst{\ref{aff120},\ref{aff52},\ref{aff121}}
\and M.~Bethermin\orcid{0000-0002-3915-2015}\inst{\ref{aff147}}
\and A.~Blanchard\orcid{0000-0001-8555-9003}\inst{\ref{aff123}}
\and L.~Blot\orcid{0000-0002-9622-7167}\inst{\ref{aff148},\ref{aff145}}
\and M.~L.~Brown\orcid{0000-0002-0370-8077}\inst{\ref{aff3}}
\and S.~Bruton\orcid{0000-0002-6503-5218}\inst{\ref{aff149}}
\and A.~Calabro\orcid{0000-0003-2536-1614}\inst{\ref{aff66}}
\and F.~Caro\inst{\ref{aff66}}
\and C.~S.~Carvalho\inst{\ref{aff131}}
\and T.~Castro\orcid{0000-0002-6292-3228}\inst{\ref{aff48},\ref{aff49},\ref{aff47},\ref{aff142}}
\and Y.~Charles\inst{\ref{aff18}}
\and F.~Cogato\orcid{0000-0003-4632-6113}\inst{\ref{aff5},\ref{aff6}}
\and A.~R.~Cooray\orcid{0000-0002-3892-0190}\inst{\ref{aff150}}
\and O.~Cucciati\orcid{0000-0002-9336-7551}\inst{\ref{aff6}}
\and S.~Davini\orcid{0000-0003-3269-1718}\inst{\ref{aff57}}
\and F.~De~Paolis\orcid{0000-0001-6460-7563}\inst{\ref{aff151},\ref{aff152},\ref{aff153}}
\and G.~Desprez\orcid{0000-0001-8325-1742}\inst{\ref{aff35}}
\and J.~J.~Diaz\inst{\ref{aff154}}
\and S.~Di~Domizio\orcid{0000-0003-2863-5895}\inst{\ref{aff56},\ref{aff57}}
\and J.~M.~Diego\orcid{0000-0001-9065-3926}\inst{\ref{aff12}}
\and A.~Enia\orcid{0000-0002-0200-2857}\inst{\ref{aff51},\ref{aff6}}
\and Y.~Fang\inst{\ref{aff23}}
\and A.~G.~Ferrari\orcid{0009-0005-5266-4110}\inst{\ref{aff17}}
\and A.~Finoguenov\orcid{0000-0002-4606-5403}\inst{\ref{aff98}}
\and A.~Fontana\orcid{0000-0003-3820-2823}\inst{\ref{aff66}}
\and A.~Franco\orcid{0000-0002-4761-366X}\inst{\ref{aff152},\ref{aff151},\ref{aff153}}
\and K.~Ganga\orcid{0000-0001-8159-8208}\inst{\ref{aff106}}
\and J.~Garc\'ia-Bellido\orcid{0000-0002-9370-8360}\inst{\ref{aff143}}
\and T.~Gasparetto\orcid{0000-0002-7913-4866}\inst{\ref{aff48}}
\and V.~Gautard\inst{\ref{aff155}}
\and E.~Gaztanaga\orcid{0000-0001-9632-0815}\inst{\ref{aff129},\ref{aff127},\ref{aff1}}
\and F.~Giacomini\orcid{0000-0002-3129-2814}\inst{\ref{aff17}}
\and F.~Gianotti\orcid{0000-0003-4666-119X}\inst{\ref{aff6}}
\and G.~Gozaliasl\orcid{0000-0002-0236-919X}\inst{\ref{aff156},\ref{aff98}}
\and M.~Guidi\orcid{0000-0001-9408-1101}\inst{\ref{aff51},\ref{aff6}}
\and C.~M.~Gutierrez\orcid{0000-0001-7854-783X}\inst{\ref{aff157}}
\and A.~Hall\orcid{0000-0002-3139-8651}\inst{\ref{aff75}}
\and W.~G.~Hartley\inst{\ref{aff22}}
\and C.~Hern\'andez-Monteagudo\orcid{0000-0001-5471-9166}\inst{\ref{aff119},\ref{aff74}}
\and H.~Hildebrandt\orcid{0000-0002-9814-3338}\inst{\ref{aff158}}
\and J.~Hjorth\orcid{0000-0002-4571-2306}\inst{\ref{aff112}}
\and J.~J.~E.~Kajava\orcid{0000-0002-3010-8333}\inst{\ref{aff159},\ref{aff160}}
\and Y.~Kang\orcid{0009-0000-8588-7250}\inst{\ref{aff22}}
\and V.~Kansal\orcid{0000-0002-4008-6078}\inst{\ref{aff161},\ref{aff162}}
\and D.~Karagiannis\orcid{0000-0002-4927-0816}\inst{\ref{aff136},\ref{aff163}}
\and K.~Kiiveri\inst{\ref{aff96}}
\and C.~C.~Kirkpatrick\inst{\ref{aff96}}
\and J.~Le~Graet\orcid{0000-0001-6523-7971}\inst{\ref{aff83}}
\and L.~Legrand\orcid{0000-0003-0610-5252}\inst{\ref{aff164},\ref{aff165}}
\and M.~Lembo\orcid{0000-0002-5271-5070}\inst{\ref{aff136},\ref{aff137}}
\and F.~Lepori\orcid{0009-0000-5061-7138}\inst{\ref{aff166}}
\and G.~Leroy\orcid{0009-0004-2523-4425}\inst{\ref{aff30},\ref{aff31}}
\and G.~F.~Lesci\orcid{0000-0002-4607-2830}\inst{\ref{aff5},\ref{aff6}}
\and J.~Lesgourgues\orcid{0000-0001-7627-353X}\inst{\ref{aff70}}
\and T.~I.~Liaudat\orcid{0000-0002-9104-314X}\inst{\ref{aff167}}
\and S.~J.~Liu\orcid{0000-0001-7680-2139}\inst{\ref{aff82}}
\and A.~Loureiro\orcid{0000-0002-4371-0876}\inst{\ref{aff168},\ref{aff169}}
\and J.~Macias-Perez\orcid{0000-0002-5385-2763}\inst{\ref{aff170}}
\and G.~Maggio\orcid{0000-0003-4020-4836}\inst{\ref{aff48}}
\and M.~Magliocchetti\orcid{0000-0001-9158-4838}\inst{\ref{aff82}}
\and E.~A.~Magnier\orcid{0000-0002-7965-2815}\inst{\ref{aff72}}
\and F.~Mannucci\orcid{0000-0002-4803-2381}\inst{\ref{aff41}}
\and R.~Maoli\orcid{0000-0002-6065-3025}\inst{\ref{aff171},\ref{aff66}}
\and C.~J.~A.~P.~Martins\orcid{0000-0002-4886-9261}\inst{\ref{aff172},\ref{aff59}}
\and L.~Maurin\orcid{0000-0002-8406-0857}\inst{\ref{aff43}}
\and M.~Miluzio\inst{\ref{aff28},\ref{aff173}}
\and P.~Monaco\orcid{0000-0003-2083-7564}\inst{\ref{aff174},\ref{aff48},\ref{aff49},\ref{aff47}}
\and C.~Moretti\orcid{0000-0003-3314-8936}\inst{\ref{aff50},\ref{aff142},\ref{aff48},\ref{aff47},\ref{aff49}}
\and G.~Morgante\inst{\ref{aff6}}
\and S.~Nadathur\orcid{0000-0001-9070-3102}\inst{\ref{aff1}}
\and K.~Naidoo\orcid{0000-0002-9182-1802}\inst{\ref{aff1}}
\and A.~Navarro-Alsina\orcid{0000-0002-3173-2592}\inst{\ref{aff104}}
\and S.~Nesseris\orcid{0000-0002-0567-0324}\inst{\ref{aff143}}
\and F.~Passalacqua\orcid{0000-0002-8606-4093}\inst{\ref{aff120},\ref{aff121}}
\and K.~Paterson\orcid{0000-0001-8340-3486}\inst{\ref{aff27}}
\and L.~Patrizii\inst{\ref{aff17}}
\and A.~Pisani\orcid{0000-0002-6146-4437}\inst{\ref{aff83},\ref{aff175}}
\and D.~Potter\orcid{0000-0002-0757-5195}\inst{\ref{aff166}}
\and S.~Quai\orcid{0000-0002-0449-8163}\inst{\ref{aff5},\ref{aff6}}
\and M.~Radovich\orcid{0000-0002-3585-866X}\inst{\ref{aff52}}
\and P.-F.~Rocci\inst{\ref{aff43}}
\and S.~Sacquegna\orcid{0000-0002-8433-6630}\inst{\ref{aff151},\ref{aff152},\ref{aff153}}
\and M.~Sahl\'en\orcid{0000-0003-0973-4804}\inst{\ref{aff176}}
\and D.~B.~Sanders\orcid{0000-0002-1233-9998}\inst{\ref{aff72}}
\and E.~Sarpa\orcid{0000-0002-1256-655X}\inst{\ref{aff50},\ref{aff142},\ref{aff49}}
\and A.~Schneider\orcid{0000-0001-7055-8104}\inst{\ref{aff166}}
\and D.~Sciotti\orcid{0009-0008-4519-2620}\inst{\ref{aff66},\ref{aff67}}
\and E.~Sellentin\inst{\ref{aff177},\ref{aff65}}
\and L.~C.~Smith\orcid{0000-0002-3259-2771}\inst{\ref{aff178}}
\and K.~Tanidis\orcid{0000-0001-9843-5130}\inst{\ref{aff9}}
\and G.~Testera\inst{\ref{aff57}}
\and R.~Teyssier\orcid{0000-0001-7689-0933}\inst{\ref{aff175}}
\and S.~Tosi\orcid{0000-0002-7275-9193}\inst{\ref{aff56},\ref{aff57},\ref{aff45}}
\and A.~Troja\orcid{0000-0003-0239-4595}\inst{\ref{aff120},\ref{aff121}}
\and M.~Tucci\inst{\ref{aff22}}
\and C.~Valieri\inst{\ref{aff17}}
\and A.~Venhola\orcid{0000-0001-6071-4564}\inst{\ref{aff179}}
\and D.~Vergani\orcid{0000-0003-0898-2216}\inst{\ref{aff6}}
\and G.~Vernardos\orcid{0000-0001-8554-7248}\inst{\ref{aff180},\ref{aff181}}
\and G.~Verza\orcid{0000-0002-1886-8348}\inst{\ref{aff182}}
\and P.~Vielzeuf\orcid{0000-0003-2035-9339}\inst{\ref{aff83}}
\and N.~A.~Walton\orcid{0000-0003-3983-8778}\inst{\ref{aff178}}
\and D.~Scott\orcid{0000-0002-6878-9840}\inst{\ref{aff183}}}
										   
\institute{Institute of Cosmology and Gravitation, University of Portsmouth, Portsmouth PO1 3FX, UK\label{aff1}
\and
David A. Dunlap Department of Astronomy \& Astrophysics, University of Toronto, 50 St George Street, Toronto, Ontario M5S 3H4, Canada\label{aff2}
\and
Jodrell Bank Centre for Astrophysics, Department of Physics and Astronomy, University of Manchester, Oxford Road, Manchester M13 9PL, UK\label{aff3}
\and
University of Applied Sciences and Arts of Northwestern Switzerland, School of Engineering, 5210 Windisch, Switzerland\label{aff4}
\and
Dipartimento di Fisica e Astronomia "Augusto Righi" - Alma Mater Studiorum Universit\`a di Bologna, via Piero Gobetti 93/2, 40129 Bologna, Italy\label{aff5}
\and
INAF-Osservatorio di Astrofisica e Scienza dello Spazio di Bologna, Via Piero Gobetti 93/3, 40129 Bologna, Italy\label{aff6}
\and
Departamento F\'isica Aplicada, Universidad Polit\'ecnica de Cartagena, Campus Muralla del Mar, 30202 Cartagena, Murcia, Spain\label{aff7}
\and
Institut de Ci\`{e}ncies del Cosmos (ICCUB), Universitat de Barcelona (IEEC-UB), Mart\'{i} i Franqu\`{e}s 1, 08028 Barcelona, Spain\label{aff8}
\and
Department of Physics, Oxford University, Keble Road, Oxford OX1 3RH, UK\label{aff9}
\and
Technical University of Munich, TUM School of Natural Sciences, Physics Department, James-Franck-Str.~1, 85748 Garching, Germany\label{aff10}
\and
Max-Planck-Institut f\"ur Astrophysik, Karl-Schwarzschild-Str.~1, 85748 Garching, Germany\label{aff11}
\and
Instituto de F\'isica de Cantabria, Edificio Juan Jord\'a, Avenida de los Castros, 39005 Santander, Spain\label{aff12}
\and
Institute of Physics, Laboratory of Astrophysics, Ecole Polytechnique F\'ed\'erale de Lausanne (EPFL), Observatoire de Sauverny, 1290 Versoix, Switzerland\label{aff13}
\and
SCITAS, Ecole Polytechnique F\'ed\'erale de Lausanne (EPFL), 1015 Lausanne, Switzerland\label{aff14}
\and
School of Physical Sciences, The Open University, Milton Keynes, MK7 6AA, UK\label{aff15}
\and
Instituci\'o Catalana de Recerca i Estudis Avan\c{c}ats (ICREA), Passeig de Llu\'{\i}s Companys 23, 08010 Barcelona, Spain\label{aff16}
\and
INFN-Sezione di Bologna, Viale Berti Pichat 6/2, 40127 Bologna, Italy\label{aff17}
\and
Aix-Marseille Universit\'e, CNRS, CNES, LAM, Marseille, France\label{aff18}
\and
Institut d'Astrophysique de Paris, UMR 7095, CNRS, and Sorbonne Universit\'e, 98 bis boulevard Arago, 75014 Paris, France\label{aff19}
\and
Dipartimento di Fisica "Aldo Pontremoli", Universit\`a degli Studi di Milano, Via Celoria 16, 20133 Milano, Italy\label{aff20}
\and
INAF-IASF Milano, Via Alfonso Corti 12, 20133 Milano, Italy\label{aff21}
\and
Department of Astronomy, University of Geneva, ch. d'Ecogia 16, 1290 Versoix, Switzerland\label{aff22}
\and
Universit\"ats-Sternwarte M\"unchen, Fakult\"at f\"ur Physik, Ludwig-Maximilians-Universit\"at M\"unchen, Scheinerstrasse 1, 81679 M\"unchen, Germany\label{aff23}
\and
Max Planck Institute for Extraterrestrial Physics, Giessenbachstr. 1, 85748 Garching, Germany\label{aff24}
\and
Caltech/IPAC, 1200 E. California Blvd., Pasadena, CA 91125, USA\label{aff25}
\and
Laboratoire univers et particules de Montpellier, Universit\'e de Montpellier, CNRS, 34090 Montpellier, France\label{aff26}
\and
Max-Planck-Institut f\"ur Astronomie, K\"onigstuhl 17, 69117 Heidelberg, Germany\label{aff27}
\and
ESAC/ESA, Camino Bajo del Castillo, s/n., Urb. Villafranca del Castillo, 28692 Villanueva de la Ca\~nada, Madrid, Spain\label{aff28}
\and
STAR Institute, University of Li{\`e}ge, Quartier Agora, All\'ee du six Ao\^ut 19c, 4000 Li\`ege, Belgium\label{aff29}
\and
Department of Physics, Centre for Extragalactic Astronomy, Durham University, South Road, Durham, DH1 3LE, UK\label{aff30}
\and
Department of Physics, Institute for Computational Cosmology, Durham University, South Road, Durham, DH1 3LE, UK\label{aff31}
\and
The Inter-University Centre for Astronomy and Astrophysics, Post Bag 4, Ganeshkhind, Pune 411007, India\label{aff32}
\and
Kavli Institute for the Physics and Mathematics of the Universe (WPI), University of Tokyo, Kashiwa, Chiba 277-8583, Japan\label{aff33}
\and
Minnesota Institute for Astrophysics, University of Minnesota, 116 Church St SE, Minneapolis, MN 55455, USA\label{aff34}
\and
Kapteyn Astronomical Institute, University of Groningen, PO Box 800, 9700 AV Groningen, The Netherlands\label{aff35}
\and
Department of Astronomy, School of Physics and Astronomy, Shanghai Jiao Tong University, Shanghai 200240, China\label{aff36}
\and
National Astronomical Observatory of Japan, 2-21-1 Osawa, Mitaka, Tokyo 181-8588, Japan\label{aff37}
\and
INAF-Osservatorio Astronomico di Capodimonte, Via Moiariello 16, 80131 Napoli, Italy\label{aff38}
\and
University of Trento, Via Sommarive 14, I-38123 Trento, Italy\label{aff39}
\and
Dipartimento di Fisica e Astronomia, Universit\`{a} di Firenze, via G. Sansone 1, 50019 Sesto Fiorentino, Firenze, Italy\label{aff40}
\and
INAF-Osservatorio Astrofisico di Arcetri, Largo E. Fermi 5, 50125, Firenze, Italy\label{aff41}
\and
Max Planck Institute for Gravitational Physics (Albert Einstein Institute), Am Muhlenberg 1, D-14476 Potsdam-Golm, Germany\label{aff42}
\and
Universit\'e Paris-Saclay, CNRS, Institut d'astrophysique spatiale, 91405, Orsay, France\label{aff43}
\and
School of Mathematics and Physics, University of Surrey, Guildford, Surrey, GU2 7XH, UK\label{aff44}
\and
INAF-Osservatorio Astronomico di Brera, Via Brera 28, 20122 Milano, Italy\label{aff45}
\and
Universit\'e Paris-Saclay, Universit\'e Paris Cit\'e, CEA, CNRS, AIM, 91191, Gif-sur-Yvette, France\label{aff46}
\and
IFPU, Institute for Fundamental Physics of the Universe, via Beirut 2, 34151 Trieste, Italy\label{aff47}
\and
INAF-Osservatorio Astronomico di Trieste, Via G. B. Tiepolo 11, 34143 Trieste, Italy\label{aff48}
\and
INFN, Sezione di Trieste, Via Valerio 2, 34127 Trieste TS, Italy\label{aff49}
\and
SISSA, International School for Advanced Studies, Via Bonomea 265, 34136 Trieste TS, Italy\label{aff50}
\and
Dipartimento di Fisica e Astronomia, Universit\`a di Bologna, Via Gobetti 93/2, 40129 Bologna, Italy\label{aff51}
\and
INAF-Osservatorio Astronomico di Padova, Via dell'Osservatorio 5, 35122 Padova, Italy\label{aff52}
\and
Institut de Physique Th\'eorique, CEA, CNRS, Universit\'e Paris-Saclay 91191 Gif-sur-Yvette Cedex, France\label{aff53}
\and
Space Science Data Center, Italian Space Agency, via del Politecnico snc, 00133 Roma, Italy\label{aff54}
\and
INAF-Osservatorio Astrofisico di Torino, Via Osservatorio 20, 10025 Pino Torinese (TO), Italy\label{aff55}
\and
Dipartimento di Fisica, Universit\`a di Genova, Via Dodecaneso 33, 16146, Genova, Italy\label{aff56}
\and
INFN-Sezione di Genova, Via Dodecaneso 33, 16146, Genova, Italy\label{aff57}
\and
Department of Physics "E. Pancini", University Federico II, Via Cinthia 6, 80126, Napoli, Italy\label{aff58}
\and
Instituto de Astrof\'isica e Ci\^encias do Espa\c{c}o, Universidade do Porto, CAUP, Rua das Estrelas, PT4150-762 Porto, Portugal\label{aff59}
\and
Faculdade de Ci\^encias da Universidade do Porto, Rua do Campo de Alegre, 4150-007 Porto, Portugal\label{aff60}
\and
Dipartimento di Fisica, Universit\`a degli Studi di Torino, Via P. Giuria 1, 10125 Torino, Italy\label{aff61}
\and
INFN-Sezione di Torino, Via P. Giuria 1, 10125 Torino, Italy\label{aff62}
\and
European Space Agency/ESTEC, Keplerlaan 1, 2201 AZ Noordwijk, The Netherlands\label{aff63}
\and
Institute Lorentz, Leiden University, Niels Bohrweg 2, 2333 CA Leiden, The Netherlands\label{aff64}
\and
Leiden Observatory, Leiden University, Einsteinweg 55, 2333 CC Leiden, The Netherlands\label{aff65}
\and
INAF-Osservatorio Astronomico di Roma, Via Frascati 33, 00078 Monteporzio Catone, Italy\label{aff66}
\and
INFN-Sezione di Roma, Piazzale Aldo Moro, 2 - c/o Dipartimento di Fisica, Edificio G. Marconi, 00185 Roma, Italy\label{aff67}
\and
Centro de Investigaciones Energ\'eticas, Medioambientales y Tecnol\'ogicas (CIEMAT), Avenida Complutense 40, 28040 Madrid, Spain\label{aff68}
\and
Port d'Informaci\'{o} Cient\'{i}fica, Campus UAB, C. Albareda s/n, 08193 Bellaterra (Barcelona), Spain\label{aff69}
\and
Institute for Theoretical Particle Physics and Cosmology (TTK), RWTH Aachen University, 52056 Aachen, Germany\label{aff70}
\and
INFN section of Naples, Via Cinthia 6, 80126, Napoli, Italy\label{aff71}
\and
Institute for Astronomy, University of Hawaii, 2680 Woodlawn Drive, Honolulu, HI 96822, USA\label{aff72}
\and
Dipartimento di Fisica e Astronomia "Augusto Righi" - Alma Mater Studiorum Universit\`a di Bologna, Viale Berti Pichat 6/2, 40127 Bologna, Italy\label{aff73}
\and
Instituto de Astrof\'{\i}sica de Canarias, V\'{\i}a L\'actea, 38205 La Laguna, Tenerife, Spain\label{aff74}
\and
Institute for Astronomy, University of Edinburgh, Royal Observatory, Blackford Hill, Edinburgh EH9 3HJ, UK\label{aff75}
\and
European Space Agency/ESRIN, Largo Galileo Galilei 1, 00044 Frascati, Roma, Italy\label{aff76}
\and
Universit\'e Claude Bernard Lyon 1, CNRS/IN2P3, IP2I Lyon, UMR 5822, Villeurbanne, F-69100, France\label{aff77}
\and
UCB Lyon 1, CNRS/IN2P3, IUF, IP2I Lyon, 4 rue Enrico Fermi, 69622 Villeurbanne, France\label{aff78}
\and
Mullard Space Science Laboratory, University College London, Holmbury St Mary, Dorking, Surrey RH5 6NT, UK\label{aff79}
\and
Departamento de F\'isica, Faculdade de Ci\^encias, Universidade de Lisboa, Edif\'icio C8, Campo Grande, PT1749-016 Lisboa, Portugal\label{aff80}
\and
Instituto de Astrof\'isica e Ci\^encias do Espa\c{c}o, Faculdade de Ci\^encias, Universidade de Lisboa, Campo Grande, 1749-016 Lisboa, Portugal\label{aff81}
\and
INAF-Istituto di Astrofisica e Planetologia Spaziali, via del Fosso del Cavaliere, 100, 00100 Roma, Italy\label{aff82}
\and
Aix-Marseille Universit\'e, CNRS/IN2P3, CPPM, Marseille, France\label{aff83}
\and
INFN-Bologna, Via Irnerio 46, 40126 Bologna, Italy\label{aff84}
\and
School of Physics, HH Wills Physics Laboratory, University of Bristol, Tyndall Avenue, Bristol, BS8 1TL, UK\label{aff85}
\and
FRACTAL S.L.N.E., calle Tulip\'an 2, Portal 13 1A, 28231, Las Rozas de Madrid, Spain\label{aff86}
\and
INFN-Sezione di Milano, Via Celoria 16, 20133 Milano, Italy\label{aff87}
\and
NRC Herzberg, 5071 West Saanich Rd, Victoria, BC V9E 2E7, Canada\label{aff88}
\and
Institute of Theoretical Astrophysics, University of Oslo, P.O. Box 1029 Blindern, 0315 Oslo, Norway\label{aff89}
\and
Jet Propulsion Laboratory, California Institute of Technology, 4800 Oak Grove Drive, Pasadena, CA, 91109, USA\label{aff90}
\and
Department of Physics, Lancaster University, Lancaster, LA1 4YB, UK\label{aff91}
\and
Felix Hormuth Engineering, Goethestr. 17, 69181 Leimen, Germany\label{aff92}
\and
Technical University of Denmark, Elektrovej 327, 2800 Kgs. Lyngby, Denmark\label{aff93}
\and
Cosmic Dawn Center (DAWN), Denmark\label{aff94}
\and
NASA Goddard Space Flight Center, Greenbelt, MD 20771, USA\label{aff95}
\and
Department of Physics and Helsinki Institute of Physics, Gustaf H\"allstr\"omin katu 2, 00014 University of Helsinki, Finland\label{aff96}
\and
Universit\'e de Gen\`eve, D\'epartement de Physique Th\'eorique and Centre for Astroparticle Physics, 24 quai Ernest-Ansermet, CH-1211 Gen\`eve 4, Switzerland\label{aff97}
\and
Department of Physics, P.O. Box 64, 00014 University of Helsinki, Finland\label{aff98}
\and
Helsinki Institute of Physics, Gustaf H{\"a}llstr{\"o}min katu 2, University of Helsinki, Helsinki, Finland\label{aff99}
\and
Centre de Calcul de l'IN2P3/CNRS, 21 avenue Pierre de Coubertin 69627 Villeurbanne Cedex, France\label{aff100}
\and
Laboratoire d'etude de l'Univers et des phenomenes eXtremes, Observatoire de Paris, Universit\'e PSL, Sorbonne Universit\'e, CNRS, 92190 Meudon, France\label{aff101}
\and
SKA Observatory, Jodrell Bank, Lower Withington, Macclesfield, Cheshire SK11 9FT, UK\label{aff102}
\and
University of Applied Sciences and Arts of Northwestern Switzerland, School of Computer Science, 5210 Windisch, Switzerland\label{aff103}
\and
Universit\"at Bonn, Argelander-Institut f\"ur Astronomie, Auf dem H\"ugel 71, 53121 Bonn, Germany\label{aff104}
\and
Universit\'e C\^{o}te d'Azur, Observatoire de la C\^{o}te d'Azur, CNRS, Laboratoire Lagrange, Bd de l'Observatoire, CS 34229, 06304 Nice cedex 4, France\label{aff105}
\and
Universit\'e Paris Cit\'e, CNRS, Astroparticule et Cosmologie, 75013 Paris, France\label{aff106}
\and
CNRS-UCB International Research Laboratory, Centre Pierre Binetruy, IRL2007, CPB-IN2P3, Berkeley, USA\label{aff107}
\and
Institut d'Astrophysique de Paris, 98bis Boulevard Arago, 75014, Paris, France\label{aff108}
\and
Aurora Technology for European Space Agency (ESA), Camino bajo del Castillo, s/n, Urbanizacion Villafranca del Castillo, Villanueva de la Ca\~nada, 28692 Madrid, Spain\label{aff109}
\and
Institut de F\'{i}sica d'Altes Energies (IFAE), The Barcelona Institute of Science and Technology, Campus UAB, 08193 Bellaterra (Barcelona), Spain\label{aff110}
\and
School of Mathematics, Statistics and Physics, Newcastle University, Herschel Building, Newcastle-upon-Tyne, NE1 7RU, UK\label{aff111}
\and
DARK, Niels Bohr Institute, University of Copenhagen, Jagtvej 155, 2200 Copenhagen, Denmark\label{aff112}
\and
Waterloo Centre for Astrophysics, University of Waterloo, Waterloo, Ontario N2L 3G1, Canada\label{aff113}
\and
Department of Physics and Astronomy, University of Waterloo, Waterloo, Ontario N2L 3G1, Canada\label{aff114}
\and
Perimeter Institute for Theoretical Physics, Waterloo, Ontario N2L 2Y5, Canada\label{aff115}
\and
Centre National d'Etudes Spatiales -- Centre spatial de Toulouse, 18 avenue Edouard Belin, 31401 Toulouse Cedex 9, France\label{aff116}
\and
Institute of Space Science, Str. Atomistilor, nr. 409 M\u{a}gurele, Ilfov, 077125, Romania\label{aff117}
\and
Consejo Superior de Investigaciones Cientificas, Calle Serrano 117, 28006 Madrid, Spain\label{aff118}
\and
Universidad de La Laguna, Departamento de Astrof\'{\i}sica, 38206 La Laguna, Tenerife, Spain\label{aff119}
\and
Dipartimento di Fisica e Astronomia "G. Galilei", Universit\`a di Padova, Via Marzolo 8, 35131 Padova, Italy\label{aff120}
\and
INFN-Padova, Via Marzolo 8, 35131 Padova, Italy\label{aff121}
\and
Institut f\"ur Theoretische Physik, University of Heidelberg, Philosophenweg 16, 69120 Heidelberg, Germany\label{aff122}
\and
Institut de Recherche en Astrophysique et Plan\'etologie (IRAP), Universit\'e de Toulouse, CNRS, UPS, CNES, 14 Av. Edouard Belin, 31400 Toulouse, France\label{aff123}
\and
Universit\'e St Joseph; Faculty of Sciences, Beirut, Lebanon\label{aff124}
\and
Departamento de F\'isica, FCFM, Universidad de Chile, Blanco Encalada 2008, Santiago, Chile\label{aff125}
\and
Universit\"at Innsbruck, Institut f\"ur Astro- und Teilchenphysik, Technikerstr. 25/8, 6020 Innsbruck, Austria\label{aff126}
\and
Institut d'Estudis Espacials de Catalunya (IEEC),  Edifici RDIT, Campus UPC, 08860 Castelldefels, Barcelona, Spain\label{aff127}
\and
Satlantis, University Science Park, Sede Bld 48940, Leioa-Bilbao, Spain\label{aff128}
\and
Institute of Space Sciences (ICE, CSIC), Campus UAB, Carrer de Can Magrans, s/n, 08193 Barcelona, Spain\label{aff129}
\and
Department of Physics, Royal Holloway, University of London, TW20 0EX, UK\label{aff130}
\and
Instituto de Astrof\'isica e Ci\^encias do Espa\c{c}o, Faculdade de Ci\^encias, Universidade de Lisboa, Tapada da Ajuda, 1349-018 Lisboa, Portugal\label{aff131}
\and
Cosmic Dawn Center (DAWN)\label{aff132}
\and
Niels Bohr Institute, University of Copenhagen, Jagtvej 128, 2200 Copenhagen, Denmark\label{aff133}
\and
Universidad Polit\'ecnica de Cartagena, Departamento de Electr\'onica y Tecnolog\'ia de Computadoras,  Plaza del Hospital 1, 30202 Cartagena, Spain\label{aff134}
\and
Infrared Processing and Analysis Center, California Institute of Technology, Pasadena, CA 91125, USA\label{aff135}
\and
Dipartimento di Fisica e Scienze della Terra, Universit\`a degli Studi di Ferrara, Via Giuseppe Saragat 1, 44122 Ferrara, Italy\label{aff136}
\and
Istituto Nazionale di Fisica Nucleare, Sezione di Ferrara, Via Giuseppe Saragat 1, 44122 Ferrara, Italy\label{aff137}
\and
INAF, Istituto di Radioastronomia, Via Piero Gobetti 101, 40129 Bologna, Italy\label{aff138}
\and
INAF - Osservatorio Astronomico di Brera, via Emilio Bianchi 46, 23807 Merate, Italy\label{aff139}
\and
INAF-Osservatorio Astronomico di Brera, Via Brera 28, 20122 Milano, Italy, and INFN-Sezione di Genova, Via Dodecaneso 33, 16146, Genova, Italy\label{aff140}
\and
ICL, Junia, Universit\'e Catholique de Lille, LITL, 59000 Lille, France\label{aff141}
\and
ICSC - Centro Nazionale di Ricerca in High Performance Computing, Big Data e Quantum Computing, Via Magnanelli 2, Bologna, Italy\label{aff142}
\and
Instituto de F\'isica Te\'orica UAM-CSIC, Campus de Cantoblanco, 28049 Madrid, Spain\label{aff143}
\and
CERCA/ISO, Department of Physics, Case Western Reserve University, 10900 Euclid Avenue, Cleveland, OH 44106, USA\label{aff144}
\and
Laboratoire Univers et Th\'eorie, Observatoire de Paris, Universit\'e PSL, Universit\'e Paris Cit\'e, CNRS, 92190 Meudon, France\label{aff145}
\and
Departamento de F{\'\i}sica Fundamental. Universidad de Salamanca. Plaza de la Merced s/n. 37008 Salamanca, Spain\label{aff146}
\and
Universit\'e de Strasbourg, CNRS, Observatoire astronomique de Strasbourg, UMR 7550, 67000 Strasbourg, France\label{aff147}
\and
Center for Data-Driven Discovery, Kavli IPMU (WPI), UTIAS, The University of Tokyo, Kashiwa, Chiba 277-8583, Japan\label{aff148}
\and
California Institute of Technology, 1200 E California Blvd, Pasadena, CA 91125, USA\label{aff149}
\and
Department of Physics \& Astronomy, University of California Irvine, Irvine CA 92697, USA\label{aff150}
\and
Department of Mathematics and Physics E. De Giorgi, University of Salento, Via per Arnesano, CP-I93, 73100, Lecce, Italy\label{aff151}
\and
INFN, Sezione di Lecce, Via per Arnesano, CP-193, 73100, Lecce, Italy\label{aff152}
\and
INAF-Sezione di Lecce, c/o Dipartimento Matematica e Fisica, Via per Arnesano, 73100, Lecce, Italy\label{aff153}
\and
Instituto de Astrof\'isica de Canarias (IAC); Departamento de Astrof\'isica, Universidad de La Laguna (ULL), 38200, La Laguna, Tenerife, Spain\label{aff154}
\and
CEA Saclay, DFR/IRFU, Service d'Astrophysique, Bat. 709, 91191 Gif-sur-Yvette, France\label{aff155}
\and
Department of Computer Science, Aalto University, PO Box 15400, Espoo, FI-00 076, Finland\label{aff156}
\and
Instituto de Astrof\'\i sica de Canarias, c/ Via Lactea s/n, La Laguna 38200, Spain. Departamento de Astrof\'\i sica de la Universidad de La Laguna, Avda. Francisco Sanchez, La Laguna, 38200, Spain\label{aff157}
\and
Ruhr University Bochum, Faculty of Physics and Astronomy, Astronomical Institute (AIRUB), German Centre for Cosmological Lensing (GCCL), 44780 Bochum, Germany\label{aff158}
\and
Department of Physics and Astronomy, Vesilinnantie 5, 20014 University of Turku, Finland\label{aff159}
\and
Serco for European Space Agency (ESA), Camino bajo del Castillo, s/n, Urbanizacion Villafranca del Castillo, Villanueva de la Ca\~nada, 28692 Madrid, Spain\label{aff160}
\and
ARC Centre of Excellence for Dark Matter Particle Physics, Melbourne, Australia\label{aff161}
\and
Centre for Astrophysics \& Supercomputing, Swinburne University of Technology,  Hawthorn, Victoria 3122, Australia\label{aff162}
\and
Department of Physics and Astronomy, University of the Western Cape, Bellville, Cape Town, 7535, South Africa\label{aff163}
\and
DAMTP, Centre for Mathematical Sciences, Wilberforce Road, Cambridge CB3 0WA, UK\label{aff164}
\and
Kavli Institute for Cosmology Cambridge, Madingley Road, Cambridge, CB3 0HA, UK\label{aff165}
\and
Department of Astrophysics, University of Zurich, Winterthurerstrasse 190, 8057 Zurich, Switzerland\label{aff166}
\and
IRFU, CEA, Universit\'e Paris-Saclay 91191 Gif-sur-Yvette Cedex, France\label{aff167}
\and
Oskar Klein Centre for Cosmoparticle Physics, Department of Physics, Stockholm University, Stockholm, SE-106 91, Sweden\label{aff168}
\and
Astrophysics Group, Blackett Laboratory, Imperial College London, London SW7 2AZ, UK\label{aff169}
\and
Univ. Grenoble Alpes, CNRS, Grenoble INP, LPSC-IN2P3, 53, Avenue des Martyrs, 38000, Grenoble, France\label{aff170}
\and
Dipartimento di Fisica, Sapienza Universit\`a di Roma, Piazzale Aldo Moro 2, 00185 Roma, Italy\label{aff171}
\and
Centro de Astrof\'{\i}sica da Universidade do Porto, Rua das Estrelas, 4150-762 Porto, Portugal\label{aff172}
\and
HE Space for European Space Agency (ESA), Camino bajo del Castillo, s/n, Urbanizacion Villafranca del Castillo, Villanueva de la Ca\~nada, 28692 Madrid, Spain\label{aff173}
\and
Dipartimento di Fisica - Sezione di Astronomia, Universit\`a di Trieste, Via Tiepolo 11, 34131 Trieste, Italy\label{aff174}
\and
Department of Astrophysical Sciences, Peyton Hall, Princeton University, Princeton, NJ 08544, USA\label{aff175}
\and
Theoretical astrophysics, Department of Physics and Astronomy, Uppsala University, Box 515, 751 20 Uppsala, Sweden\label{aff176}
\and
Mathematical Institute, University of Leiden, Einsteinweg 55, 2333 CA Leiden, The Netherlands\label{aff177}
\and
Institute of Astronomy, University of Cambridge, Madingley Road, Cambridge CB3 0HA, UK\label{aff178}
\and
Space physics and astronomy research unit, University of Oulu, Pentti Kaiteran katu 1, FI-90014 Oulu, Finland\label{aff179}
\and
Department of Physics and Astronomy, Lehman College of the CUNY, Bronx, NY 10468, USA\label{aff180}
\and
American Museum of Natural History, Department of Astrophysics, New York, NY 10024, USA\label{aff181}
\and
Center for Computational Astrophysics, Flatiron Institute, 162 5th Avenue, 10010, New York, NY, USA\label{aff182}
\and
Department of Physics and Astronomy, University of British Columbia, Vancouver, BC V6T 1Z1, Canada\label{aff183}}    
%
%
%
%
%
%
\abstract{
Strong gravitational lensing has the potential to provide a powerful probe of astrophysics and cosmology, but fewer than 1000 strong lenses have been confirmed so far. With a \ang{;;0.16} resolution covering a third of the sky, the \Euclid telescope will revolutionise the identification of strong lenses, with \num{170000} lenses forecasted to be discovered amongst the 1.5 billion galaxies it will observe. We present an analysis of the performance of five machine-learning models at finding strong gravitational lenses in the quick release of \Euclid data (Q1) covering 63\,deg$^{2}$. The models have been validated by citizen scientists and expert visual inspection. We focus on the best-performing network: a fine-tuned version of the \texttt{Zoobot} pretrained model originally trained to classify galaxy morphologies in heterogeneous astronomical imaging surveys. Of the one million Q1 objects that \texttt{Zoobot} was tasked to find strong lenses within, the top 1000 ranked objects contain 122 grade A lenses (almost-certain lenses) and 41 grade B lenses (probable lenses). A deeper search with the five networks combined with visual inspection yielded 250 (247) grade A (B) lenses, of which 224 (182) are ranked in the top \num{20000} by \texttt{Zoobot}. When extrapolated to the full \Euclid survey, the highest ranked one million images will contain \num{75000} grade A or B strong gravitational lenses. 
}

%
%
%
   \titlerunning{Euclid Q1: Finding strong lenses with machine learning}
   \authorrunning{Euclid Collaboration: Lines et. al.}
   
   \maketitle
%
%
%
%
   
\section{Introduction}
{Covering approximately} \num{14000}\,deg$^2$, the Euclid Wide Survey 
{\citep[{EWS;}][]{EuclidSkyOverview}} has the power to revolutionise strong gravitational lensing by galaxies: the  \num{170000} strong lenses \Euclid is forecasted to discover will provide a sample two orders of magnitude larger than what was previously known \citep{collett15}. Whilst many telescopes on the ground have surveyed large areas of sky, no previous wide area survey has ever had sufficient resolution to resolve multiple images of a single background source around a typical lens galaxy. The Einstein radius distribution of galaxy-galaxy lenses in the Universe is expected to peak at \mbox{\ang{;;0.4}} \citep{collett15}, and many of these lenses are within reach of \Euclid with its {\ang{;;0.16}} full width at half maximum (FWHM) in the \IE filter \citep{Q1-TP002}.

With exceptional resolution comes exceptional challenges of data volume. \Euclid will {observe} around 1.5 billion unlensed galaxies \citep{Scaramella-EP1} in its \num{14000}\,deg$^2$, and manually searching these is intractable. Even the largest citizen science lens search projects have only inspected up to around one million images \citep{SpaceWarpsI, SpaceWarpsII, SpaceWarpsHSC, jimena}. To deliver on the promise of \Euclid's strong lens finding capabilities, it is clear that automated methods must be employed to pre-screen candidates.

Recently, machine learning has gained substantial traction in the field of strong lens finding, with many teams developing machine-learning-based classifiers \citep{Metcalf2018}. \citet{Lanusse} and \citet{Avestruz} developed deep convolutional neural network (CNN) architectures that perform extremely well on simulated lenses {\citep{Metcalf2018}}, whilst \citet{Jacobs17} trained a CNN to find $14$ lens candidates per square degree in the Canada--France--Hawaii Telescope (CFHT) data. CNNs have now been successfully employed to find lenses in the Dark Energy Survey (DES; \citealt{Jacobs2018, Jacobs2019,rojas22}), \textcolor{black}{the Dark Energy Spectroscopic Instrument (DESI) Legacy Survey \citep{desihuang2020,desihuang2021,desistorfer2024}}, the Kilo-Degree Survey (KiDS; \citealt{petrillo2017,petrillo--19,li2020,li2021,nagam23,nagam24,Grespan24}), 
Panoramic Survey Telescope and Rapid Response System (PanSTARRS; \citealt{Canameras20}), the Ultraviolet Near-Infrared Optical Northern Survey (UNIONS; \citealt{Savary2021}), Hyper Suprime-Cam (HSC; \citealt{shu2022,Jaelani2024}), \textcolor{black}{and \HST(HST) data \citep{hstlensflow}}. More recently, \citet{jimena} showed that multi-class vision transformers are powerful alternatives to CNNs.

Machine-learning approaches to strong lens finding are data intensive, requiring large samples of positive and negative example sets. Negative training sets are relatively easy to construct, since almost all astronomical objects are not strongly lensed. Constructing a positive training set is much more challenging, since the known lens sample is small and the observations heterogeneous. Faced with this problem, almost all strong lens finders have relied on simulations of lenses to act as positive training sets, using either pure simulations or lensed sources painted on top of real non-lensing luminous red galaxies (LRGs). These approaches allow for the creation of very large positive training sets, but carry a substantial risk that the machine-learning models may only learn peculiarities {from the simulation process,} rather than the true features that distinguish a lens from a non-lens.

The Euclid Quick {Data} Release 1 (Q1; \citealt{Q1cite}) data set provides the first opportunity to search for lenses in a homogeneous data set covering a large area (63\,deg$^2$) with {\mbox{\ang{;;0.16}}} resolution.
Prior to this, the only \Euclid data available for finding lenses has been the \Euclid Early Release Observations (ERO; \citealt{EROcite}) data, which covered an area of 8.12\,${\rm deg}^2$. With {these} data, 
an area of 0.7\,${\rm deg}^2$ was {entirely visually inspected down to $\IE = 22.5$} by experts, and three grade A and 13 grade B lens candidates were found \citep{ELSE24}. These lenses were used to test machine-learning models at finding lenses in \Euclid \citep{pearce-casey24}, and these models were then applied to the rest of the ERO data, resulting in 14 grade A and 31 grade B strong lens candidates being found \citep{Nagam25}. 

This paper describes the efforts within the Euclid Consortium to train machine-learning models to discover strong lenses in Q1, which is part of the five-paper series `The Strong Lensing Discovery Engine'. The series consists of the overview paper \citet{Q1-SP048}, which includes a description of all the lenses found in Q1;
\citet{Q1-SP052}, which describes our effort to build a strong lens training set for Q1; this paper, which describes the machine-learning lens finding effort; \citet{Q1-SP054}, which analyses the double-source-plane lens (DSPL) candidates {that} were discovered; and \citet{Q1-SP059}, which explores how to combine multiple lens-finding classifiers into an ensemble classifier for future lens searches.

{The present} paper is structured as follows. In Sect. \ref{sec:data} we discuss the data used in this work, which consists of the \Euclid Q1 data (Sect. \ref{ssec:q1-data}) and the machine-learning training data (Sect. \ref{ssec:training-data}). In Sect. {\ref{sec:zoobot}} we describe the \texttt{Zoobot} machine-learning model {and present what was learnt from optimising \texttt{Zoobot} to find lenses}. In Sect. {\ref{sec:other-models}} we explore the other machine-learning models used in this work. The results are presented in Sect. {\ref{sec:results}}, which include the results from a preliminary inspection of each model's top 1000 ranked objects (Sect. \ref{ssec:GJ}), the results from combining machine learning with citizen science to find lenses in Q1 (Sect. \ref{ssec:SW}), and a statistical analysis of how many lenses there are in Q1 beyond the sample that was visually inspected (Sect. \ref{ssec:rest-of-Q1}). A discussion of what was found is presented in Sect. {\ref{sec:discussion}}, and the conclusions are presented in Sect. \ref{sec:conclusions}.

\begin{figure}
    \centering
    \includegraphics[width=1\linewidth]{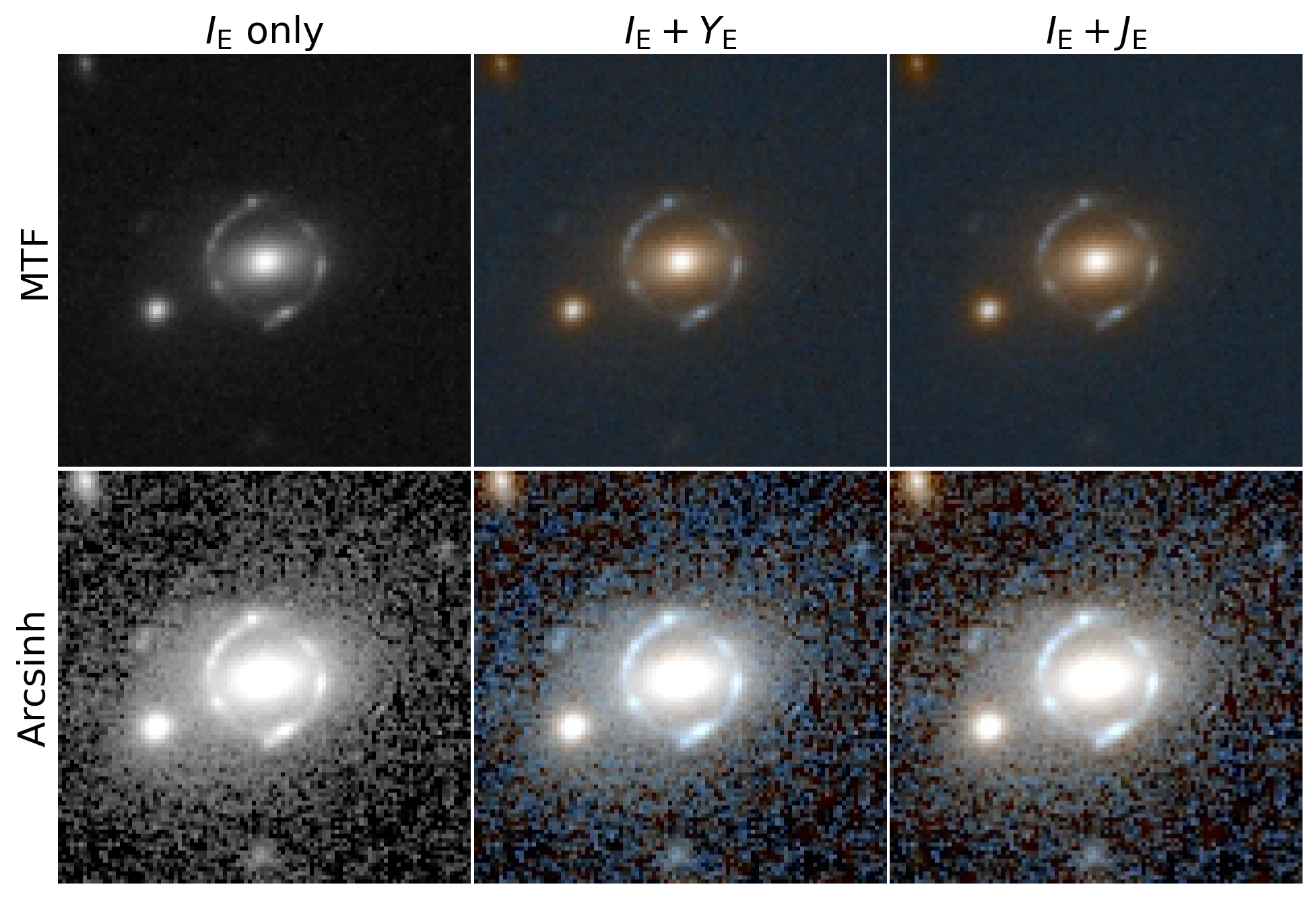}
    \caption{Different scaling and colour options for one sample cutout in Q1 processed into JPEG format. The top row is MTF scaling and the bottom row is $\arcsinh$ scaling. {From left to right, the columns show}  \IE-only, \IE+\YE, and \IE+\JE images.
    {The cutout size} is $100 \times 100$ pixels, or \mbox{$10'' \times 10''$} angular scale.}
    \label{fig:scalings}
\end{figure}

\section{Data} \label{sec:data}
\Euclid collects data with two instruments, the visible imager {\citep[VIS;][]{EuclidSkyVIS}} and the near-infrared spectrometer and photometer {\citep[NISP;][]{EuclidSkyNISP}}. 
There are four bands in total, the VIS \IE band and the NISP \YE, \JE, and \HE bands. 
Although \Euclid images are originally in Flexible Image Transport System (FITS) format, the quantity of  data meant that it was necessary to work with 
JPEG (Joint Photographic Experts Group)
files, which required a choice of scaling and filters. Three choices of filter combinations with two image scalings were chosen, producing a total of six combinations. The filter combinations used were \IE-only greyscale images, \IE+\YE colour images, and \IE+\JE colour images. 
{In the case of the colour images, the red (blue) channels correspond to the longest (shortest) wavelength images, and the green channel is the median of the two. All images have been produced at the highest resolution of the \Euclid bands, that of VIS \IE, at \mbox{\ang{;;0.101}} per pixel (see \citealt{Q1-SP048} for details).}
The two scalings chosen were $\arcsinh$, which provides a high contrast that amplifies both faint arcs and background noise, and the midtone-transfer function (MTF; see \citealt{Q1-SP048}), which produces a more homogeneous image {without saturating bright pixels}.
An example of these images for one cutout is shown in Fig. \ref{fig:scalings}.
{Although converting images from FITS to JPEG files necessarily means some information is lost, 
the full 16 bits of information per pixel in FITS files results in large quantities of mostly redundant information. Providing that the scaling to the 8-bit format of JPEG images is such that it preserves most of the dynamic range, the amount of information lost is made up for by the significantly faster processing time. We find that for both the brightest and the faintest of lenses (whether these be the lenses we find in Q1 or lenses found in previous searches), the lenses are generally as easy to distinguish in the JPEG images as in the FITS images.} 
These options were chosen based on the appearance of the Q1 data, but the same scalings were applied to simulations and other training and validation data to ensure homogeneity. 
All images have a size of $100\times 100$ pixels, corresponding to an angular scale of $10'' \times 10''$ in the \IE band. {This size allows the full Einstein ring of all galaxy-scale lenses to be visible but does not necessarily encompass group-scale lenses.}

\subsection{Q1 data} \label{ssec:q1-data}
\Euclid collects data for two main surveys: the {EWS} and the Euclid Deep Survey {(EDS)}. The former aims to survey around a third of the sky, covering the area that excludes the Galactic and ecliptic plane, amounting to \num{14000}\,${\rm deg}^2$, while the latter surveys three selected areas (Euclid Deep Field North, Euclid Deep Field South, and Euclid Deep Field Fornax) totalling $63\textrm{\,deg}^2$, but with repeated exposures producing {images that are up to two magnitudes deeper}. The \Euclid Q1 release covers the Deep Field areas but at the depth of the EWS. Q1 provides the first look into what we will be able to find within \Euclid's lifetime and provides us with the first sample of \Euclid lenses.

{The \Euclid data products we used were the background-subtracted merged mosaic tiles, in which the four images from the different \Euclid bands are matched to the resolution of VIS \IE. While \IE has a PSF FWHM of \ang{;;0.16} \citep{EuclidSkyVIS}, for \YE, \JE, and \HE the PSF FWHM is \ang{;;0.35}, \ang{;;0.34}, and \ang{;;0.35}, respectively \citep{EuclidSkyNISP} and the details of the interpolation are described in \citet{Q1-TP004}.}
Of all the objects in Q1, \num{1086554} passed our selection criteria of \IE $< 22.5$, \textcolor{black}{which was motivated by the findings from the lens search in the ERO data \citep{ELSE24}.} There was an additional quality check to remove objects such as stars (for further details see \citealt{Q1-SP048}). It was these remaining objects that were analysed {by the machine-learning models}. Using cutouts with size $10'' \times 10''$ means there will inevitably be overlap between images, but we ensure that the numbers of lenses reported has duplicate lenses removed.

\subsection{Training and validation data sets} \label{ssec:training-data}
For the Q1 lens finding challenge, any member of the Euclid Collaboration strong lensing working group was invited to develop a machine-learning model. 
Many of the models used in this work were built on models developed for the strong lens search in the \Euclid ERO data \citep{ELSE24, pearce-casey24,Nagam25}, which were the first investigations of the lens-finding ability of \Euclid.
Given the aim of finding as many lenses as possible, there were few restrictions placed on how the models should be developed; rather, they were all developed independently and evaluated on their Q1 lens-finding performance. 

{Machine-learning models are best able to recognise images that appear similar to the images they have been trained on, meaning ideally we want to train the models on real \Euclid lenses.}
However, the number of objects needed to train deep learning models (typically of the order $10^{4}$--$10^{5}$) greatly exceeds the number of known lenses. While pretrained models can greatly reduce the number of images that are needed to be trained on (of the order $10^{3}$), the {number of available} lenses is still too low to span the range of plausible strong lensing configurations. Therefore, we have to augment our set of known lenses with simulations on which to train the machine-learning models, which are then applied to the Q1 data.

{At the time of training the machine-learning models in preparation for Q1, there were multiple options of what images to use. The different data sets available consisted of} simulated lenses, known real lenses, and non-lenses. The different machine-learning models were developed independently and as such were free to choose any available data sets for training and validation of the models. All of these data sets were available as JPEG files in all the six combinations of scalings and filters, and it was left to the developers of each model to choose which of these to use for their final model. 
By allowing a diversity of options in terms of choice of training data and image processing, we are more likely to be able to find an optimal approach for lens finding in future \Euclid data releases, allowing us to later refine these options to be able to make more direct comparisons between models. {In the future when a large number of real \Euclid lenses are available, we will be able to quantify different machine-learning approaches in a fully data-driven way.}

\subsubsection{Simulated lenses} \label{ssec:sims}
The first set of simulations (S1) {was} based on the work presented in \citet{Q1-SP052} and {was} created as follows.
They simulated strong lensing systems using all four \Euclid bands, following the procedure outlined in \citet{rojas22} and utilising the \texttt{Lenstronomy} package \citep{birrer18,birrer21}. Deflectors were selected as LRGs with known redshifts and velocity dispersions, modelled using a Sérsic profile in the \JE band to estimate ellipticities and central positions. These parameters were optimised with a {downhill simplex} algorithm. Background sources were drawn from the HST Advanced Camera for Surveys F814W high-resolution catalogue \citep{Leauthaud2007,scoville2007,Koekemoer2007}, combined with HSC colour information \citep{Canameras20}. VIS magnitudes were {computed} by combining $r$-band + $i$-band images, while infrared magnitudes were matched to COSMOS 2020 VISTA (Visible and Infrared Survey Telescope for Astronomy) $YJH$ bands \citep{weaver2022}. They paired {lenses and sources} such that they produce an Einstein radius in the range \mbox{\ang{;;0.5}--\ang{;;2.0}}. A singular isothermal ellipsoid (SIE) mass model was adopted, with parameters derived from the Sérsic fit and Einstein radii calculated using the redshifts and velocity dispersions of the lens and source. Source images were lensed based on the mass model, and then downsampled to match \Euclid~{VIS \IE} image resolution. This last step included convolution with the telescope point-spread function (PSF) and scaling to the lens image flux. Finally, the resulting simulated lensing features were added into the VIS and NISP lens images. 
\textcolor{black}{The simulation pipeline did not include a deconvolution with HST PSF. This may produce a small impact in the smoothness and shape of the simulated arc that under Euclid PSF would be negligible, and would never result in unrealistic lensing configurations.}
Through augmenting the number of LRGs available for simulating by using four {\ang{90;;}} rotations of each LRG, the number of simulations was \num{11057}. {There was an additional earlier version of the same simulations containing \num{10311} simulations, making a total number of \num{21368} simulations available.}

{\color{black}
A second set of simulations (S2) was created using the \texttt{GLAMER} lensing code \citep{glamerm,glamerp}. These simulations started by cutting out $200\times 200$ pixel images of all the observed galaxies with \IE < 22, with some additional cuts to remove stars and reduce the number of face on spirals. 
The resulting sample of galaxies exhibited a range of morphologies, including elliptical and disc galaxies, and reflect the diversity of galaxy morphologies in the Universe.} Each image was matched to an object in the Flagship simulation \citep{EuclidSkyFlagship} with a nearest-neighbour algorithm in the space of all magnitudes in all four bands, ellipticity, and redshift, when available for the observed galaxy. The parameters of the Flagship galaxy and dark matter halo were then used to construct a mass model for the lens. The mass model consisted of a baryonic part in the form of a Sérsic profile and a dark matter halo in the form of a Navarro--Frenk--White profile \citep{nfw}. The position angle was set to that measured in the data. A source redshift was drawn from a distribution fit to the COSMOS 2020 \texttt{FARMER} catalogue $I$-band number counts 
\citep{weaver2022}. Rays were then shot through the mass profile to the source redshift. The caustics and critical curves were found. The lens was rejected if the Einstein radius was less than \ang{;;0.5}. A source was then selected and placed near or in the tangential caustic. {\color{black} The source's surface brightness was represented by between one and four Sérsic profiles. To emulate some of the complex morphologies of high-redshift sources, the number of Sérsic profiles, their relative brightnesses, and their positions were generated randomly from distributions tuned by eye. Their total brightnesses and overall sizes were based on randomly selected sources from the Hubble Ultra Deep Field with similar redshifts (see \citealt{2008AandA…482..403M,2010AandA…514A..93M} for a discussion of this sample). The source redshifts were set to five, allowing the distribution of lens redshifts and mass parameters as well as source sizes and brightnesses to provide the variation in lens properties.} The image of the lensed object was then convolved with the local PSF, and Poisson noise was added. {The simulations were performed at $4 \times$ the resolution of VIS \IE, then downsampled to the respective resolutions for VIS and NISP before being convolved with the PSFs. The NISP images were then interpolated to the higher VIS \IE resolution using a \texttt{BILINEAR} interpolation kernel, similar to the processing of the real \Euclid data.} 
Another rejection step was taken after combining the lensed image with the original data image. Cases where the lensed image was too low in signal-to-noise or in contrast to the lens galaxy light were removed {when visually inspecting the images}. In these simulations, the mass of neighbouring objects was not included. See Metcalf {et al.} (2025. in prep.) for more details on these and related simulations. {5363} images were created without any augmentation. 

A third set of {\IE-only} simulations (S3) was created, implementing the code presented by \citet{schuldt21}.
To create galaxy-scale lens simulations, \textcolor{black}{the same LRGs as in simulation set S1 were used.} The lens mass distribution was modelled with {an} SIE profile, with the following input parameters: lens redshift, source redshift, and velocity dispersion. The lens centroid, axis ratio, and position angle were derived from the first and second brightness moments of the \Euclid VIS images. \textcolor{black}{The source galaxies originated from the catalogue presented in \citet{schuldt21}, using images of high-redshift galaxies from the Hubble Ultra Deep Field \citep{inami2017}.}
Sources were randomly placed in the source plane, restricting the simulations to those with an Einstein radius between \mbox{\ang{;;0.5}} and \ang{;;2.5}. Subsequently, the sources were lensed in the image plane using the {\tt GLEE} software \citep{suyu10,suyu12}. Finally, the simulated lensed source image was convolved with the \Euclid PSF, rescaled according to \Euclid’s zero points, and combined with the original LRG image. The final simulations were given as images of \mbox{\ang{;;19}$\times$ \ang{;;19}}, with $2143$ simulations produced.\footnote{We note that due to the nature of the internal Euclid Collaboration Q1 timeline, this set of simulations was made available after many machine-learning models had started to be developed and hence was employed by fewer models.}

\begin{figure}
    \centering
    \includegraphics[width=1\linewidth]{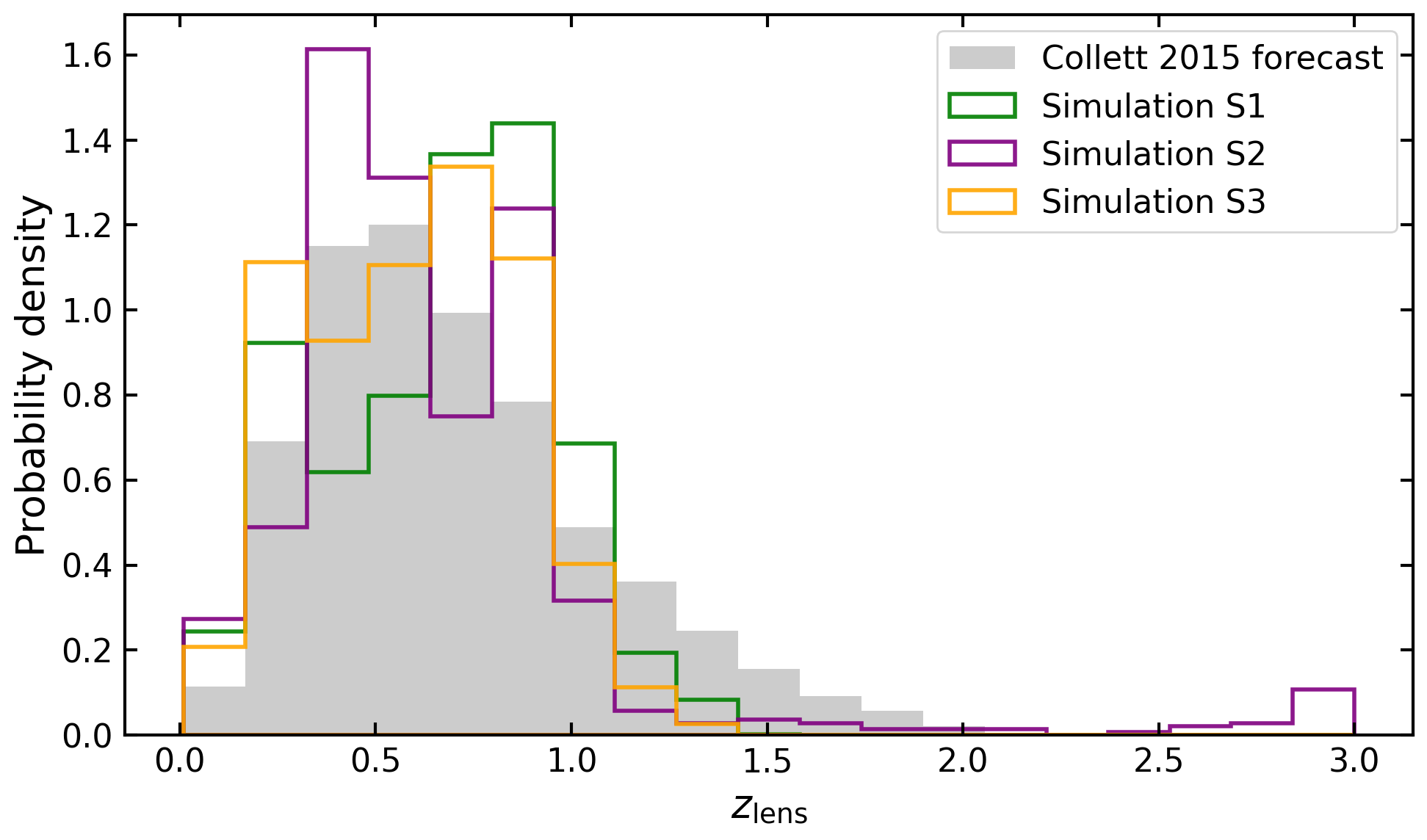}
    \caption{\textcolor{black}{Distribution of the redshifts of the lens galaxies used in simulation sets S1, S2, and S3, along with the distribution forecasted by \citet{collett15}.}}
    \label{fig:zl-hist}
\end{figure}

\begin{figure}
    \centering
    \includegraphics[width=1\linewidth]{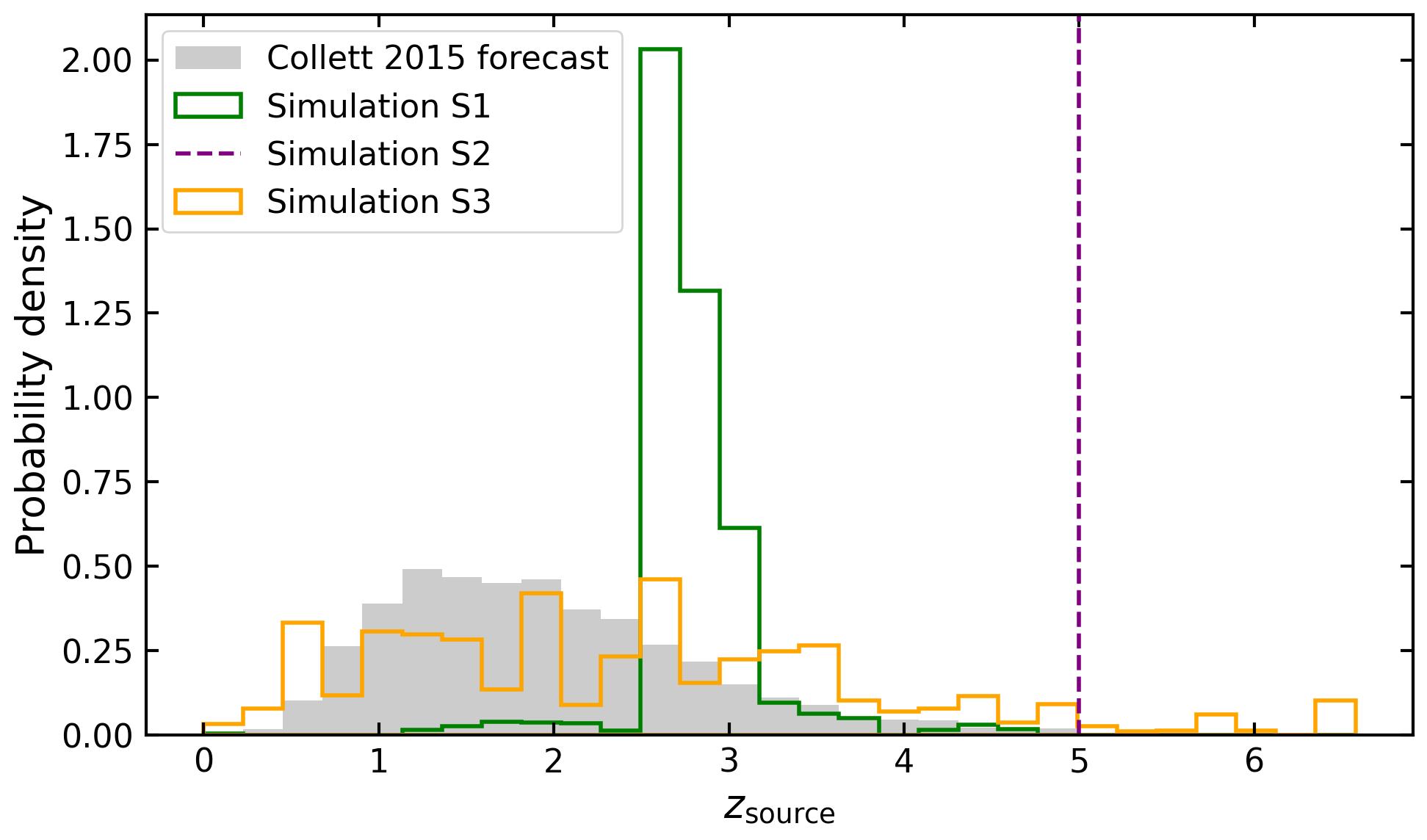}
    \caption{\textcolor{black}{Distribution of the redshifts of the source galaxies used in simulation sets S1, S2, and S3, along with the distribution forecasted by \citet{collett15}.}}
    \label{fig:zs-hist}
\end{figure}

\begin{figure}
    \centering
    \includegraphics[width=1\linewidth]{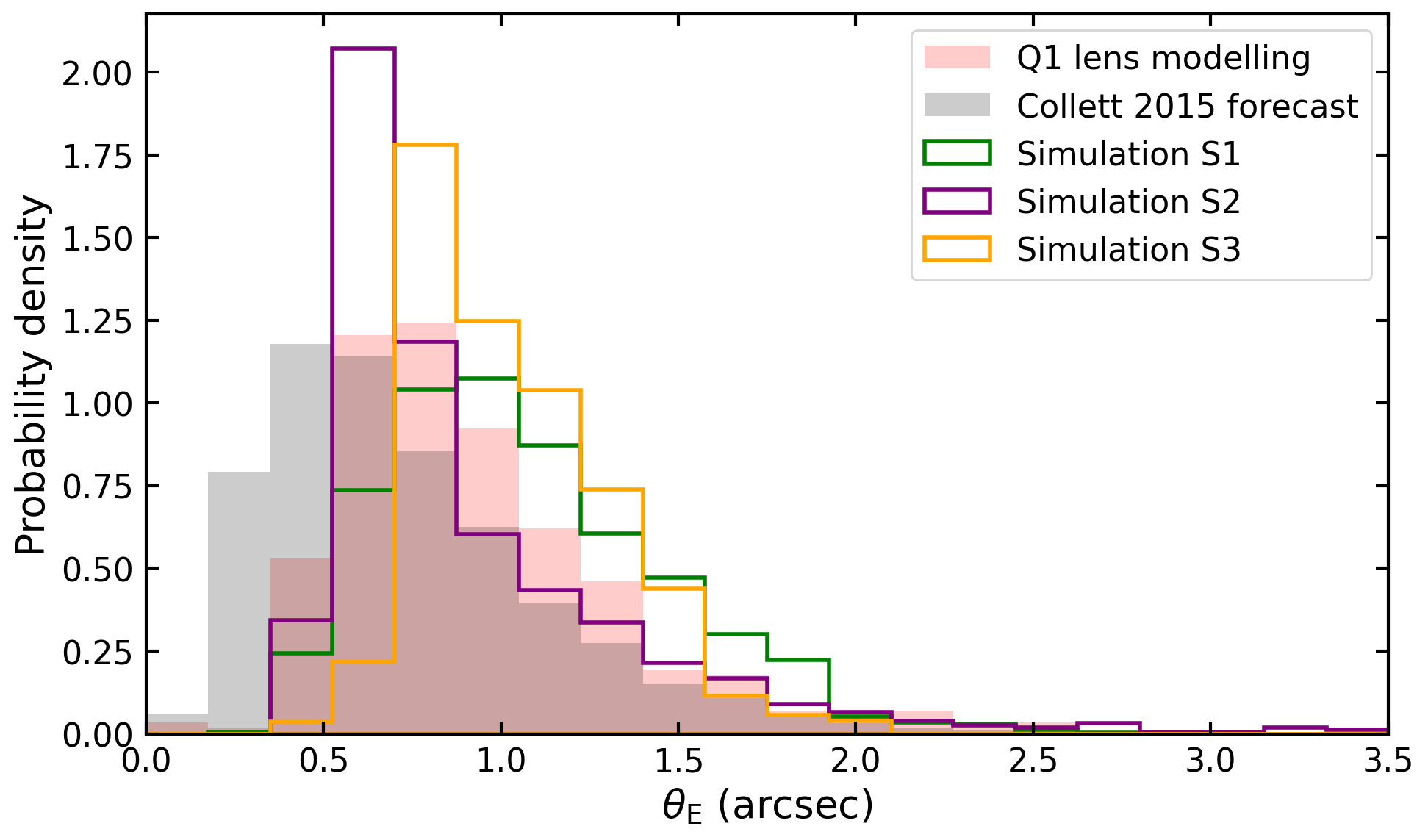}
    \caption{\textcolor{black}{Distribution of the Einstein radii in simulation sets S1, S2, and S3, along with the distribution forecasted by \citet{collett15}, and the Einstein radii inferred from the successful automated modelling of 336 Q1 lenses.}}
    \label{fig:thetaE-hist}
\end{figure}

All of the simulations used for training employed an approach of painting simulated {lensed sources} on to real galaxies, as opposed to a fully simulated approach. We used three sets of simulations to increase training diversity. 
\textcolor{black}{The distributions of redshifts of the lenses and sources from the simulations are shown in Fig. \ref{fig:zl-hist} and Fig. \ref{fig:zs-hist} respectively, where they are compared to the distributions predicted by \citet{collett15}. These distributions mostly align for the lens redshifts, while there is much more variation in source redshift distribution.}
Two simulation sets (S1 and S3) painted {lensed sources} on to galaxies with spectroscopically measured velocity dispersions, improving realism at the cost of using only bright LRGs as lenses (\citealt{Q1-SP052}, Melo et al. in prep.). As high velocity dispersion LRGs were selected for this, the resulting simulations had disproportionately large Einstein radii.
\textcolor{black}{The distributions of Einstein radii of the simulations are compared to the results from the automated modelling of the Q1 lenses and the forecast by \citet{collett15} in Fig. \ref{fig:thetaE-hist}. Simulations S1 and S3 tend to produce larger Einstein radii than S2, and all simulation sets have larger radii than predicted by the forecast. The distribution from Q1 lens modelling aligns reasonably well with the simulations but is shifted to larger radii compared to the forecast. This suggests that the simulations could be biasing the machine learning models towards detecting lenses with larger Einstein radii. However, there are the additional effects that humans are better at recognising lenses with larger Einstein radii \citep{Q1-SP047}, and that automated lens modelling may be more likely to succeed for such systems, both of which could also contribute to a bias in the distribution.}
The simulation set S2 used proxies to avoid {using} spectra, resulting in potentially less physically realistic simulations but meaning that \textcolor{black}{simulations could be produced around a wider range of galaxy morphologies and masses without being constrained by spectroscopic data availability} (Metcalf 2025 in prep.).
{Only S1 and S2 produced simulations in the NISP bands. S1 produced the NISP images directly at the higher VIS \IE resolution, while S2 produced the NISP simulations at the lower resolution and then interpolated this to the resolution of VIS \IE, taking into account the correlated noise that exists in the real \Euclid NISP images.}

\subsubsection{{Previously} known real lenses}
Prior to this lens search, there were a number of previously known lenses coinciding with areas of \Euclid coverage,
{which were ideal for training on as they most closely resemble the lenses in Q1.}
A data set of {approximately $150$} previously known grade A and B lenses \textcolor{black}{that had been visually inspected with available \Euclid data} was used in training and validation of the machine-learning models, but {was not counted in the number of newly discovered lenses reported in this paper}. This collection of {$150$} lenses includes lenses discovered in the \Euclid ERO data \citep{ELSE24}, the expert visual inspection of galaxies with spectroscopic velocity dispersion \citep{Q1-SP052}, the Galaxy Zoo Cosmic Dawn project, other surveys that overlapped with Q1, and other serendipitous discoveries. Outside Q1, there were around 50 additional known grade A+B lenses in \Euclid discovered from the Galaxy Zoo \Euclid project, where volunteers searched the EWS, {giving a total number of approximately $200$ real \Euclid lenses available for training}. Around the time of the internal Q1 release, more lenses were discovered serendipitously, which were not discovered in time to add to our training sample, but which are collated in the Q1 lens data set presented in \citet{Q1-SP048}. 

As the number of known lenses is small and heterogeneous {due to the number of different searches they originate from}, training on these few known lenses alone may amplify any selection bias in the pre-existing lens sets. Therefore, these lenses were primarily used to supplement the simulated lenses, which are designed to cover a larger range of lensing configurations. {Some machine-learning models were trained on the combinations of simulations and real lenses, while others were trained on simulations alone and the real lenses were used to validate their performance. For details of how each machine-learning model was trained see Sect. \ref{sec:other-models}.}

\subsubsection{Non-lenses} \label{ssec:non-lenses}
When curating a data set of non-lenses for machine-learning training, there are two general approaches. Firstly, one can use a collection of random non-lenses, typically with a similar selection function {as} foreground galaxies used for the simulations, where the idea is that by training on data that resembles the population of objects in real data, it will extrapolate well to the data to which it is being applied. An alternative approach is to recognise objects that are more likely to be mistaken for lenses (such as spiral and ring galaxies) and train the models on these specifically. In \citet{canameras24}, it was found that the lens-finding performance of a machine-learning model depends significantly on the contents of the negative set, favouring the use of non-random negative sets. 

There were two sets of random cutouts, one of {objects} in Q1 with the same selection function as the \textcolor{black}{deflectors} used for S2 {(see Sect. \ref{ssec:sims})} containing \num{4106} objects, and one of {objects} with the same selection function as Q1 but from EWS tiles that lie outside the Q1 area.  
There was a set of labelled non-lenses, which originated from a visual inspection of objects with DESI data available \citep{Q1-SP052}. \textcolor{black}{These were a by-product of the visual inspection of high velocity dispersion galaxies that primarily functioned as a lens search, since lensing cross section scales as velocity dispersion to the fourth power, and additionally collected LRGs that could be used for lens simulations.} This provided a data set of {approximately} \num{2300} spiral galaxies, $250$ mergers, $60$ ring galaxies, \num{2700} LRGs, and $2300$ others. The DESI LRGs were the LRGs used for S1 and S3. \textcolor{black}{In the future, other catalogues of visually inspected \Euclid objects will also be available for machine learning training, such as that from \citet{Q1-SP047}, or the objects classified as non-lenses from this work.} Table \ref{tab:ml-data-sets} shows all the data sets available along with which machine-learning models used which sets.

\begin{table}
\centering
\caption{{Different available training data sets and which machine-learning models used them.}}
\label{tab:ml-data-sets}
{%
\begin{tabular}{llllll}
\hline
\hline
\multirow{2}{*}{Data set (size)} &            &            & \multicolumn{2}{l}{Model} &            \\ \cline{2-6} 
                                 & 1          & 2          & 3           & 4           & 5          \\ \hline
S1 (\num{11057} + \num{10311})   & \checkmark & \checkmark & \checkmark  & \checkmark  & \checkmark \\
S2 (\num{5363})                  & \checkmark & \checkmark & \checkmark  & \checkmark  & \checkmark \\
S3 (\num{2143})                  & \checkmark &            &             &             &            \\
Real lenses                      & \checkmark & \checkmark & \checkmark* & \checkmark* & \checkmark \\ \hline
DESI spirals ($\sim$ 2300)       & \checkmark & \checkmark & \checkmark  & \checkmark  & \checkmark \\
DESI mergers ($\sim$ 250)        & \checkmark & \checkmark & \checkmark  & \checkmark  & \checkmark \\
DESI rings ($\sim$ 60)           & \checkmark & \checkmark & \checkmark  & \checkmark  & \checkmark \\
DESI LRGs ($\sim$ 2700)          & \checkmark & \checkmark &             & \checkmark  & \checkmark \\
DESI others ($\sim$ 2300)        & \checkmark & \checkmark & \checkmark  &             &            \\
S2 random counterpart (4106)     & \checkmark & \checkmark & \checkmark  &             & \checkmark \\
EWS random cutouts               & \checkmark &            &             & \checkmark* &            \\ \hline
\end{tabular}%
}
\tablefoot{The size of each data set is shown in brackets, other than the EWS random cutouts and the real lenses, since the different models used different numbers of these.
Data sets marked with an asterisk were only used for validation and not training by the model.}
\end{table}

\section{\texttt{Zoobot}: Pretrained Bayesian CNN} \label{sec:zoobot}

\subsection{Model description} 
\texttt{Zoobot}\footnote{{https://github.com/mwalmsley/zoobot}} \citep{Walmsley2023} is a Bayesian CNN that is pretrained on 92 million morphological classifications of galaxies as part of the Galaxy Zoo citizen science project \citep{galzoo}. 
Using pretrained models such as \texttt{Zoobot} and fine-tuning them is an example of `transfer learning', whereby taking a tool that is already primed to solve a different but related problem and then adapting only a part of the model, it is possible to create a model that is optimised to solve a new problem in a way that is quicker and easier than creating a model from scratch \citep{walmsley24}. 
{In our case, we take the pretrained base \zb~CNN model with pre-determined weights and fine-tune it by freezing all the parameters except those in the final layers}. \textcolor{black}{These parameters reside in the last layer (the classification head) and the ConvNeXt blocks that precede it, and} are allowed to vary through training the CNN on new data such as images of strong lenses. The idea of using \texttt{Zoobot} as a base model for lens finding was first presented and tested by \citet{pearce-casey24}, where it was found to be the best{-}performing machine-learning model of the 20 models tested on \Euclid ERO data; this motivated the use of \texttt{Zoobot} for lens finding in Q1.

The base \texttt{Zoobot} model was not built to detect lenses originally, nor was it developed on \Euclid data. However, lens finding in \Euclid is still a morphological classification task using astronomical images. Many of the features of the latent space that \texttt{Zoobot} uses to distinguish the features of galaxy morphology are likely to be relevant to lens finding. While \texttt{Zoobot} will have seen very few images of lenses, it has been pretrained on a large number of non-lenses that are likely to be common in the \Euclid data set, making it primed to detect these.
Pretrained models are especially powerful in tasks such as strong lens finding where the lack of known lenses to train on is a significant issue; fine-tuning a model requires significantly less training data to learn from than a usual model would, and while the number of simulated lenses available is much greater than the number of real lenses, it is still limited. 
Using a pretrained model also greatly reduces the chance of overfitting as the model is already primed to pick out useful morphological features; \texttt{Zoobot} should already be able to infer what parts of an image are relevant {for} classification and what to ignore as background objects, artefacts, or stars. By keeping the first layers from the pretrained \texttt{Zoobot}, this also ensures that the model is encouraged to learn true morphological properties rather than other patterns that may arise in simulated lenses such as PSF orientation \citep{wilde22}. 

There are many different pretrained architectures of \texttt{Zoobot} available. The version employed in this work is the ConvNeXT-Nano version, since it performs well while being quick to train, with 15.6 million trainable parameters. The ConvNeXT architecture is based on the foundational ResNet-50 model, but `modernised', incorporating elements from vision transformers to improve performance while retaining the strengths of CNNs such as simplicity (see \citealt{convnext} for more details). There were both greyscale (one channel input) and colour (three channel input) pretrained versions of this model available and we used the respective version to match the type of images with which we fine-tune \texttt{Zoobot}. 

\subsection{{Fine-tuning approach}}
\begin{figure}
    \centering
    \includegraphics[width=1\linewidth]{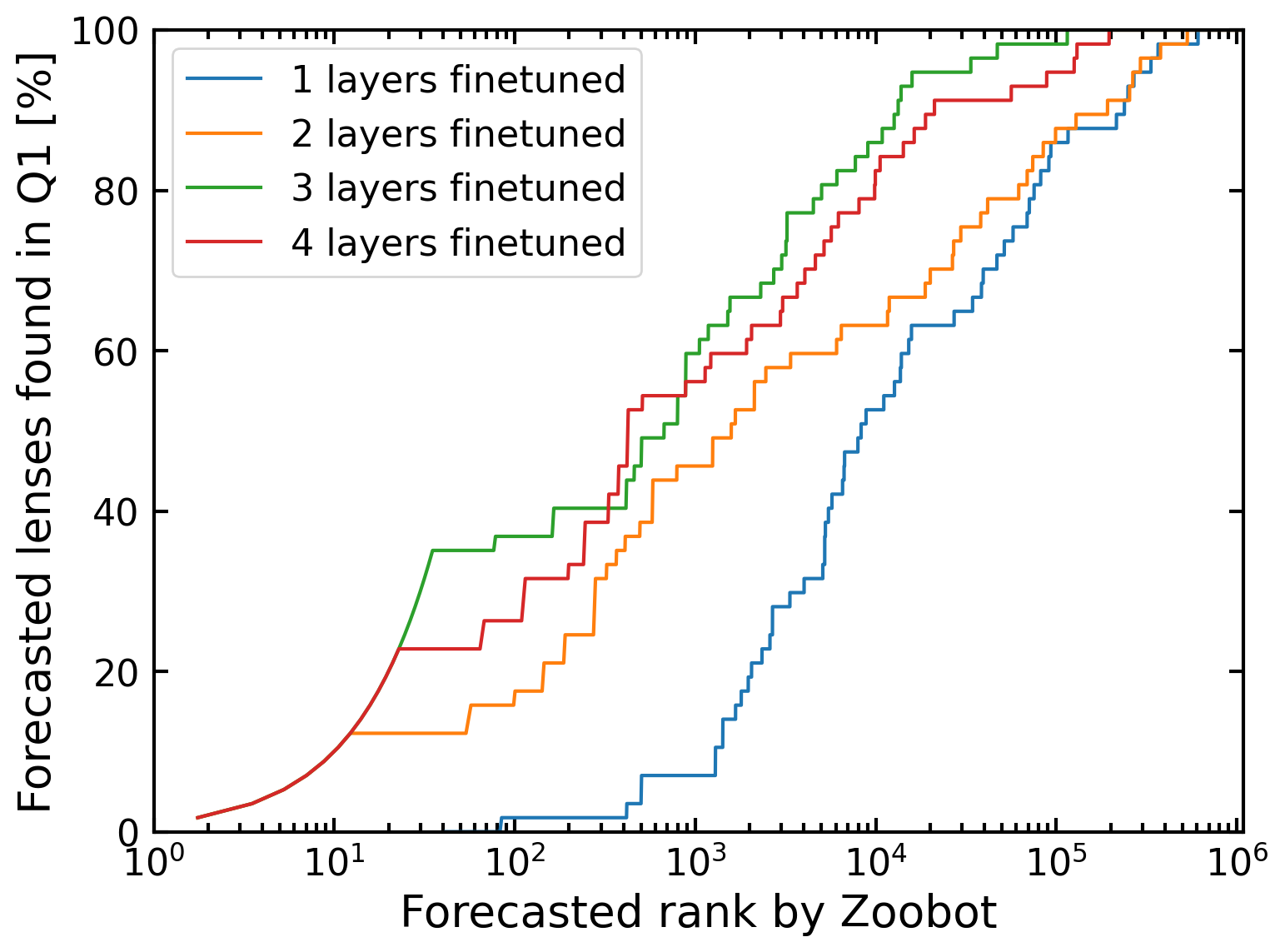}
    \caption{Forecasted performance of different versions of \texttt{Zoobot} when changing the number of layers that were fine-tuned. Numbers are based on the performance on the sample of known lenses in \Euclid and {cutouts from} random {EWS} tiles outside Q1, with the statistics extrapolated to predict the performance in Q1.}
    \label{fig:finetuned-layers}
\end{figure}

There were many different variables that could be adjusted to improve the performance of \texttt{Zoobot}{when fine-tuning it from a general morphology classifier to a strong lens finder}. These included the choice of training data, the choice of what types of images (scalings and filters) to use, and machine-learning hyperparameters (such as batch size and the number of layers fine-tuned). The framework that we used to build an optimal version of \texttt{Zoobot} involved selecting one choice for each of the mentioned options, fine-tuning this version of the model on the choice of training data, and then evaluating each version by its performance on real lenses and non-lenses. 
This consisted of 97 known grade A \Euclid lenses and 
a data set of {\num{26222} non-lens} \Euclid cutouts from {EWS} tiles outside Q1 (see Sect. \ref{ssec:non-lenses}).
Since the ratio of 97:{\num{26222}} does not reflect the proportion of lenses expected in real \Euclid data, we scale the statistics to match the ratio of expected lenses according to \citet{collett15}. For each fine-tuned version of \texttt{Zoobot}, we rank order all of the ${\num{26222}}+97$ cutouts. {To compare the different versions, we plot (e.g. see Fig.} \ref{fig:finetuned-layers}) the cumulative fraction of the 97 lenses recovered against the total number of images up to that \texttt{Zoobot} rank. This allows us to forecast the number of lenses we should find in Q1 as a function of how many objects we have to inspect.

The different versions of the models were compared according to the fraction of lenses recovered after inspecting \mbox{$0.01\%$--$0.1\%$} of the data. This corresponds to $1000$--$\num{10000}$ images over the entirety of Q1, roughly our expected expert and citizen scientist inspection capacity. {Typically, machine-learning models have their performance evaluated by zero-dimensional quantifiers, such as the area under receiver operating characteristic (ROC) curve. However, because we have a limited visual inspection budget we are only interested in the performance in a specific range; if we are only able to visually inspect $n$ images, it makes no difference if the lenses that were missed were ranked at $n+1$ or $n+\num{1000000}$. Therefore, using metrics that value the performance across the whole range of ranks equally is less useful than inspecting the performance in a range of ranks that corresponds to what can be visually inspected.}

\subsubsection{{Hyperparameter tuning}}
One hyperparameter that was specific to the pretrained model which was found to make a significant impact on performance was the number of layers that were fine-tuned. The more layers that are fine-tuned, the less the model relies on what was learnt during pretraining, but the more it can adapt to the new data. 
{In order to find the optimal number of layers to fine-tune, we allowed this to vary while keeping the other hyperparameters constant, and the performance of the different versions are plotted in Fig. \ref{fig:finetuned-layers}.}
We found that the number of lenses recovered at a certain rank increases as the number of fine-tuned layers increases up to three, but increasing the number of fine-tuned layers from three to four resulted in a decrease in the number of lenses recovered in the range of ranks we are interested in (specifically $10^{3}$--$10^{4}$). As such, we chose the version with the last three layers fine-tuned as the best model. This fine-tuning performance makes sense; fine-tuning too few layers requires the classifier to closely resemble the original \texttt{Zoobot} classifier, which is not designed for lens finding, whereas fine-tuning too many layers erases more of the useful features of astrophysical data that \texttt{Zoobot} has learnt from millions of Galaxy Zoo classifications.

\begin{figure}
    \centering
    \includegraphics[width=1\linewidth]{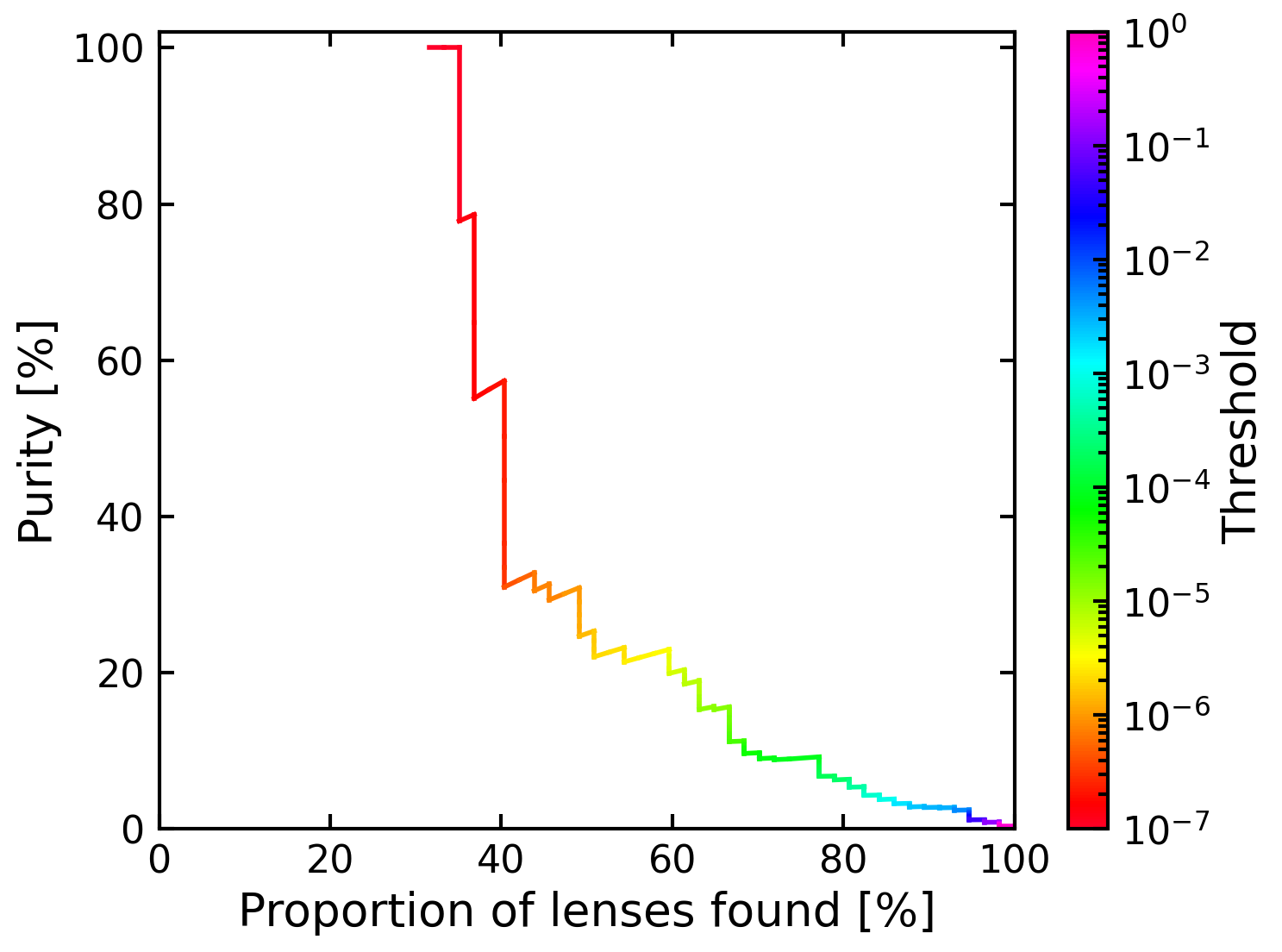}
    \caption{Purity and proportion of lenses found (or TPR) for lenses scored by \texttt{Zoobot} at different thresholds, as evaluated on the \textcolor{black}{test} data set of known lenses after being rescaled to the forecasted lens rate.}
    \label{fig:purity-tpr}
\end{figure}

Other choices such as batch size, learning rate, and choice of image scalings were also found to affect performance. {Similarly to the number of layers fine-tuned, we varied these hyperparameters and compared the performance of the different versions of the model to determine the optimal choice for each.}
In terms of image scalings, it was found that for \IE-only greyscale images, the $\arcsinh$ scaling resulted in a better performance than the MTF scaling, and the \IE-only greyscale images generally performed better than the multi-band colour images. {It was found that the optimal values of batch size and learning rate were 64 and $10^{-4}$ respectively, and these were used for the final version of the model for the Q1 lens search. However, it is likely that the optimal hyperparameters found here are specific to this training set and as such are likely to be different to what is optimal in future \Euclid lens searches when we will be able to train new versions of the machine learning models using the lenses found in Q1.}

\subsubsection{{Training data set selection}}
While the possibilities of combinations of data sets to use in training was large, we found that the best performance came when training \texttt{Zoobot} on a combination of S1 and S2 as the positives, and the best set of negatives was the labelled DESI non-lenses.
We also implemented the default \texttt{Zoobot} augmentation of images, which includes random rotations, vertical flips, and off-centre cropping and resizing.

We notably found that adding more sets of simulations to the training data did not always improve performance, demonstrating that for pretrained models, the performance is not necessarily constrained by the quantity of training data as much as the quality of it. However, for other models that are not pretrained and therefore typically require larger training sets, this may not be the case.

Additionally, we found that using a smaller but more {selective} data set of negatives, where each object had been labelled and the proportion of commonly misclassified false positives (such as ring galaxies and spiral galaxies) had been inflated, resulted in a performance that appeared to generalise better than a model trained on random cutouts. It was also found that training on these {labelled} non-lenses alone resulted in better performance than a combination of {labelled} non-lenses and random cutouts, again suggesting that a quality-over-quantity approach is preferable for pretrained models.

\begin{figure*}
    \centering
    \includegraphics[width=1\textwidth]{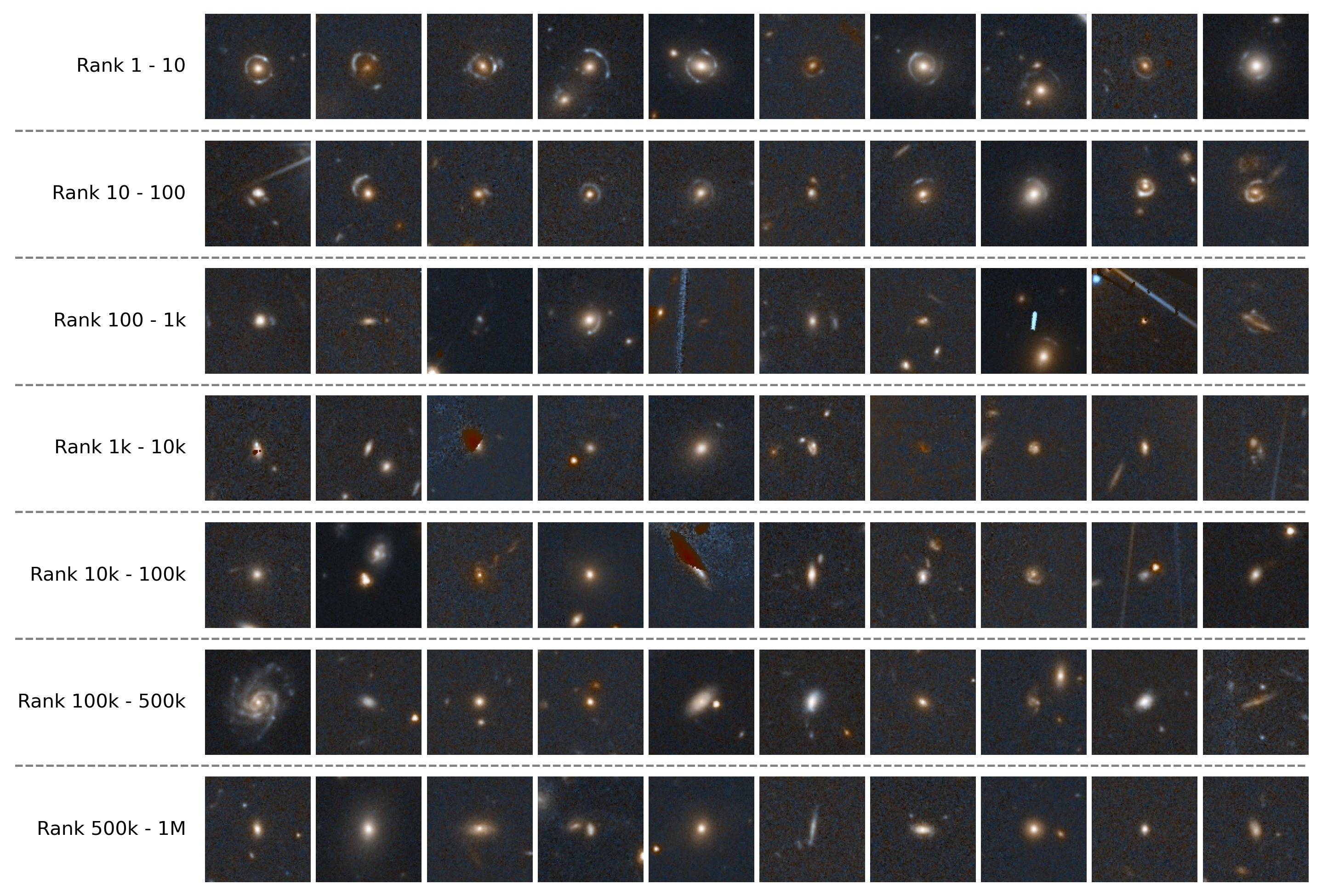}
    \caption{Random images ranked by \texttt{Zoobot} at different {ranges of ranks}.}
    \label{fig:ranked_grid}
\end{figure*}

\subsection{Results from fine-tuning}

To quantify the performance of a model, statistics that are of particular interest are {the} true positive rate (TPR), false positive rate (FPR), and purity {($P$)},\footnote{The TPR is also known as recall or sensitivity;
purity is also known as precision.
} given in terms of true positives ({$N_\sfont{TP}$}, number of {lenses that were predicted to be lenses}), false positives ({$N_\sfont{FP}$}, number of non-lenses {that were predicted to be lenses}), true negatives ({$N_\sfont{TN}$},  {number of non-lenses that were predicted to be non-lenses}), and false negatives ({$N_\sfont{FN}$}, {number of lenses that were predicted to be non-lenses}):

\begin{equation}
{\rm TPR} = \frac{{N_\sfont{TP}}}{{N_\sfont{TP} + N_\sfont{FN}}}~,
\end{equation}
\begin{equation}
{\rm FPR} = \frac{{N_\sfont{FP}}}{{N_\sfont{FP} + N_\sfont{TN}}}~,
\end{equation}
\begin{equation}
{{P}} = \frac{{N_\sfont{TP}}}{{N_\sfont{TP} + N_\sfont{FP}}}~.
\end{equation}
The TPR can be thought of as the fraction of real lenses that {the machine-learning model predicted to be} lenses, FPR is the fraction of non-lenses that {the machine-learning model predicted to be} lenses, and purity is the fraction of things that {the machine-learning model predicted to be} lenses that are real lenses.
Since machine-learning models generally do not produce a binary (lens or non-lens) output, but rather give objects a continuous score from 0 to 1 of how likely an object is to be a lens, the purity, TPR, and FPR vary according to a score threshold at which to split the objects into lenses and non-lenses. 
There is generally a trade-off between purity and TPR; by setting a {high score} threshold, {fewer} objects are classified as lenses and so the proportion of total lenses {found} is lower, {but as the score threshold is lower the purity decreases}. The purity at different TPR values for the best performing version of \texttt{Zoobot} is displayed in Fig. \ref{fig:purity-tpr}. As before, these statistics are calculated based on the performance on the known lenses and a set of random Q1 cutouts, with the purity scaled to take into consideration the expected ratio of lenses to non-lenses in such a sample.
We recovered an almost entirely pure sample of $35 \%$ of the lenses, but the purity falls to 20\% when recovering up to 60\% of the known lenses. As can be seen in Fig. \ref{fig:finetuned-layers}, the best version of \texttt{Zoobot} requires the first 10\% of all the data (score threshold in the range $0.2$--$1.0$) to be inspected to reach a TPR of 100\%. \textcolor{black}{Although the model was allowed to train for up to 100 epochs, early stopping was implemented if there was no improvement for ten consecutive epochs. The best performing version reached an optimal performance after 13 epochs, with each version taking under an hour to train on an NVIDIA A100 80GB GPU.}

When \texttt{Zoobot} was applied to the \Euclid ERO data in \citet{pearce-casey24}, \texttt{Zoobot} was found to perform better than the other models tested. However, they found that the purity of the lenses {discovered} was low, which may be partly due to the fact that the number of fine-tuning layers was set to one and was not allowed to vary. Additionally, since the \Euclid ERO lens-finding experiment there has been significant development in simulations of \Euclid lenses (e.g. \citealt{Q1-SP052}), which is likely to also contribute to a discrepancy in performance between \Euclid ERO and Q1 lens searches. Other factors, such as the limited amount of available \Euclid data at the time, training only on random cutouts as negatives instead of a curated set of non-lenses, as well as subsequent improvements to the \texttt{Zoobot} base models \citep{walmsley24}, may also contribute to this.

Figure \ref{fig:ranked_grid} displays random Q1 images at different samples of rankings according to \texttt{Zoobot}. The model is unable to perfectly split the sample into lenses and non-lenses, with non-lenses still contaminating the highly ranked images, meaning that visual inspection is still an essential part of finding strong lenses. However, it is reassuring that at the higher ranks, not only is the density of lenses higher, but it also appears that the lenses that are found are typically more obvious lenses, whether that is due to a high signal-to-noise ratio or a large Einstein radii creating a clear distinction between the lens and arc light. We do not find that the false positives are dominated by spirals, which has historically been a problem for lens finding. This is likely due to the fact that \texttt{Zoobot} had been pretrained to recognise a diversity of spiral morphologies, \textcolor{black}{along with the fine-tuning on high-resolution space-based imaging}. Rather, we find that other than a few artefacts, most of the false positives are chance alignment of galaxies; examples of highly ranked false positives by \texttt{Zoobot} are shown in Appendix \ref{sec:false-positives}.

\section{Other models} \label{sec:other-models}
In addition to \texttt{Zoobot}, four other models were developed to find lenses in \Euclid Q1.
In this section we provide a brief overview of each model and refer the reader to the original papers in which they were developed. The models are numbered as follows: Model 1 is a four-layer CNN; Model 2 is OU-100; Model 3 is IncNet; Model 4 is \texttt{Zoobot} (previously described); and Model 5 is LensVision. \textcolor{black}{There were no performance criteria for a model to be used for Q1 since it was hard to extrapolate from any pre-Q1 data how well these models would perform in practice. Rather, these models represent all the models that had been trained and were available at the time.} This does not provide an exhaustive list of all the machine-learning models that are employed to find lenses in \Euclid in general; additional networks were used in the lens search within the ERO data \citep{pearce-casey24}, but were not applied to the Q1 data due to time limitations. 

\subsection{Model 1: Four-layer CNN}
Model 1 is a shallow four-layer CNN, with the architecture used being an adaptation of the network developed for the morphological classification of galaxies in \citet{dominguezsanchez18}. It has already been tested with \Euclid-like simulations \citep{manjongarcia21} and with the ERO fields \citep{pearce-casey24,Nagam25}.
The network has four convolutional layers of different spacings ($6\times 6,~5\times5,~2\times2$, and $3\times3$,
respectively), and two fully connected layers. ReLU activation functions are applied after
every convolutional layer, and a $2\times2$ max-pooling is applied after the second and third convolutional layers. 

The model was trained on a combination of S1, S2, and S3, and also {80} known grade A and B \Euclid lenses as the set of positives. For the negatives, we used a combination of random Q1 galaxies matching the selection function of S2, all the LRGs selected for S1, the labelled DESI non-lenses, and also {7378} cutouts of random \Euclid galaxies from non-Q1 tiles. Two versions of the model were trained, both using $100\times 100$ \IE-only images with fluxes normalised from 0 to 1, but one used the MTF scaling and the other used the $\arcsinh$ scaling, with the final scores being the average of these two outputs. 

The training of the network was carried out following a supervised learning approach. {Decisions regarding} pre-processing, architecture, and learning parameters were influenced by what was found to work when the same model was applied to the ERO data in \citet{pearce-casey24}.
In the learning process, a binary cross-entropy loss function was used and was optimised by the Adam stochastic optimisation method \citep{Adam}, with a learning rate of $10^{-3}$. 
The models were trained for 40 epochs, using a batch size of 32. In order to prevent overfitting, \textcolor{black}{several real-time data augmentation techniques were performed randomly at each epoch during training}. These augmentations included allowing the images to be zoomed in and out (0.75 to 1.3 times the original size), rotated, and flipped and shifted both vertically and horizontally. The output of the final convolution layer was flattened and was inputted to a fully connected layer with dropout probability $p=0.5$. The final fully connected layer used a sigmoid activation to turn the output scores into lens probabilities.

\subsection{Model 2: OU-100}
The OU-100 model is another CNN-based model.
The network is a modification of OU-200 \citep{wilde22}, removing the final two convolutional layers due to the change in input size between the two models ($200 \times 200$ versus $100\times 100$ pixels). 
The architecture consists of four convolutional layers \textcolor{black}{with kernels of sizes} $5 \times 5,~5 \times 5,~2\times2$, and $2\times 2$, each followed by a $2\times2$ max pooling layer, a dropout layer with dropout probability $p=0.2$, and a ReLU activation function. This is followed by a fully connected layer with 350 neurons, followed by an output layer with a softmax activation function to produce a lens prediction.  

The positive training set included S1, S2 and all the {$200$} known grade A and B lenses. The negative training set consisted of the labelled DESI non-lenses, the remaining LRGs used for S1, and the random galaxies counterpart from S2.
A random selection of \num{40000} images were selected from the resulting \num{171478} images, using the four image options: \IE-only $\arcsinh$; \IE-only MTF; \IE+\YE MTF; and \IE+\JE MTF. 
Of each set of \num{40000} images used for training the different versions, \num{32000} images were used for training, \num{2000} used for validation, and \num{6000} for testing. The network was trained with a batch size of 100 and a learning rate of $3\times 10^{-4}$. A categorical loss is used to classify the images into lens and non-lens classes. Since the training data contains $12\%$ lenses, a weight was applied to give each class an equal weighting during training.

\textcolor{black}{The top 100 ranked images of the four versions of the model were visually inspected. For the \IE+\YE MTF and \IE+\JE MTF versions, the model picked out a few stars, as well as other features such as diffraction spikes and artefacts. The rest of the images were spirals or lens candidates.}
When comparing the top 100 ranked candidates from the \IE-only MTF and \IE-only $\arcsinh$ images, it was found that the MTF version appeared better at finding lenses. \textcolor{black}{The \IE-only $\arcsinh$ version had 34 stars in the top 100 images, as well as many very clear spirals. Based on the number of lens candidates and false positives in the top 100 images of each version, the MTF \IE-only version of the model performed the best, and this was used as the final model.}

\subsection{Model 3: IncNet}
The IncNet architecture is based on the `inception module' \citep{szegedy15} that applies convolutions of different size on the images in parallel, to extract features on different scales simultaneously and then combines them. The images are passed through two convolutional layers, alternated with max-pooling layers. After this, they go through seven inception modules. The fifth of them is connected to a classifier whose score is combined with the score of the final classifier (i.e. that at the end of the seven inception modules) to guide the weights update (refer to \citealt{2szegedy14} for a more detailed discussion). The output of the classifiers are weighted differently: 0.3 for the intermediate one; and 1.0 for the final one. This baseline architecture, presented in \citet{leuzzi24}, was further modified for the Q1 lens finding challenge to account for the input image sizes of $100\times 100$ pixels compared to the original $200\times 200$ image sizes it was optimised for, requiring a reduction in the number of filters per layer, while keeping the same number of layers.

For network training, a combination of S1 and S2 {was} used for the positive class. In the case of S2, the images were augmented by applying either a random rotation or flip to every image. For the negative class, a combination of random galaxies matching S2 and labelled DESI non-lenses were used, with the total number of training images being approximately $\num{40000}$.
For validating the network, a combination of known grade A and B lenses were used, in addition to a non-overlapping set of negatives from the training data. The images used were the \IE-only images with the $\arcsinh$ scaling.
During training, an ADAM optimiser was used with an initial learning rate of $10^{-4}$, \textcolor{black}{the \texttt{ReduceLROnPlateau} method was implemented to reduce the learning rate in case of overfitting,} and binary cross-entropy was used as the loss function. The model was trained for 100 epochs, and although early stopping was enabled in the training loop to prevent overfitting, it was not needed.

\subsection{Model 4: \texttt{Zoobot}}
Model 4 is \texttt{Zoobot}, which has been described previously. (Refer to Sect. \ref{sec:zoobot} for more details.)

\subsection{Model 5: LensVision}
The final model is the only non-convolutional network developed, instead opting for a vision transformer-based architecture, and similarly to \texttt{Zoobot} employs pretraining. It uses multi-step supervised contrastive learning and a classification head. Input images are divided into $16\times 16$ patches, which are flattened and projected into 768-dimensional embeddings. These embeddings are passed through a transformer encoder that processes the relationships between the patches.
The transformer encoder is pretrained using a two-step supervised contrastive learning technique. This pretraining helps the model to learn to separate lensing features from non-lensing features by pulling embeddings of similar images closer together and pushing embeddings of dissimilar images apart in the feature space to create a similarity matrix. 
After the encoder, a fully connected classification head takes the learned image representations and predicts whether the image contains a gravitational lens or not, represented by a score between 0 and 1. The classification head includes dense layers with activation functions and dropout to prevent overfitting. During training, the transformer layers are gradually unfrozen, allowing the model to fine-tune the earlier layers without losing the features learned during pretraining.

The data used for training includes S1 and S2 as well as a set of known grade A and B lenses for positives and the random galaxies matching S2 and the labelled DESI non-lenses as the negatives. This data set was augmented to increase the size by including rotations, \textcolor{black}{horizontal and vertical flips,} and corner cutouts \textcolor{black}{cropping the original $150\times150$ images to $100\times100$. These augmentations resulted in \num{116835} images in both the positive and negative class for training.}
The images used the $\arcsinh$ scaling on the \IE-only images.

\section{Results} \label{sec:results}
\begin{table}
\centering
\caption{Number of lenses of each grade and non-lenses found in each model's top 1000 ranked objects, along with the total number of lenses of each grade found in Q1 in total.}
\label{tab:model-rankings}
\resizebox{\columnwidth}{!}{%
\begin{tabular}{lllllll}
\hline
\hline
               &     &     & \multicolumn{2}{l}{Model} &     & \multirow{2}{*}{\begin{tabular}[c]{@{}l@{}}Total found\\ in Q1\end{tabular}} \\ \cline{2-6}
               & 1   & 2   & 3           & 4           & 5   &                                                                              \\ \hline
Grade A lenses & 69  & 24  & 8           & 122         & 3   & 250                                                                          \\
Grade B lenses & 34  & 12  & 5           & 41          & 4   & 247                                                                          \\
Grade C lenses & 46  & 22  & 1           & 67          & 10  & 585                                                                          \\
Non-lenses     & 851 & 942 & 986         & 770         & 983 & $\sim 10^{6}$                                                                \\ \hline
\end{tabular}%
}
\end{table}

Originally there were approximately $10^{6}$ Q1 cutouts which may have contained lenses, and only $10^{4}$--$10^{5}$ of these could be visually inspected based on the typical engagement in citizen-science projects.
Therefore we carried out visual grading of lenses in two rounds.
A first preliminary inspection allowed us to evaluate each model's performance in order to prioritise high-performing models in the selection of images for the second, larger visual inspection round. 

First, a group of strong lensing experts within the Euclid {C}ollaboration visually inspected the top \num{1000} ranked objects from each of the models 1--5, in {what is referred to as} the Galaxy Judges (GJ) project \citep{Q1-SP048}. This was to gain an initial understanding of how the models compared in lens finding, and was to inform which models should be prioritised in selecting which candidates were shown to citizens. 
Following this was the citizen science visual inspection, for which we created a \Euclid-specific version of the Space Warps (SW) project ran on the Zooniverse platform \citep{SpaceWarpsI, SpaceWarpsII}. All objects that were ranked highly by these citizens were then added to the GJ project and were ranked by experts, giving the final grades for all the interesting lens-like objects discovered.
In both the SW citizen science project and the GJ expert classifications, each image was inspected by multiple humans, with the results being aggregated in order to reduce noise and the level of subjectivity. 
For SW, the citizens were given the option of lens or not a lens, while in GJ the experts assigned a grade to quantify the quality of a lens. These grades were: X (not a lens); C (contains lens-like features but not obviously a lens); B (probably a lens); and A (confidently a lens). Examples of lenses with these grades are shown in Appendix \ref{sec:grade-abc}.
In this paper we focus on using the results from the visual inspection to quantify model performance; for a detailed analysis of the visual inspection process see \citet{Q1-SP048}.

Machine-learning models typically output scores which range from 0 to 1, with 1 corresponding to a high likelihood of being a lens (though not strictly a probability). Similarly to grades, the absolute value of scores across different machine-learning models is arbitrary, so we rank the objects by the scores of each model and compare the ranks.

\subsection{Preliminary expert classifications for model comparison} \label{ssec:GJ}

\begin{figure}
    \centering
    \includegraphics[width=1\linewidth]{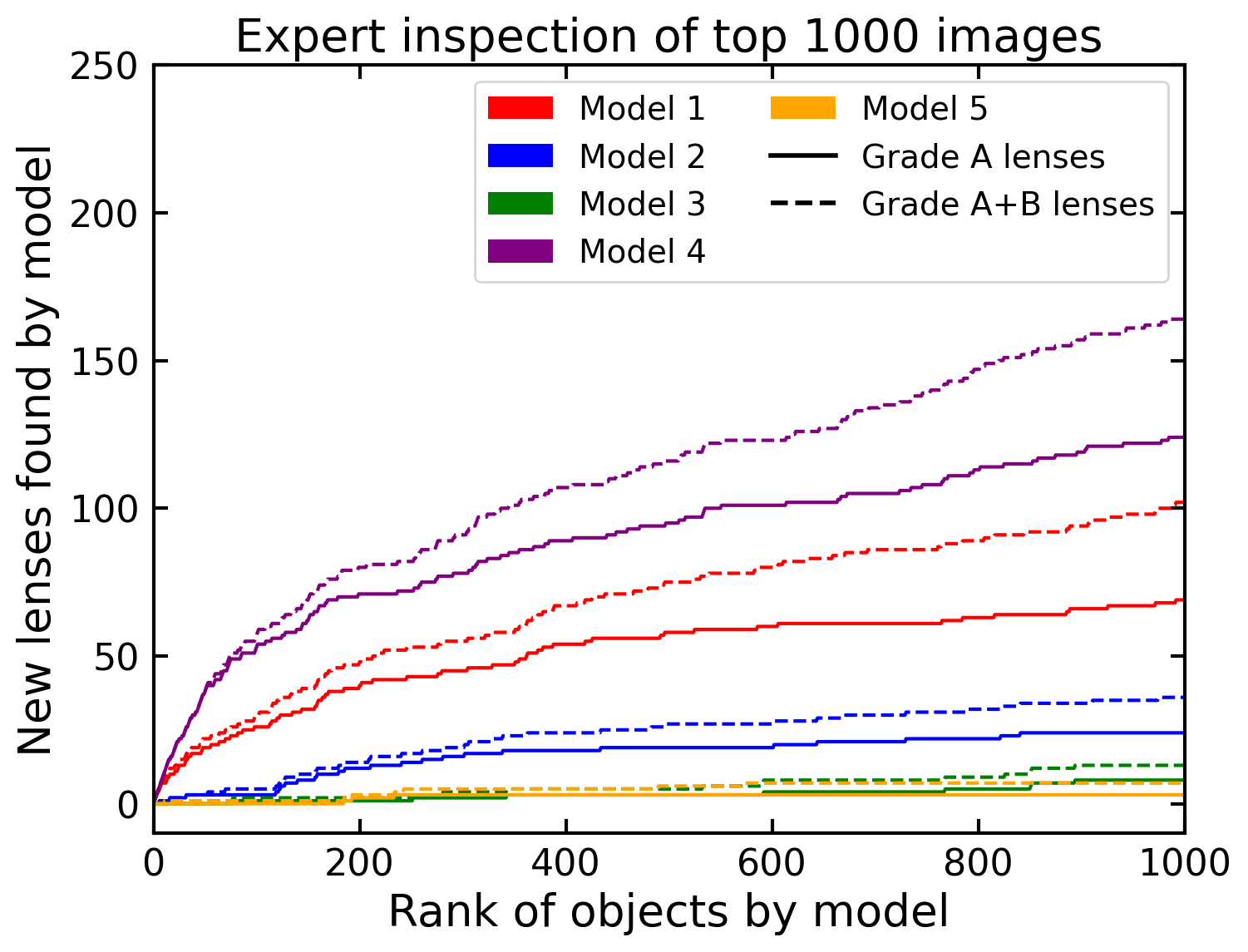}
    \caption{Number of \textcolor{black}{new} lenses found in the top{-}ranked objects according to models 1--5. This is shown for grade A lenses (confident strong lenses) and the combination of grade A and grade B lenses (likely strong lenses).}
    \label{fig:lenses-found-all-models}
\end{figure}

Each of the five models' top \num{1000} objects were ranked by experts, with each cutout receiving approximately ten votes which were then aggregated into one grade. 
We note that a rank of \num{1000} corresponds to different threshold scores for each machine-learning model: the corresponding scores for Models 1--5 are  0.992, 0.999, 0.983, 0.999, and 0.967, respectively. 
The aim of this was to evaluate each model's ability to find lenses. The two main areas of consideration in this analysis were: (1) how many lenses of each grade were found by each model; and (2) of the lenses that one model did not rank in its top \num{1000} object but were highly ranked by another model, where did these missed lenses sit in its score distribution.
The first of these is addressed in Fig. \ref{fig:lenses-found-all-models}, displaying the number of grade A and the number of grade A + B lenses found by each of the Models 1--5. The breakdown of the number of lenses and non-lenses in each model's top 1000 ranked objects is shown in Table \ref{tab:model-rankings}.
The best performing model is Model 4 (\texttt{Zoobot}), {discovering} almost twice the number of grade A lenses as the second{-}best performing model, Model 1 (four-layer CNN). Although the other models {discover} fewer lenses, the proportion of lenses picked out by every model in the top 1000 is significantly higher than the random lens to non-lens ratio {(around 1:$10^3$)}. However, the top 1000 ranked images from every model are still non-lenses by majority, meaning visual inspection by humans is essential.

\subsection{Citizen scientist classifications} \label{ssec:SW}
Once we had verified that all the models were able to find lenses, we picked the images that were ranked highly by any of the models and uploaded these to a public citizen science SW project on the Zooniverse platform.
Here, citizens classified each cutout based on whether they thought it contained a lens or not. {This took place over the span of about six weeks, with \num{857278} total classifications being made by approximately $1\,800$ users.} Images that were flagged as possible lenses by the citizens were then passed to expert inspection in the GJ project to get a final grade. 
It was decided to use SW to visually inspect the top \num{10000} objects from each model, and once this was completed the next \num{10000} from Model 1 and Model 4 were added, along with a subset of the next \num{10000} images from the other models.\footnote{The specifics of SW are outlined in the system overview paper \citep{Q1-SP048}.} A random sample of \num{40000} images was also added for calibration.

\begin{figure}
    \centering
    \includegraphics[width=1\linewidth]{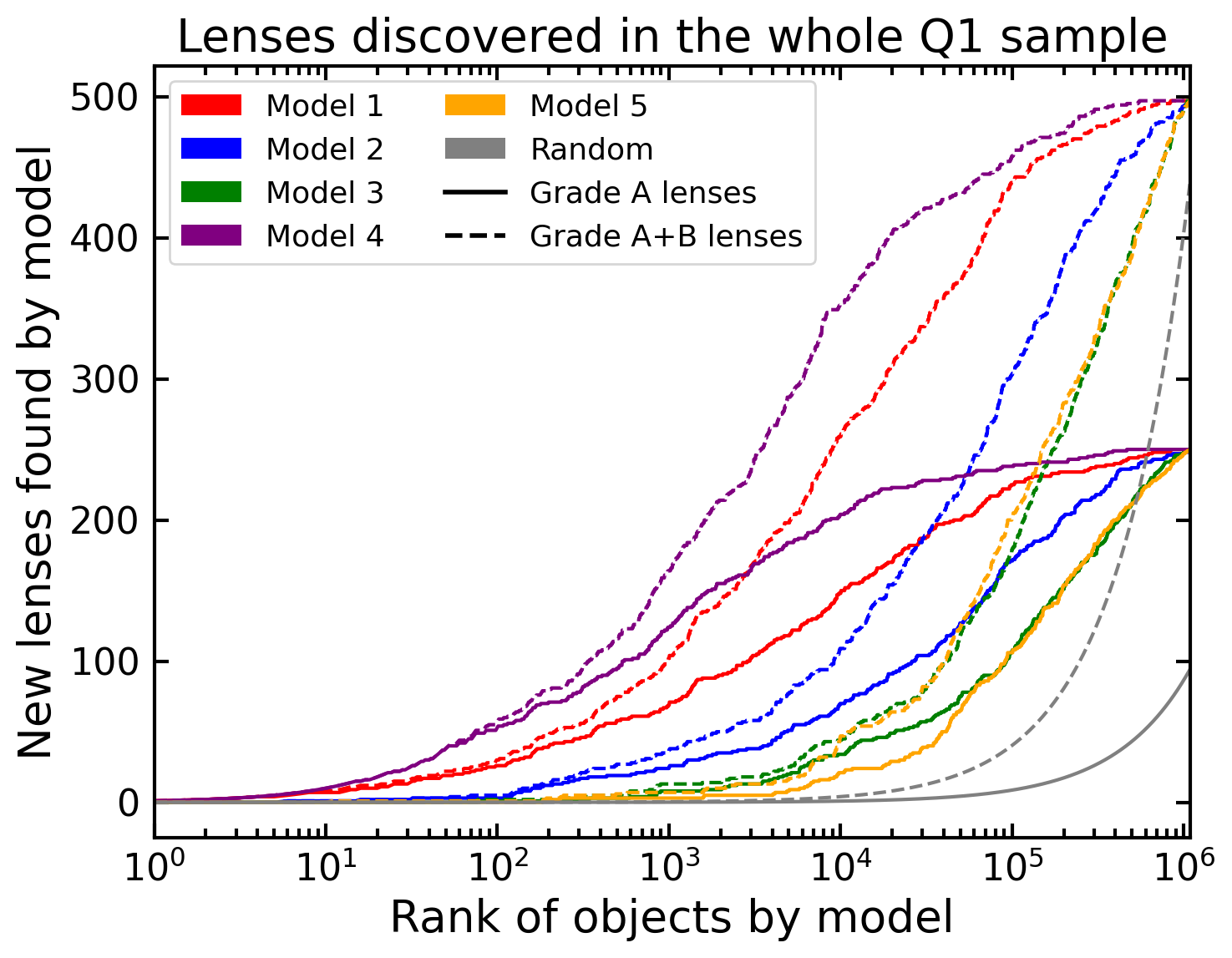}
    \caption{Number of \textcolor{black}{new} grade A and grade A + B lenses {discovered} by each model as a function of rank. This is similar to Fig. \ref{fig:lenses-found-all-models}, except extended to the entirety of the million Q1 objects and plotted with a {log} rank scale. A random distribution of lenses is shown for comparison. We note that unlike Fig. \ref{fig:lenses-found-all-models}, not all the objects at every rank {have} been graded as lens or non-lens, so this only reports the distribution of lenses given that they have been recognised as a lens.}
    \label{fig:all-lens-all-models}
\end{figure}

Of the \num{1086554} cutouts that we started with, a total of \num{115329} objects were inspected by SW, of which a subset of \num{7362} were graded highly enough to make it to the follow-up expert classification stage. Of these, there were 250 grade A, 247 grade B, 585 grade C lenses, and 6280 non-lenses.
{The top grade A lenses are displayed in Sect. \ref{ssec:grade-a}.
Of these, only four grade B and five grade C Q1 lenses appeared in the `known real lenses' data set that some models used for training.}
Almost all the lenses discovered were from images that were uploaded to SW due to being highly ranked by at least one of {the} machine-learning models: $2/250$ grade A lenses and  $8/247$ grade B lenses were found in the random images that were not selected for SW by any of the machine-learning models. A distribution of how all these lenses were ranked by each model is shown in Fig. \ref{fig:all-lens-all-models}. The catalogue of all the lenses we found is publicly available on Zenodo\footnote{\url{https://doi.org/10.5281/zenodo.15003116}}.

\subsection{Completeness of our lens sample} \label{ssec:rest-of-Q1}

In addition to uploading highly ranked images that had been selected by the machine-learning models, we also uploaded \num{40000} randomly selected images from the original \num{1086554} Q1 images. This allowed us to quantify how good the machine-learning models are at finding lenses and forecast how many lenses we would have found had we visually inspected the full Q1 sample. We note that this random sample necessarily had a small overlap with the machine{-}learning-selected sample, since otherwise it would have contained fewer lenses than average. Within this random sample we found nine grade A lenses and 22 grade B lenses, although many of these random images had also been selected by the machine-learning models. 

\begin{figure}
    \centering
    \includegraphics[width=1\linewidth]{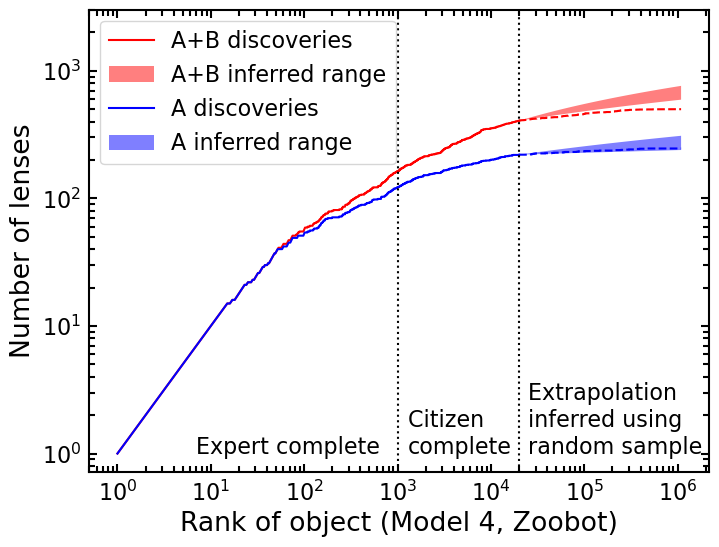}
    \caption{Performance of \texttt{Zoobot}, accounting for uninspected images. The first \num{20000} objects have been inspected by citizen scientists. Beyond \num{20000} the performance is extrapolated based on the prevalence of grade A and B lenses on a randomly selected, citizen{-}inspected sample of images. The dashed lines show lenses {discovered} by the inspection of multiple networks detailed in \citet{Q1-SP048}. The gap between the dashed line and the band shows the likely rank of {undiscovered} lenses in Q1.}
    \label{fig:zoobotwithrandoms}
\end{figure}

We used Bayes' theorem to infer the rate of lenses in the total sample, given that we found nine grade A and 22 grade B lenses in a sample of \num{40000}. Assuming a log-uniform prior on the rate of lenses between 0 and 1, we inferred $240^{+90}_{-70}$ grade A lenses and $540^{+130}_{-120}$ grade B lenses (68\% confidence). The overall system found 250 grade A lenses and 247 grade B lenses (497 grade A + B), indicating that some grade B lenses may have been missed, but it is likely that we {discovered} almost all grade A lenses. 
A substantially better estimate of the total number of lenses in Q1 can be made by including the fact that we have complete knowledge of how the system performs up to a \texttt{Zoobot} rank of \num{20000}. This is the regime where most lenses exist, but it is sparsely probed by the random sample. Given that the random sample included two grade A lenses beyond this point, we can infer that \texttt{Zoobot}  missed $45^{+43}_{-26}$ grade A lenses (68\% confidence). For grade B, \texttt{Zoobot}'s top \num{20000} misses $289^{+97}_{-79}$ lenses. The combination of the other four machine-learning models found 26 of the grade A lenses and 68 of the grade B lenses that \texttt{Zoobot} missed. 

The binomial estimates in the previous paragraph allowed us to estimate where the \texttt{Zoobot} ROC curve should end. We do not know how the 
 ROC curve should evolve between \num{20000} and the final image, but it is likely that the rate of missing grade A and B lenses decreases with \texttt{Zoobot} rank. 
 Figure \ref{fig:zoobotwithrandoms} shows the range of ROC-like curves for \texttt{Zoobot} that are consistent with the binomial estimates and assuming the probability of finding a lens is proportional to the \texttt{Zoobot} rank between \num{20000} and the final image. 
 
 {Furthermore, it is important to understand the completeness of this sample as a function of properties such as brightness and Einstein radius. An analysis of this for the full Q1 lens catalogue is presented in \citet{Q1-SP048}, but we will be able to do this for specific machine learning models in the future when we have enough real lenses to sufficiently span this parameter space.}

\section{Discussion} \label{sec:discussion}

\subsection{Machine-learning takeaways} \label{ssec:ML}
Of the models trained, all of them used both S1, S2, and the labelled DESI non-lenses The remaining data sets used by some but not all models included the real known lenses and other random cutouts. While it was found during the training of \texttt{Zoobot} that adding additional simulations and sets of random cutouts to the training often decreased the performance, the second{-}best performing model (Model 1) notably used all the available data sets for training. While Model 1 and \texttt{Zoobot} both are CNN-based models, \texttt{Zoobot} implemented pretraining which likely changes how the training data affects performance, and as such it is hard to draw any conclusions about what training data are best for lens finding {in general}. 

A similar case is that of image scalings, which again varied across models and did not appear to directly correlate with machine-learning performance. \texttt{Zoobot} found that when using \IE-only images, the $\arcsinh$ scaling resulted in a better performance than MTF, 
while Model 2 found that the $\arcsinh$ scaling {picked out} more stars and performed worse than the MTF scaling. Models 3 and 5 used the $\arcsinh$ \IE-only images, and Model 1 trained \IE-only models for both the $\arcsinh$ and MTF scaling and averaged the results. Although colour images were available, all teams decided to use the \IE-only greyscale images in {the} best versions of their model{s}, despite many teams also testing colour-based models. This is of interest as during human visual inspection, it is often found that colour aids in the detection of lenses, since there is typically a red--blue gradient between the lens and source galaxy that is helpful for distinguishing lenses. The fact that no team found there to be a significant improvement in performance when adding colour data may suggest that machine-learning models prioritise different aspects of images when classifying lenses, or alternatively that these models are not able to pick up on properties such as colour gradients as well as they are able to pick up on morphological information. The latter may make sense in the case of \texttt{Zoobot}; through being pretrained on galaxies of different morphologies, it is not clear that there would be any pattern in colour information that could be extrapolated to lenses in the way that patterns picked up in morphology may be useful. Patterns picked up in the difference between intensities in different filter bands likely might be something that happens in the first few layers of a network and therefore is not something that fine-tuning can address.

\begin{figure}
    \centering
    \includegraphics[width=1\linewidth]{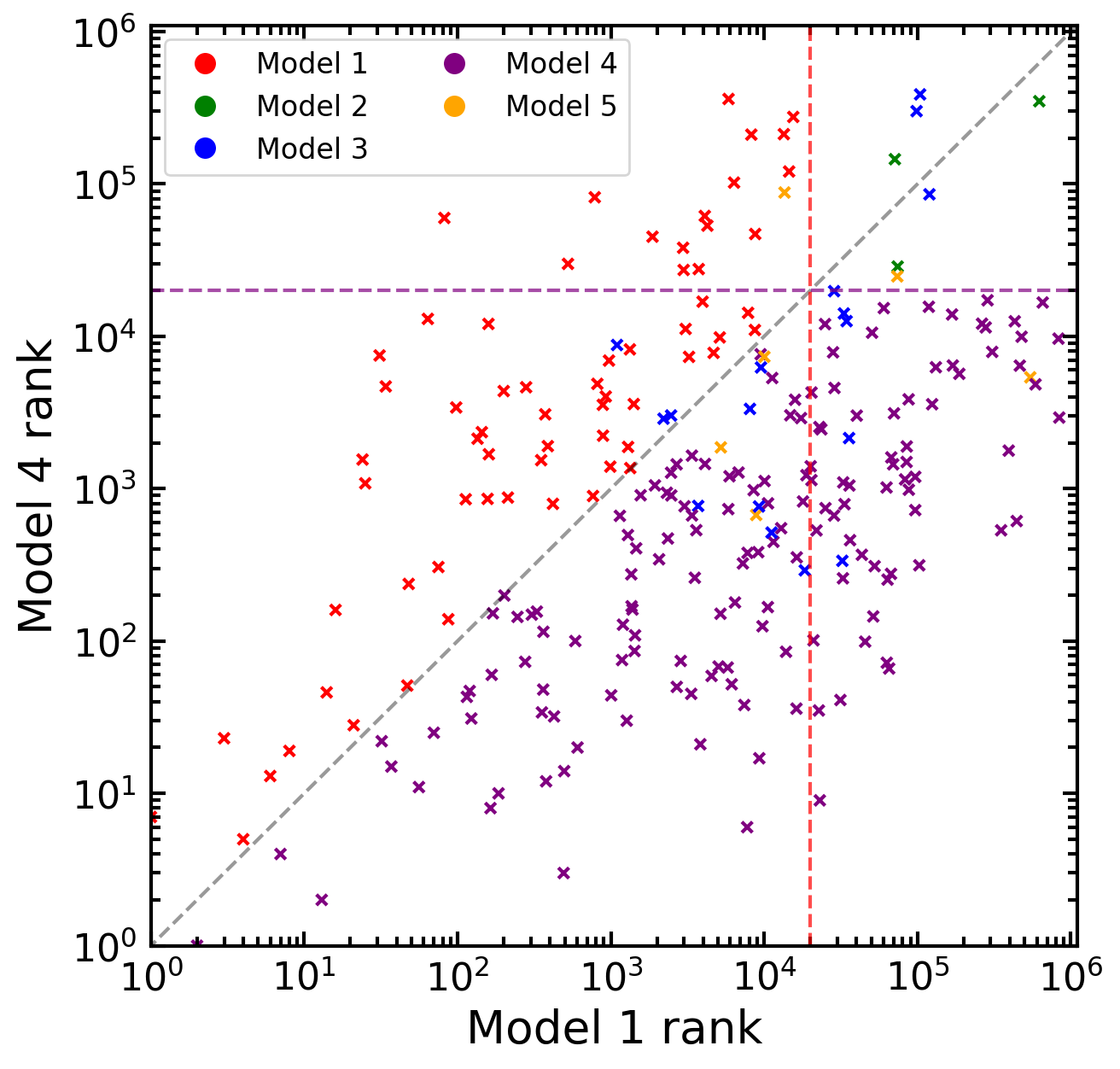}
    \caption{Rank of the grade A lenses by Models 1 and 4. The colours indicate which of the five models ranked each grade A lens the highest. The vertical and horizontal lines at rank \num{20000} mark the limit of the visual inspection for Models 1 and 4.}
    \label{fig:rank-scatter}
\end{figure}

Figure \ref{fig:rank-scatter} shows the position of grade A lenses as a function of Model 1 and Model 4 rank. The two ranks are broadly correlated, but with substantial scatter. It is not clear if the scatter is due to the networks learning different features, or due to the intrinsic uncertainty of each model. We colour each of the points according to the model that best ranked each grade A lens. This can be seen as a marker for which network would have discovered it first. Models 1 and 4 combined make up the large majority of the grade A lens first discoveries (88$\%$) and the combination of them results in a powerful lens finder.  Many grade A lenses beyond Model 1's top \num{20000} were highly ranked by Model 4 and vice versa. While Models 2, 3, and 5 were able to provide additional lenses, only seven were beyond the \num{20000} inspection threshold of Models 1 and 4. These results suggest that Models 1 and 4 provide the bulk of the discovery information. Future searches should use a verification set to check machine-learning performance and use that performance to prioritise the visual inspection effort between networks, or to build an ensemble classifier \citep{Q1-SP059}.

\subsection{The future of lens finding in \Euclid} \label{ssec:future}
With the number of lenses discovered thus far, it is evident that a large number of strong lenses in \Euclid are {discoverable}, and the future releases of \Euclid data are going to contain a wealth of interesting data for strong lensing. However, while the number of lenses will scale with the amount of data \Euclid collects, the number of objects that {can} be visually inspected by humans will remain roughly constant, providing a challenge for lens finding as long as humans remain better at recognising strong gravitational lenses than machine-learning algorithms. \textcolor{black}{Compared to visual inspection, the machine learning computation time is negligible, taking \zb~less than an hour to run on Q1 using an NVIDIA A100 80GB GPU.}

The number of objects able to be classified by humans in the SW project was around \num{100000}, meaning 10\% of the Q1 data was visually inspected. However, the \Euclid DR1 data will cover around $2000$ deg$^2$ of the EWS ($40$ times the area of Q1), and hence visually inspecting \num{100000} objects will only make up 0.25\% of the total DR1 data. Previous lens-finding citizen-science projects have been able to search through a larger number of images (e.g. \num{430000} in the CFHT Legacy Survey SW project{; see \citealt{SpaceWarpsII}}), so assuming an upper limit of visually inspecting one million objects, it is therefore informative to investigate what lenses fall inside and outside the top 0.2\%--2\% of Q1 (corresponding to the top \num{2000}--\num{20000} objects). 

In the left panel of Fig. \ref{fig:lens-density} we present the density of lenses found in Q1 as a function of \texttt{Zoobot} rank. Although there will be lenses in Q1 that we did not {discover}, as explained in Sect. \ref{ssec:rest-of-Q1}, almost all grade A lenses were likely discovered, and hence we can take the frequency of the discovered grade A lenses as being the true frequency. The {worst} rank of a grade A lens according to \texttt{Zoobot} is \num{466371}, meaning \texttt{Zoobot} alone is unable to {discover} 100\% of the grade A lenses in DR1. However, 152 (220) of the grade A lenses were ranked in the top 2000 (\num{20000}) by \texttt{Zoobot}, meaning the number of grade A lenses in DR1 that we will be able to {discover} with the help of citizen science is around $7500$--$\num{11000}$, which will be by far the largest collection of strong lens candidates at the time. Looking beyond DR1, in the full \Euclid 6-year survey the top million ranked \texttt{Zoobot} objects will contain \num{75000} grade A+B lenses, providing a revolution for strong lensing. 

These DR1 forecasts assume that the models used here are the best we can do at lens finding. However, different machine-learning models have different strengths and weaknesses when it comes to lens finding, so by analysing the abilities of different models it is possible to create an ensemble classifier that is better than the sum of its parts. In \citet{Q1-SP059}, it was found that by combining \texttt{Zoobot} with Models 1 and 2, the resulting ensemble classifier outperforms \texttt{Zoobot} alone at lens finding in Q1.
Additionally, as the number of known lenses grows, we will be compiling an increasingly large sample of real \Euclid strong lenses that we can include in our training, which is likely to improve the lens-finding abilities of all the machine-learning models. As such, the numbers forecasted here should provide a lower limit of performance.

\subsection{{Discovery} of lenses of particular interest} \label{funny-lenses}

An ensemble approach to lens finding will still benefit from the individual machine-learning models being optimised as much as possible, and one aspect that is worth considering is the ability to {discover} lenses of individual scientific value. Figure \ref{fig:lens-density} shows the rank and the corresponding images of objects that may be more interesting than the typical single-galaxy-lensed-by-an-LRG lensing configuration that dominate the strong lens samples and are typically the only type of lens system that is simulated. It is interesting to investigate how these objects are scored by machine-learning models which are not trained to recognise lenses of this type, \textcolor{black}{such as DSPLs and non-LRG lens galaxies. We note that one population of strong lenses that are underrepresented in our sample are lensed quasars, and hence a designated search is required to discover the $2000$--$3000$ lensed quasars expected in the full EWS.}

\subsubsection{{DSPLs}}

DSPLs are useful cosmological probes; an analysis and modelling of the DSPLs found in \Euclid Q1 is presented by \citet{Q1-SP054}. There it was found that four lenses in Q1 could be convincingly modelled as DSPLs, and there were another four candidate DSPLs. The ranking of the four DSPLs by each of the models is presented in Table \ref{tab:dspl-ranks}, {and images of these four DSPLs are shown in Sect. \ref{fig:dspls}}. Although four DSPLs is not enough to confidently extrapolate about our ability to {discover} DSPLs in the EWS, it is reassuring that many of them were ranked highly. Model 1 performs exceedingly well at picking out the DSPLs, with all four of them being ranked in the top 500. With the combined performance of the five models it is unlikely that we missed many high-quality DSPLs, with only $6\pm{3}$ forecasted to be detectable within Q1 to begin with \citep{Q1-SP054}.

Although worse than Model 1, \texttt{Zoobot} also performs well at recovering DSPLs, with two of the four DSPLs being in the top 10 ranked objects (${\rm top}\,0.001\%$), at ranks 3 and 7. The ranks of the other two DSPLs were 1902 and 2349, which although sitting at the lower end of the grade A lens ranks, still ranked highly enough for all of them to likely be recovered if they were in DR1. Of the four DSPLs, the two that were ranked in the top 10 both have a larger Einstein radii, with  brighter arcs that are more easily distinguishable than the two lower ranked. The DSPL candidates on average rank lower, with ranks of 622, 1540, 4277, and \num{24853}, and also have smaller Einstein radii, fainter arcs or lens light blending with the arc light. Although only candidate DSPLs, these are still systems that we are interested in, and the fact that these are all ranked on the lower end amongst the lenses ranked by \texttt{Zoobot} (though still in the top 2.5\% of the whole sample), this may be something we want to address in future machine-learning training.

\begin{table}
\centering
\caption{How the four DSPLs in Q1 were ranked by each of the models.}
\label{tab:dspl-ranks}
\resizebox{\columnwidth}{!}{%
\begin{tabular}{rrrrrr}
\hline
\hline
                 &     &           & \multicolumn{2}{l}{Model} &            \\ \cline{2-6} 
                 & 1   & 2         & 3              & 4        & 5          \\ \hline
Teapot lens      & 144 & \num{15214} & \num{310534}     & 2349     & \num{417140} \\
Cosmic dartboard & 388 & \num{36156} & 894            & 1902     & \num{56723}  \\
Galileo's lens   & 492 & 302       & 6452           & 3        & \num{45497}  \\
Cosmic ammonite  & 1   & 241       & 6141           & 7        & \num{78550}  \\ \hline
\end{tabular}%
}
\end{table}

\subsubsection{{Edge-on} lenses}
\begin{figure*}
    \centering
    \includegraphics[width=1\linewidth]{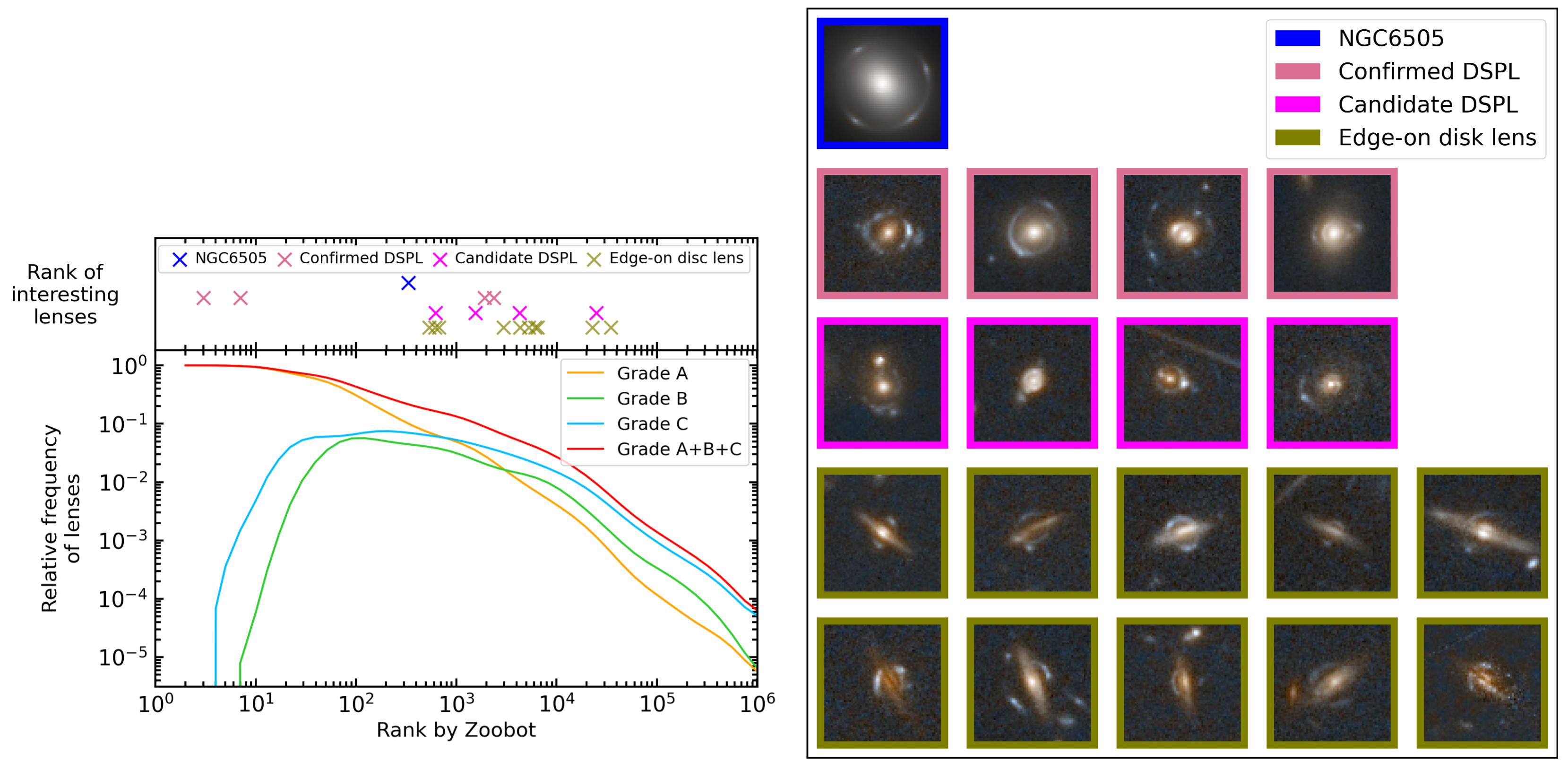}
    \caption{(a) {Left panel, bottom}: Density of grade A, B, and C lenses found in Q1, as well as the three combined, as a function of \texttt{Zoobot} rank. {Above}: Ranks of lenses of particular interest, including the only discovered lens around an NGC galaxy, {four} confirmed DSPLs, four candidate DSPLs, and ten of the more visually obvious edge-on disc lenses. (b) {Right panel}: Images of the interesting lenses corresponding to objects that are plotted on the left. The images are ordered by increasing rank from left to right (left to right on the penultimate row then left to right on the last row for the edge-on disc lenses).}
    \label{fig:lens-density}
\end{figure*}

A similar case can be made for the edge-on disc lenses. As LRGs make up the majority of lens galaxies and typically are easier to simulate accurately, machine-learning models are not generally trained on other types of lens galaxies. 
Edge-on disc lenses are rarer amongst strong lenses, since disc galaxies typically have lower masses and therefore smaller Einstein radii, and they visually appear quite different to the average lens, especially at high inclination. In the right panel of Fig. \ref{fig:lens-density} we show a sample of eight of the more visually obvious edge-on disc lenses. Because there is not a clear distinction between edge-on disc lenses and regular LRG lenses, especially in the case of lenticular galaxies, it is hard to report the exact number of these in our Q1 lens sample. However, we estimate it to be around $20$--$30$, given an inspection by two experts of the top 1000 GJ-graded lenses, which is consistent with the number from SW (during SW, citizens were given the option to provide tags to images of interest, and 29 had been tagged as `\#edge-on'). 
{In Sect. {\ref{ssec:edge-on-fig}} we display some of the grade A+B lenses where the lens galaxy is an edge-on disc or lenticular galaxy.}
Given how different these can appear to the simulated lenses, it is perhaps not surprising that edge-on disc galaxies are ranked low by \texttt{Zoobot}, even compared to the grade C lens frequency, though they are still picked out at a significantly higher rate than random. 

Previous lens finding efforts have in most part not focused on edge-on discs, largely because the number previously known
was very small: the  Sloan Wide Field Camera Edge-on Latetype Lens Survey (SWELLS) found 16 grade A and one grade B strong lensing disc
galaxies \citep{treu11}, while \textcolor{black}{the first instance of a CNN being trained specifically on these systems found four grade A and 20 grade B edge-on lenses within UNIONS \citep{javieredgeon2}}. However, edge-on disc lenses are particularly good probes of dark matter, being able to examine properties such as the mass density profile slope and axial ratio of the dark matter halo \citep{treu11}, and the number density of edge-on disc lenses alone is able to distinguish between theories of dark matter and modified gravity \citep{harvey24}. Since disc galaxies have to be much more massive than average to act as strong gravitational lenses, disc lenses provide a probe of a rare population of galaxies \citep{tortora2019}.
Other more complex lensing configurations 
such as Einstein zig-zags (theorised by \citealt{collettbacon} and first observed by \citealt{zigzag}) are also likely to exist within the data that \Euclid collects, and it therefore may be required for us to include such rarer lensing configurations in our training sets in the future.

\subsubsection{NGC 6505}
Although the image configuration is typical, the NGC 6505 lens discovered in \citet{ORiordan25} is notable as a very low{-}redshift lens. The lens is thus much brighter than is typical in our training set, with the Einstein ring in the very core. Such low-redshift lenses are valuable because they provide a unique probe of the mass profile of the centre of a galaxy and therefore the stellar initial mass function, and because they provide good constraints on the kinematic properties of the galaxy, they can provide a test of general relativity \citep{collett18}. The NGC 6505 lens was discovered in \Euclid data serendipitously \citep{ORiordan25}, but it is reassuring that had this not happened it would have still been discovered by the strong lensing discovery engine, being ranked 332 by \texttt{Zoobot}.

\section{Conclusions} \label{sec:conclusions}
In this work we have presented an analysis of the performance of different machine-learning models at finding strong lenses in the \Euclid Q1 data {covering 63\,deg$^2$}. We compared five different models and investigated some of the factors that can improve the lens finding abilities of the machine-learning models. In combination with expert visual inspection and citizen science efforts, we were able to obtain a large number of lenses and evaluated the performance of the machine-learning models against this. We also explored the abilities of machine-learning models to find different types of lenses and explored how this may scale to the next \Euclid data releases. 
However, we have tested the ability of machine-learning models to identify images that an expert human would classify as a grade A or B lens. We have ignored the fallibility of expert humans at lens finding \citep{rojas23}. It is possible that some real lenses were identified as such by the machine-learning models yet graded poorly by humans, or that some non-lenses were identified as lens candidates by the human inspectors.

\textcolor{black}{The machine-learning models presented in this work identify lenses at rates significantly higher than random, which is significant given the rarity of strong lensing, though the models vary in their overall performance.}
The best performing machine-learning model is \texttt{Zoobot}, a model pretrained on general (non-lens) galaxy morphologies in non-\Euclid data. Of the million Q1 objects that were searched through, \texttt{Zoobot} found 122 grade A, 41 grade B and 67 grade C lenses in the top 1000 ranked objects. By fine-tuning the pretrained model, we can transfer the features it has learned about galaxies in general to the specific problem of lens finding, producing a model that required less training data than training a model from scratch (needing only around $10^{3}$ images to train). Using a less data-hungry model meant that we could more deliberately curate the training data set, which was found to be beneficial.

{We used the machine-learning models to inform the selection of approximately \num{100000} cutouts that were visually inspected by citizens and validated by experts.
These images contain the top \num{20000} ranked by \texttt{Zoobot}. Within the visually inspected images, we found 250 (247) grade A (B) lenses. Of these lenses, 26 (65) are ranked outside of \texttt{Zoobot}'s top \num{20000}. However, since we did not visually inspect the full Q1 sample, some inference is required to estimate the number of lenses that were ranked poorly by \texttt{Zoobot} that had not necessarily been found during visual inspection. By analysing a randomly selected subset of the full Q1 sample}, we inferred that \texttt{Zoobot} missed $45^{+43}_{-26}$ grade A lenses (68\% confidence) {within the full million Q1 sample}. For grade B, \texttt{Zoobot}'s top \num{20000} misses $289^{+97}_{-79}$ lenses. 
{The} other four networks combined found 26 of the missing grade A lenses and 68 of the missing grade B lenses, meaning we found that the other classifiers add compl{e}mentary information which explains why an ensemble of classifiers \citep{Q1-SP059} is more accurate than \texttt{Zoobot} alone.

While machine-learning models are able to pick out atypical lenses such edge-on disc lenses, they are likely to do worse at these than classic configurations of a single source lensed by an elliptical galaxy. We {discovered} four likely DSPLs, but it is possible that further examples have been missed. 
{Additionally, we find that the best model at discovering DSPLs was different from the model that performed best at finding more regular lensing configurations, further highlighting the benefits of combining multiple machine-learning models.}
A bespoke search for high value atypical lenses is likely to be beneficial for future lens searches over larger areas.

\citet{collett15} forecasted that \Euclid would have around \num{170000} strong lenses discoverable within its footprint. However, at the time it was unknown if there was a tractable method of discovering these; a brute-force method of visually inspecting all 1.5 billion galaxies would take a cumulative inspection time of 48 years at a rate of one image classification per second. Whilst there is plenty of scope for improvement, the fine-tuned version of \texttt{Zoobot} presented in this paper is already sufficiently accurate to find most of the lenses in \Euclid in a tractable amount of time{: by} visually inspecting the top million \texttt{Zoobot} ranked objects in the full EWS, we can already expect to find \num{75000} grade A+B lenses in just 12 days of cumulative inspection time. {The lenses found in Q1 will allow further improvements in the machine-learning models, and by combining different improved machine-learning models we can expect to extract purer samples of lenses from future \Euclid data releases. These samples will be orders of magnitude larger than what is currently available and will be able to revolutionise cosmological and astrophysical constraints from strong lensing.}

\section{\textcolor{black}{Data availability}}
\textcolor{black}{All data underlying this article are available on Zenodo at \url{https://doi.org/10.5281/zenodo.15003116}. This article builds on data released during Euclid Quick Release 1, available from \cite{Q1-TP001}.}

%
%

\begin{acknowledgements}
NEPL is supported through a graduate studentship from the UKRI STFC and the University of Portsmouth.

This work has received funding from the European Research Council (ERC) under the European Union's Horizon 2020 research and innovation programme (LensEra: grant agreement No 945536). TEC is funded by the Royal Society through a University Research Fellowship.

A.M.G. acknowledges the support of project PID2022-141915NB-C22 funded by MCIU/AEI/10.13039/501100011033 and FEDER/UE.

SS has received funding from the European Union’s Horizon 2022 research and innovation programme under the Marie Skłodowska-Curie grant agreement No 101105167 — FASTIDIoUS.

GD acknowledges the funding by the European Union - NextGenerationEU, in the framework of the HPC project – “National Centre for HPC, Big Data and Quantum Computing” (PNRR - M4C2 - I1.4 - CN00000013 – CUP J33C22001170001).

C.T. acknowledges the INAF grant 2022 LEMON.

The Dunlap Institute is funded through an endowment established by the David Dunlap family and the University of Toronto.

Numerical computations were carried out on the SCIAMA High Performance Compute (HPC) cluster which is supported by the ICG and the University of Portsmouth.
  
\AckEC  
\AckQone
\AckDatalabs
\end{acknowledgements}

\bibliography{references} 

\begin{thebibliography}{81}
\expandafter\ifx\csname natexlab\endcsname\relax\def\natexlab#1{#1}\fi

\bibitem[{{Acevedo Barroso} {et~al.}(2025{\natexlab{a}}){Acevedo Barroso}, {Cl{\'e}ment}, {Courbin}, {Gavazzi}, {Lemon}, {Rojas}, {Scott}, {Gwyn}, {Hammer}, {Hudson}, \& {Magnier}}]{javieredgeon2}
{Acevedo Barroso}, J.~A., {Cl{\'e}ment}, B., {Courbin}, F., {et~al.} 2025{\natexlab{a}}, A\&A, submitted, arXiv:2503.10610

\bibitem[{{Acevedo Barroso} {et~al.}(2025{\natexlab{b}}){Acevedo Barroso}, {O'Riordan}, {Cl{\'e}ment}, {Tortora}, {Collett}, {Courbin}, {Gavazzi}, {Metcalf}, {Busillo}, {Andika}, {Cabanac}, {Courtois}, {Crook-Mansour}, {Delchambre}, {Despali}, {Ecker}, {Franco}, {Holloway}, {Jackson}, {Jahnke}, {Mahler}, {Marchetti}, {Matavulj}, {Melo}, {Meneghetti}, {Moustakas}, {M{\"u}ller}, {Nucita}, {Paulino-Afonso}, {Pearson}, {Rojas}, {Scarlata}, {Schuldt}, {Serjeant}, {Sluse}, {Suyu}, {Vaccari}, {Verma}, {Vernardos}, {Walmsley}, {Bouy}, {Walth}, {Powell}, {Bolzonella}, {Cuillandre}, {Kluge}, {Saifollahi}, {Schirmer}, {Stone}, {Acebron}, {Bazzanini}, {D{\'\i}az-S{\'a}nchez}, {Hogg}, {Koopmans}, {Kruk}, {Leuzzi}, {Manj{\'o}n-Garc{\'\i}a}, {Mannucci}, {Nagam}, {Pearce-Casey}, {Scharr{\'e}}, {Wilde}, {Altieri}, {Amara}, {Andreon}, {Auricchio}, {Baccigalupi}, {Baldi}, {Balestra}, {Bardelli}, {Basset}, {Battaglia}, {Bender}, {Bonino}, {Branchini}, {Brescia}, {Brinchmann}, {Caillat}, {Camera}, {Candini}, {Capobianco},
  {Carbone}, {Carretero}, {Casas}, {Castellano}, {Castignani}, {Cavuoti}, {Cimatti}, {Colodro-Conde}, {Congedo}, {Conselice}, {Conversi}, {Copin}, {Corcione}, {Cropper}, {Da Silva}, {Degaudenzi}, {De Lucia}, {Dinis}, {Dubath}, {Dupac}, {Dusini}, {Farina}, {Farrens}, {Ferriol}, {Frailis}, {Franceschi}, {Galeotta}, {Garilli}, {George}, {Gillard}, {Gillis}, {Giocoli}, {G{\'o}mez-Alvarez}, {Grazian}, {Grupp}, {Guzzo}, {Haugan}, {Hoekstra}, {Holmes}, {Hook}, {Hormuth}, {Hornstrup}, {Jhabvala}, {Joachimi}, {Keih{\"a}nen}, {Kermiche}, {Kiessling}, {Kubik}, {Kunz}, {Kurki-Suonio}, {Le Mignant}, {Ligori}, {Lilje}, {Lindholm}, {Lloro}, {Mainetti}, {Maiorano}, {Mansutti}, {Marcin}, {Marggraf}, {Martinelli}, {Martinet}, {Marulli}, {Massey}, {Medinaceli}, {Melchior}, {Mellier}, {Merlin}, {Meylan}, {Moresco}, {Moscardini}, {Munari}, {Nakajima}, {Neissner}, {Nichol}, {Niemi}, {Nightingale}, {Padilla}, {Paltani}, {Pasian}, {Pedersen}, {Percival}, {Pettorino}, {Pires}, {Polenta}, {Poncet}, {Popa}, {Pozzetti}, {Raison},
  {Rebolo}, {Renzi}, {Rhodes}, {Riccio}, {Romelli}, {Roncarelli}, {Rossetti}, {Saglia}, {Sakr}, {S{\'a}nchez}, {Sapone}, {Schneider}, {Schrabback}, {Secroun}, {Seidel}, {Serrano}, {Sirignano}, {Sirri}, {Skottfelt}, {Stanco}, {Steinwagner}, {Tallada-Cresp{\'\i}}, {Tavagnacco}, {Taylor}, {Tereno}, {Toledo-Moreo}, {Torradeflot}, {Tutusaus}, {Valentijn}, \& {Valenziano}}]{ELSE24}
{Acevedo Barroso}, J.~A., {O'Riordan}, C.~M., {Cl{\'e}ment}, B., {et~al.} 2025{\natexlab{b}}, \aap, 697, A14

\bibitem[{{Avestruz} {et~al.}(2019){Avestruz}, {Li}, {Zhu}, {Lightman}, {Collett}, \& {Luo}}]{Avestruz}
{Avestruz}, C., {Li}, N., {Zhu}, H., {et~al.} 2019, \apj, 877, 58

\bibitem[{Birrer \& Amara(2018)}]{birrer18}
Birrer, S. \& Amara, A. 2018, Physics of the Dark Universe, 22, 189

\bibitem[{Birrer {et~al.}(2021)Birrer, Shajib, Gilman, Galan, Aalbers, Millon, Morgan, Pagano, Park, Teodori, Tessore, Ueland, de~Vyvere, Wagner-Carena, Wempe, Yang, Ding, Schmidt, Sluse, Zhang, \& Amara}]{birrer21}
Birrer, S., Shajib, A.~J., Gilman, D., {et~al.} 2021, Journal of Open Source Software, 6, 3283

\bibitem[{{Ca{\~n}ameras} {et~al.}(2024){Ca{\~n}ameras}, {Schuldt}, {Shu}, {Suyu}, {Taubenberger}, {Andika}, {Bag}, {Inoue}, {Jaelani}, {Leal-Taix{\'e}}, {Meinhardt}, {Melo}, \& {More}}]{canameras24}
{Ca{\~n}ameras}, R., {Schuldt}, S., {Shu}, Y., {et~al.} 2024, \aap, 692, A72

\bibitem[{{Ca{\~n}ameras} {et~al.}(2020){Ca{\~n}ameras}, {Schuldt}, {Suyu}, {Taubenberger}, {Meinhardt}, {Leal-Taix{\'e}}, {Lemon}, {Rojas}, \& {Savary}}]{Canameras20}
{Ca{\~n}ameras}, R., {Schuldt}, S., {Suyu}, S.~H., {et~al.} 2020, \aap, 644, A163

\bibitem[{{Collett}(2015)}]{collett15}
{Collett}, T.~E. 2015, \apj, 811, 20

\bibitem[{{Collett} \& {Bacon}(2016)}]{collettbacon}
{Collett}, T.~E. \& {Bacon}, D.~J. 2016, \mnras, 456, 2210

\bibitem[{{Collett} {et~al.}(2018){Collett}, {Oldham}, {Smith}, {Auger}, {Westfall}, {Bacon}, {Nichol}, {Masters}, {Koyama}, \& {van den Bosch}}]{collett18}
{Collett}, T.~E., {Oldham}, L.~J., {Smith}, R.~J., {et~al.} 2018, Science, 360, 1342

\bibitem[{{Dom{\'\i}nguez S{\'a}nchez} {et~al.}(2018){Dom{\'\i}nguez S{\'a}nchez}, {Huertas-Company}, {Bernardi}, {Tuccillo}, \& {Fischer}}]{dominguezsanchez18}
{Dom{\'\i}nguez S{\'a}nchez}, H., {Huertas-Company}, M., {Bernardi}, M., {Tuccillo}, D., \& {Fischer}, J.~L. 2018, \mnras, 476, 3661

\bibitem[{{Dux} {et~al.}(2025){Dux}, {Millon}, {Lemon}, {Schmidt}, {Courbin}, {Shajib}, {Treu}, {Birrer}, {Wong}, {Agnello}, {Andrade}, {Galan}, {Hjorth}, {Paic}, {Schuldt}, {Schweinfurth}, {Sluse}, {Smette}, \& {Suyu}}]{zigzag}
{Dux}, F., {Millon}, M., {Lemon}, C., {et~al.} 2025, \aap, 694, A300

\bibitem[{{Euclid Collaboration: Aussel} {et~al.}(2025){Euclid Collaboration: Aussel}, {Tereno}, {Schirmer}, {et~al.}}]{Q1-TP001}
{Euclid Collaboration: Aussel}, H., {Tereno}, I., {Schirmer}, M., {et~al.} 2025, A\&A, submitted (Euclid Q1 SI), arXiv:2503.15302

\bibitem[{{Euclid Collaboration: Castander} {et~al.}(2025){Euclid Collaboration: Castander}, {Fosalba}, {Stadel}, {et~al.}}]{EuclidSkyFlagship}
{Euclid Collaboration: Castander}, F., {Fosalba}, P., {Stadel}, J., {et~al.} 2025, A\&A, 697, A5

\bibitem[{{Euclid Collaboration: Cropper} {et~al.}(2025){Euclid Collaboration: Cropper}, {Al-Bahlawan}, {Amiaux}, {et~al.}}]{EuclidSkyVIS}
{Euclid Collaboration: Cropper}, M., {Al-Bahlawan}, A., {Amiaux}, J., {et~al.} 2025, A\&A, 697, A2

\bibitem[{{Euclid Collaboration: Holloway} {et~al.}(2025){Euclid Collaboration: Holloway}, {Verma}, {Walmsley}, {et~al.}}]{Q1-SP059}
{Euclid Collaboration: Holloway}, P., {Verma}, A., {Walmsley}, M., {et~al.} 2025, A\&A, submitted (Euclid Q1 SI), arXiv:2503.15328

\bibitem[{{Euclid Collaboration: Jahnke} {et~al.}(2025){Euclid Collaboration: Jahnke}, {Gillard}, {Schirmer}, {et~al.}}]{EuclidSkyNISP}
{Euclid Collaboration: Jahnke}, K., {Gillard}, W., {Schirmer}, M., {et~al.} 2025, A\&A, 697, A3

\bibitem[{{Euclid Collaboration}:~{Leuzzi} {et~al.}(2024){Euclid Collaboration}:~{Leuzzi}, {Meneghetti}, {Angora}, {Metcalf}, {Moscardini}, {Rosati}, {Bergamini}, {Calura}, {Cl{\'e}ment}, {Gavazzi}, {Gentile}, {Lochner}, {Grillo}, {Vernardos}, {Aghanim}, {Amara}, {Amendola}, {Auricchio}, {Bodendorf}, {Bonino}, {Branchini}, {Brescia}, {Brinchmann}, {Camera}, {Capobianco}, {Carbone}, {Carretero}, {Castellano}, {Cavuoti}, {Cimatti}, {Cledassou}, {Congedo}, {Conselice}, {Conversi}, {Copin}, {Corcione}, {Courbin}, {Cropper}, {Da Silva}, {Degaudenzi}, {Dinis}, {Dubath}, {Dupac}, {Dusini}, {Farrens}, {Ferriol}, {Frailis}, {Franceschi}, {Fumana}, {Galeotta}, {Gillis}, {Giocoli}, {Grazian}, {Grupp}, {Guzzo}, {Haugan}, {Holmes}, {Hormuth}, {Hornstrup}, {Hudelot}, {Jahnke}, {K{\"u}mmel}, {Kermiche}, {Kiessling}, {Kitching}, {Kunz}, {Kurki-Suonio}, {Lilje}, {Lloro}, {Maiorano}, {Mansutti}, {Marggraf}, {Markovic}, {Marulli}, {Massey}, {Medinaceli}, {Mei}, {Melchior}, {Mellier}, {Merlin}, {Meylan}, {Moresco}, {Munari},
  {Niemi}, {Nightingale}, {Nutma}, {Padilla}, {Paltani}, {Pasian}, {Pedersen}, {Pettorino}, {Pires}, {Polenta}, {Poncet}, {Raison}, {Renzi}, {Rhodes}, {Riccio}, {Romelli}, {Roncarelli}, {Rossetti}, {Saglia}, {Sapone}, {Sartoris}, {Schneider}, {Secroun}, {Seidel}, {Serrano}, {Sirignano}, {Sirri}, {Stanco}, {Tallada-Cresp{\'\i}}, {Taylor}, {Tereno}, {Toledo-Moreo}, {Torradeflot}, {Tutusaus}, {Valenziano}, {Vassallo}, {Wang}, {Weller}, {Zamorani}, {Zoubian}, {Andreon}, {Bardelli}, {Boucaud}, {Bozzo}, {Colodro-Conde}, {Di Ferdinando}, {Farina}, {Farinelli}, {Graci{\'a}-Carpio}, {Keih{\"a}nen}, {Lindholm}, {Maino}, {Mauri}, {Neissner}, {Schirmer}, {Scottez}, {Tenti}, {Tramacere}, {Veropalumbo}, {Zucca}, {Akrami}, {Allevato}, {Baccigalupi}, {Ballardini}, {Bernardeau}, {Biviano}, {Borgani}, {Borlaff}, {Bretonni{\`e}re}, {Burigana}, {Cabanac}, {Cappi}, {Carvalho}, {Casas}, {Castignani}, {Castro}, {Chambers}, {Cooray}, {Coupon}, {Courtois}, {Davini}, {de la Torre}, {De Lucia}, {Desprez}, {Di Domizio}, {Dole},
  {Escartin Vigo}, {Escoffier}, {Ferrero}, {Gabarra}, {Ganga}, {Garcia-Bellido}, {Gaztanaga}, {George}, {Gozaliasl}, {Hildebrandt}, {Hook}, {Huertas-Company}, {Joachimi}, {Kajava}, {Kansal}, {Kirkpatrick}, {Legrand}, {Loureiro}, {Magliocchetti}, {Mainetti}, {Maoli}, {Martinelli}, {Martinet}, {Martins}, {Matthew}, {Maurin}, {Monaco}, {Morgante}, {Nadathur}, {Nucita}, {Patrizii}, {Popa}, {Porciani}, {Potter}, {P{\"o}ntinen}, {Reimberg}, {S{\'a}nchez}, {Sakr}, {Schneider}, {Sereno}, {Simon}, {Spurio Mancini}, {Stadel}, {Steinwagner}, {Teyssier}, {Valiviita}, {Viel}, {Zinchenko}, \& {Dom{\'\i}nguez S{\'a}nchez}}]{leuzzi24}
{Euclid Collaboration}:~{Leuzzi}, L., {Meneghetti}, M., {Angora}, G., {et~al.} 2024, \aap, 681, A68

\bibitem[{{Euclid Collaboration: Li} {et~al.}(2025){Euclid Collaboration: Li}, {Collett}, {Walmsley}, {et~al.}}]{Q1-SP054}
{Euclid Collaboration: Li}, T., {Collett}, T.~E., {Walmsley}, M., {et~al.} 2025, A\&A, submitted (Euclid Q1 SI), arXiv:2503.15327

\bibitem[{{Euclid Collaboration: McCracken} {et~al.}(2025){Euclid Collaboration: McCracken}, {Benson}, {Dolding}, {et~al.}}]{Q1-TP002}
{Euclid Collaboration: McCracken}, H.~J., {Benson}, K., {Dolding}, C., {et~al.} 2025, A\&A, submitted (Euclid Q1 SI), arXiv:2503.15303

\bibitem[{{Euclid Collaboration: Mellier} {et~al.}(2025){Euclid Collaboration: Mellier}, {Abdurro'uf}, {Acevedo~Barroso}, {et~al.}}]{EuclidSkyOverview}
{Euclid Collaboration: Mellier}, Y., {Abdurro'uf}, {Acevedo~Barroso}, J., {et~al.} 2025, A\&A, 697, A1

\bibitem[{{Euclid Collaboration: Rojas} {et~al.}(2025){Euclid Collaboration: Rojas}, {Collett}, {Acevedo Barroso}, {et~al.}}]{Q1-SP052}
{Euclid Collaboration: Rojas}, K., {Collett}, T.~E., {Acevedo Barroso}, J.~A., {et~al.} 2025, A\&A, submitted (Euclid Q1 SI), arXiv:2503.15325

\bibitem[{{Euclid Collaboration: Romelli} {et~al.}(2025){Euclid Collaboration: Romelli}, {K\"ummel}, {Dole}, {et~al.}}]{Q1-TP004}
{Euclid Collaboration: Romelli}, E., {K\"ummel}, M., {Dole}, H., {et~al.} 2025, A\&A, accepted (Euclid Q1 SI), arXiv:2503.15305

\bibitem[{{Euclid Collaboration: Scaramella} {et~al.}(2022){Euclid Collaboration: Scaramella}, {Amiaux}, {Mellier}, {et~al.}}]{Scaramella-EP1}
{Euclid Collaboration: Scaramella}, R., {Amiaux}, J., {Mellier}, Y., {et~al.} 2022, \aap, 662, A112

\bibitem[{{Euclid Collaboration: Walmsley} {et~al.}(2025{\natexlab{a}}){Euclid Collaboration: Walmsley}, {Holloway}, {Lines}, {et~al.}}]{Q1-SP048}
{Euclid Collaboration: Walmsley}, M., {Holloway}, P., {Lines}, N.~E.~P., {et~al.} 2025{\natexlab{a}}, A\&A, submitted (Euclid Q1 SI), arXiv:2503.15324

\bibitem[{{Euclid Collaboration: Walmsley} {et~al.}(2025{\natexlab{b}}){Euclid Collaboration: Walmsley}, {Huertas-Company}, {Quilley}, {et~al.}}]{Q1-SP047}
{Euclid Collaboration: Walmsley}, M., {Huertas-Company}, M., {Quilley}, L., {et~al.} 2025{\natexlab{b}}, A\&A, submitted (Euclid Q1 SI), arXiv:2503.15310

\bibitem[{{Euclid Early Release Observations}(2024)}]{EROcite}
{Euclid Early Release Observations}. 2024, \url{https://doi.org/10.57780/esa-qmocze3}

\bibitem[{{Euclid Quick Release Q1}(2025)}]{Q1cite}
{Euclid Quick Release Q1}. 2025, \url{https://doi.org/10.57780/esa-2853f3b}

\bibitem[{{Gonz{\'a}lez} {et~al.}(2025){Gonz{\'a}lez}, {Holloway}, {Collett}, {Verma}, {Bechtol}, {Marshall}, {More}, {Acevedo Barroso}, {Cartwright}, {Martinez}, {Li}, {Rojas}, {Schuldt}, {Birrer}, {Diehl}, {Morgan}, {Drlica-Wagner}, {O'Donnell}, {Zaborowski}, {Nord}, {Baeten}, {Johnson}, {Macmillan}, {Roodman}, {Pieres}, {Walker}, {Plazas Malag{\'o}n}, {Carnero Rosell}, {Santiago}, {Flaugher}, {Gruen}, {Brooks}, {Burke}, {James}, {Sanchez Cid}, {Hollowood}, {Tucker}, {Buckley-Geer}, {Gaztanaga}, {Suchyta}, {Sanchez}, {Gutierrez}, {Giannini}, {Tarle}, {Sevilla-Noarbe}, {Marshall}, {Carretero}, {Frieman}, {De Vicente}, {Garc{\'\i}a-Bellido}, {Mena-Fern{\'a}ndez}, {Myles}, {Honscheid}, {Kuehn}, {Lima}, {Pereira}, {Smith}, {Aguena}, {Weaverdyck}, {Lahav}, {Doel}, {Miquel}, {Gruendl}, {Cawthon}, {Hinton}, {Allam}, {Desai}, {Samuroff}, {Everett}, {Lee}, {Davis}, {Abbott}, \& {Vikram}}]{jimena}
{Gonz{\'a}lez}, J., {Holloway}, P., {Collett}, T., {et~al.} 2025, \apj, submitted, arXiv:2501.15679

\bibitem[{{Grespan} {et~al.}(2024){Grespan}, {Thuruthipilly}, {Pollo}, {Lochner}, {Biesiada}, \& {Etsebeth}}]{Grespan24}
{Grespan}, M., {Thuruthipilly}, H., {Pollo}, A., {et~al.} 2024, \aap, 688, A34

\bibitem[{{Harvey-Hawes} \& {Galoppo}(2024)}]{harvey24}
{Harvey-Hawes}, C. \& {Galoppo}, M. 2024, arXiv e-prints, arXiv:2411.17888

\bibitem[{{Huang} {et~al.}(2021){Huang}, {Storfer}, {Gu}, {Ravi}, {Pilon}, {Sheu}, {Venguswamy}, {Banka}, {Dey}, {Landriau}, {Lang}, {Meisner}, {Moustakas}, {Myers}, {Sajith}, {Schlafly}, \& {Schlegel}}]{desihuang2021}
{Huang}, X., {Storfer}, C., {Gu}, A., {et~al.} 2021, \apj, 909, 27

\bibitem[{{Huang} {et~al.}(2020){Huang}, {Storfer}, {Ravi}, {Pilon}, {Domingo}, {Schlegel}, {Bailey}, {Dey}, {Gupta}, {Herrera}, {Juneau}, {Landriau}, {Lang}, {Meisner}, {Moustakas}, {Myers}, {Schlafly}, {Valdes}, {Weaver}, {Yang}, \& {Y{\`e}che}}]{desihuang2020}
{Huang}, X., {Storfer}, C., {Ravi}, V., {et~al.} 2020, \apj, 894, 78

\bibitem[{{Inami} {et~al.}(2017){Inami}, {Bacon}, {Brinchmann}, {Richard}, {Contini}, {Conseil}, {Hamer}, {Akhlaghi}, {Bouch{\'e}}, {Cl{\'e}ment}, {Desprez}, {Drake}, {Hashimoto}, {Leclercq}, {Maseda}, {Michel-Dansac}, {Paalvast}, {Tresse}, {Ventou}, {Kollatschny}, {Boogaard}, {Finley}, {Marino}, {Schaye}, \& {Wisotzki}}]{inami2017}
{Inami}, H., {Bacon}, R., {Brinchmann}, J., {et~al.} 2017, \aap, 608, A2

\bibitem[{{Jacobs} {et~al.}(2019{\natexlab{a}}){Jacobs}, {Collett}, {Glazebrook}, {Buckley-Geer}, {Diehl}, {Lin}, {McCarthy}, {Qin}, {Odden}, {Caso Escudero}, {Dial}, {Yung}, {Gaitsch}, {Pellico}, {Lindgren}, {Abbott}, {Annis}, {Avila}, {Brooks}, {Burke}, {Carnero Rosell}, {Carrasco Kind}, {Carretero}, {da Costa}, {De Vicente}, {Fosalba}, {Frieman}, {Garc{\'\i}a-Bellido}, {Gaztanaga}, {Goldstein}, {Gruen}, {Gruendl}, {Gschwend}, {Hollowood}, {Honscheid}, {Hoyle}, {James}, {Krause}, {Kuropatkin}, {Lahav}, {Lima}, {Maia}, {Marshall}, {Miquel}, {Plazas}, {Roodman}, {Sanchez}, {Scarpine}, {Serrano}, {Sevilla-Noarbe}, {Smith}, {Sobreira}, {Suchyta}, {Swanson}, {Tarle}, {Vikram}, {Walker}, {Zhang}, \& {DES Collaboration}}]{Jacobs2019}
{Jacobs}, C., {Collett}, T., {Glazebrook}, K., {et~al.} 2019{\natexlab{a}}, \apjs, 243, 17

\bibitem[{{Jacobs} {et~al.}(2019{\natexlab{b}}){Jacobs}, {Collett}, {Glazebrook}, {McCarthy}, {Qin}, {Abbott}, {Abdalla}, {Annis}, {Avila}, {Bechtol}, {Bertin}, {Brooks}, {Buckley-Geer}, {Burke}, {Carnero Rosell}, {Carrasco Kind}, {Carretero}, {da Costa}, {Davis}, {De Vicente}, {Desai}, {Diehl}, {Doel}, {Eifler}, {Flaugher}, {Frieman}, {Garc{\'\i}a-Bellido}, {Gaztanaga}, {Gerdes}, {Goldstein}, {Gruen}, {Gruendl}, {Gschwend}, {Gutierrez}, {Hartley}, {Hollowood}, {Honscheid}, {Hoyle}, {James}, {Kuehn}, {Kuropatkin}, {Lahav}, {Li}, {Lima}, {Lin}, {Maia}, {Martini}, {Miller}, {Miquel}, {Nord}, {Plazas}, {Sanchez}, {Scarpine}, {Schubnell}, {Serrano}, {Sevilla-Noarbe}, {Smith}, {Soares-Santos}, {Sobreira}, {Suchyta}, {Swanson}, {Tarle}, {Vikram}, {Walker}, {Zhang}, {Zuntz}, \& {DES Collaboration}}]{Jacobs2018}
{Jacobs}, C., {Collett}, T., {Glazebrook}, K., {et~al.} 2019{\natexlab{b}}, \mnras, 484, 5330

\bibitem[{{Jacobs} {et~al.}(2017){Jacobs}, {Glazebrook}, {Collett}, {More}, \& {McCarthy}}]{Jacobs17}
{Jacobs}, C., {Glazebrook}, K., {Collett}, T., {More}, A., \& {McCarthy}, C. 2017, \mnras, 471, 167

\bibitem[{{Jaelani} {et~al.}(2024){Jaelani}, {More}, {Wong}, {Inoue}, {Chao}, {Premadi}, \& {Ca{\~n}ameras}}]{Jaelani2024}
{Jaelani}, A.~T., {More}, A., {Wong}, K.~C., {et~al.} 2024, \mnras, 535, 1625

\bibitem[{Kingma \& Ba(2015)}]{Adam}
Kingma, D.~P. \& Ba, J.~L. 2015, in International Conference on Learning Representations

\bibitem[{{Koekemoer} {et~al.}(2007){Koekemoer}, {Aussel}, {Calzetti}, {Capak}, {Giavalisco}, {Kneib}, {Leauthaud}, {Le F{\`e}vre}, {McCracken}, {Massey}, {Mobasher}, {Rhodes}, {Scoville}, \& {Shopbell}}]{Koekemoer2007}
{Koekemoer}, A.~M., {Aussel}, H., {Calzetti}, D., {et~al.} 2007, \apjs, 172, 196

\bibitem[{{Lanusse} {et~al.}(2018){Lanusse}, {Ma}, {Li}, {Collett}, {Li}, {Ravanbakhsh}, {Mandelbaum}, \& {P{\'o}czos}}]{Lanusse}
{Lanusse}, F., {Ma}, Q., {Li}, N., {et~al.} 2018, \mnras, 473, 3895

\bibitem[{{Leauthaud} {et~al.}(2007){Leauthaud}, {Massey}, {Kneib}, {Rhodes}, {Johnston}, {Capak}, {Heymans}, {Ellis}, {Koekemoer}, {Le F{\`e}vre}, {Mellier}, {R{\'e}fr{\'e}gier}, {Robin}, {Scoville}, {Tasca}, {Taylor}, \& {Van Waerbeke}}]{Leauthaud2007}
{Leauthaud}, A., {Massey}, R., {Kneib}, J.-P., {et~al.} 2007, \apjs, 172, 219

\bibitem[{{Li} {et~al.}(2021){Li}, {Napolitano}, {Spiniello}, {Tortora}, {Kuijken}, {Koopmans}, {Schneider}, {Getman}, {Xie}, {Long}, {Shu}, {Vernardos}, {Huang}, {Covone}, {Dvornik}, {Heymans}, {Hildebrandt}, {Radovich}, \& {Wright}}]{li2021}
{Li}, R., {Napolitano}, N.~R., {Spiniello}, C., {et~al.} 2021, \apj, 923, 16

\bibitem[{{Li} {et~al.}(2020){Li}, {Napolitano}, {Tortora}, {Spiniello}, {Koopmans}, {Huang}, {Roy}, {Vernardos}, {Chatterjee}, {Giblin}, {Getman}, {Radovich}, {Covone}, \& {Kuijken}}]{li2020}
{Li}, R., {Napolitano}, N.~R., {Tortora}, C., {et~al.} 2020, \apj, 899, 30

\bibitem[{{Lintott} {et~al.}(2008){Lintott}, {Schawinski}, {Slosar}, {Land}, {Bamford}, {Thomas}, {Raddick}, {Nichol}, {Szalay}, {Andreescu}, {Murray}, \& {Vandenberg}}]{galzoo}
{Lintott}, C.~J., {Schawinski}, K., {Slosar}, A., {et~al.} 2008, \mnras, 389, 1179

\bibitem[{Liu {et~al.}(2022)Liu, Mao, Wu, Feichtenhofer, Darrell, \& Xie}]{convnext}
Liu, Z., Mao, H., Wu, C.-Y., {et~al.} 2022, in 2022 IEEE/CVF Conference on Computer Vision and Pattern Recognition (CVPR) (Los Alamitos, CA, USA: IEEE Computer Society), 11966--11976

\bibitem[{{Manjón-García}(2021)}]{manjongarcia21}
{Manjón-García}, A. 2021, {PhD thesis}, University of Cantabria (Spain)

\bibitem[{Marshall {et~al.}(2015)Marshall, Verma, More, Davis, More, Kapadia, Parrish, Snyder, Wilcox, Baeten, Macmillan, Cornen, Baumer, Simpson, Lintott, Miller, Paget, Simpson, Smith, Küng, Saha, \& Collett}]{SpaceWarpsI}
Marshall, P.~J., Verma, A., More, A., {et~al.} 2015, MNRAS, 455, 1171

\bibitem[{{Meneghetti} {et~al.}(2008){Meneghetti}, {Melchior}, {Grazian}, {De Lucia}, {Dolag}, {Bartelmann}, {Heymans}, {Moscardini}, \& {Radovich}}]{2008AandA…482..403M}
{Meneghetti}, M., {Melchior}, P., {Grazian}, A., {et~al.} 2008, \aap, 482, 403

\bibitem[{{Meneghetti} {et~al.}(2010){Meneghetti}, {Rasia}, {Merten}, {Bellagamba}, {Ettori}, {Mazzotta}, {Dolag}, \& {Marri}}]{2010AandA…514A..93M}
{Meneghetti}, M., {Rasia}, E., {Merten}, J., {et~al.} 2010, \aap, 514, A93

\bibitem[{{Metcalf} {et~al.}(2019){Metcalf}, {Meneghetti}, {Avestruz}, {Bellagamba}, {Bom}, {Bertin}, {Cabanac}, {Courbin}, {Davies}, {Decenci{\`e}re}, {Flamary}, {Gavazzi}, {Geiger}, {Hartley}, {Huertas-Company}, {Jackson}, {Jacobs}, {Jullo}, {Kneib}, {Koopmans}, {Lanusse}, {Li}, {Ma}, {Makler}, {Li}, {Lightman}, {Petrillo}, {Serjeant}, {Sch{\"a}fer}, {Sonnenfeld}, {Tagore}, {Tortora}, {Tuccillo}, {Valent{\'\i}n}, {Velasco-Forero}, {Verdoes Kleijn}, \& {Vernardos}}]{Metcalf2018}
{Metcalf}, R.~B., {Meneghetti}, M., {Avestruz}, C., {et~al.} 2019, \aap, 625, A119

\bibitem[{{Metcalf} \& {Petkova}(2014)}]{glamerm}
{Metcalf}, R.~B. \& {Petkova}, M. 2014, \mnras, 445, 1942

\bibitem[{{More} {et~al.}(2016){More}, {Verma}, {Marshall}, {More}, {Baeten}, {Wilcox}, {Macmillan}, {Cornen}, {Kapadia}, {Parrish}, {Snyder}, {Davis}, {Gavazzi}, {Lintott}, {Simpson}, {Miller}, {Smith}, {Paget}, {Saha}, {K{\"u}ng}, \& {Collett}}]{SpaceWarpsII}
{More}, A., {Verma}, A., {Marshall}, P.~J., {et~al.} 2016, \mnras, 455, 1191

\bibitem[{{Nagam} {et~al.}(2025){Nagam}, {Acevedo Barroso}, {Wilde}, {et~al.}}]{Nagam25}
{Nagam}, B.~C., {Acevedo Barroso}, J.~A., {Wilde}, J., {et~al.} 2025, A\&A, submitted, arXiv:2502.09802

\bibitem[{{Nagam} {et~al.}(2023){Nagam}, {Koopmans}, {Valentijn}, {Kleijn}, {de Jong}, {Napolitano}, {Li}, \& {Tortora}}]{nagam23}
{Nagam}, B.~C., {Koopmans}, L. V.~E., {Valentijn}, E.~A., {et~al.} 2023, \mnras, 523, 4188

\bibitem[{{Nagam} {et~al.}(2024){Nagam}, {Koopmans}, {Valentijn}, {Kleijn}, {de Jong}, {Napolitano}, {Li}, {Tortora}, {Busillo}, \& {Dong}}]{nagam24}
{Nagam}, B.~C., {Koopmans}, L. V.~E., {Valentijn}, E.~A., {et~al.} 2024, \mnras, 533, 1426

\bibitem[{{Navarro} {et~al.}(1997){Navarro}, {Frenk}, \& {White}}]{nfw}
{Navarro}, J.~F., {Frenk}, C.~S., \& {White}, S. D.~M. 1997, \apj, 490, 493

\bibitem[{{O'Riordan} {et~al.}(2025){O'Riordan}, {Oldham}, {Nersesian}, {et~al.}}]{ORiordan25}
{O'Riordan}, C.~M., {Oldham}, L.~J., {Nersesian}, A., {et~al.} 2025, \aap, 694, A145

\bibitem[{{Pearce-Casey} {et~al.}(2025){Pearce-Casey}, {Nagam}, {Wilde}, {Busillo}, {Ulivi}, {Andika}, {Manj{\'o}n-Garc{\'\i}a}, {Leuzzi}, {Matavulj}, {Serjeant}, {Walmsley}, {Acevedo Barroso}, {O'Riordan}, {Cl{\'e}ment}, {Tortora}, {Collett}, {Courbin}, {Gavazzi}, {Metcalf}, {Cabanac}, {Courtois}, {Crook-Mansour}, {Delchambre}, {Despali}, {Ecker}, {Franco}, {Holloway}, {Jahnke}, {Mahler}, {Marchetti}, {Melo}, {Meneghetti}, {M{\"u}ller}, {Nucita}, {Pearson}, {Rojas}, {Scarlata}, {Schuldt}, {Sluse}, {Suyu}, {Vaccari}, {Vegetti}, {Verma}, {Vernardos}, {Bolzonella}, {Kluge}, {Saifollahi}, {Schirmer}, {Stone}, {Paulino-Afonso}, {Bazzanini}, {Hogg}, {Koopmans}, {Kruk}, {Mannucci}, {Bromley}, {D{\'\i}az-S{\'a}nchez}, {Dickinson}, {Powell}, {Bouy}, {Laureijs}, {Altieri}, {Amara}, {Andreon}, {Baccigalupi}, {Baldi}, {Balestra}, {Bardelli}, {Battaglia}, {Bonino}, {Branchini}, {Brescia}, {Brinchmann}, {Caillat}, {Camera}, {Capobianco}, {Carbone}, {Carretero}, {Casas}, {Castellano}, {Castignani}, {Cavuoti}, {Cimatti},
  {Colodro-Conde}, {Congedo}, {Conselice}, {Conversi}, {Copin}, {Cropper}, {Da Silva}, {Degaudenzi}, {De Lucia}, {Di Giorgio}, {Dinis}, {Dubath}, {Dupac}, {Dusini}, {Farina}, {Farrens}, {Faustini}, {Ferriol}, {Frailis}, {Franceschi}, {Galeotta}, {George}, {Gillard}, {Gillis}, {Giocoli}, {G{\'o}mez-Alvarez}, {Grazian}, {Grupp}, {Haugan}, {Holmes}, {Hook}, {Hormuth}, {Hornstrup}, {Hudelot}, {Jhabvala}, {Joachimi}, {Keih{\"a}nen}, {Kermiche}, {Kiessling}, {Kilbinger}, {Kubik}, {K{\"u}mmel}, {Kunz}, {Kurki-Suonio}, {Le Mignant}, {Ligori}, {Lilje}, {Lindholm}, {Lloro}, {Maiorano}, {Mansutti}, {Marggraf}, {Markovic}, {Martinelli}, {Martinet}, {Marulli}, {Massey}, {Medinaceli}, {Mei}, {Melchior}, {Mellier}, {Merlin}, {Meylan}, {Moresco}, {Moscardini}, {Nakajima}, {Neissner}, {Nichol}, {Niemi}, {Nightingale}, {Padilla}, {Paltani}, {Pasian}, {Pedersen}, {Percival}, {Pettorino}, {Pires}, {Polenta}, {Poncet}, {Popa}, {Pozzetti}, {Raison}, {Renzi}, {Rhodes}, {Riccio}, {Romelli}, {Roncarelli}, {Rossetti}, {Saglia},
  {Sakr}, {S{\'a}nchez}, {Sapone}, {Sartoris}, {Schneider}, {Schrabback}, {Secroun}, {Seidel}, {Serrano}, {Sirignano}, {Sirri}, {Skottfelt}, {Stanco}, {Steinwagner}, {Tallada-Cresp{\'\i}}, {Tereno}, {Toledo-Moreo}, {Torradeflot}, {Tutusaus}, {Valentijn}, {Valenziano}, {Vassallo}, {Verdoes Kleijn}, {Veropalumbo}, {Wang}, {Weller}, {Zamorani}, \& {Zucca}}]{pearce-casey24}
{Pearce-Casey}, R., {Nagam}, B.~C., {Wilde}, J., {et~al.} 2025, \aap, 696, A214

\bibitem[{{Petkova} {et~al.}(2014){Petkova}, {Metcalf}, \& {Giocoli}}]{glamerp}
{Petkova}, M., {Metcalf}, R.~B., \& {Giocoli}, C. 2014, \mnras, 445, 1954

\bibitem[{{Petrillo} {et~al.}(2017){Petrillo}, {Tortora}, {Chatterjee}, {Vernardos}, {Koopmans}, {Verdoes Kleijn}, {Napolitano}, {Covone}, {Schneider}, {Grado}, \& {McFarland}}]{petrillo2017}
{Petrillo}, C.~E., {Tortora}, C., {Chatterjee}, S., {et~al.} 2017, \mnras, 472, 1129

\bibitem[{{Petrillo} {et~al.}(2019){Petrillo}, {Tortora}, {Vernardos}, {Koopmans}, {Verdoes Kleijn}, {Bilicki}, {Napolitano}, {Chatterjee}, {Covone}, {Dvornik}, {Erben}, {Getman}, {Giblin}, {Heymans}, {de Jong}, {Kuijken}, {Schneider}, {Shan}, {Spiniello}, \& {Wright}}]{petrillo--19}
{Petrillo}, C.~E., {Tortora}, C., {Vernardos}, G., {et~al.} 2019, \mnras, 484, 3879

\bibitem[{{Pourrahmani} {et~al.}(2018){Pourrahmani}, {Nayyeri}, \& {Cooray}}]{hstlensflow}
{Pourrahmani}, M., {Nayyeri}, H., \& {Cooray}, A. 2018, \apj, 856, 68

\bibitem[{{Rojas} {et~al.}(2023){Rojas}, {Collett}, {Ballard}, {Magee}, {Birrer}, {Buckley-Geer}, {Chan}, {Cl{\'e}ment}, {Diego}, {Gentile}, {Gonz{\'a}lez}, {Joseph}, {Mastache}, {Schuldt}, {Tortora}, {Verdugo}, {Verma}, {Daylan}, {Millon}, {Jackson}, {Dye}, {Melo}, {Mahler}, {Ogando}, {Courbin}, {Fritz}, {Herle}, {Acevedo Barroso}, {Ca{\~n}ameras}, {Cornen}, {Dhanasingham}, {Glazebrook}, {Martinez}, {Ryczanowski}, {Savary}, {G{\'o}is-Silva}, {Arturo Ure{\~n}a-L{\'o}pez}, {Wiesner}, {Wilde}, {Valim Cal{\c{c}}ada}, {Cabanac}, {Pan}, {Sierra}, {Despali}, {Cavalcante-Gomes}, {Macmillan}, {Maresca}, {Grudskaia}, {O'Donnell}, {Paic}, {Niemiec}, {de la Bella}, {Bromley}, {Williams}, {More}, \& {Levine}}]{rojas23}
{Rojas}, K., {Collett}, T.~E., {Ballard}, D., {et~al.} 2023, \mnras, 523, 4413

\bibitem[{{Rojas} {et~al.}(2022){Rojas}, {Savary}, {Cl{\'e}ment}, {Maus}, {Courbin}, {Lemon}, {Chan}, {Vernardos}, {Joseph}, {Ca{\~n}ameras}, \& {Galan}}]{rojas22}
{Rojas}, K., {Savary}, E., {Cl{\'e}ment}, B., {et~al.} 2022, \aap, 668, A73

\bibitem[{{Savary} {et~al.}(2022){Savary}, {Rojas}, {Maus}, {Cl{\'e}ment}, {Courbin}, {Gavazzi}, {Chan}, {Lemon}, {Vernardos}, {Ca{\~n}ameras}, {Schuldt}, {Suyu}, {Cuillandre}, {Fabbro}, {Gwyn}, {Hudson}, {Kilbinger}, {Scott}, \& {Stone}}]{Savary2021}
{Savary}, E., {Rojas}, K., {Maus}, M., {et~al.} 2022, \aap, 666, A1

\bibitem[{{Schuldt} {et~al.}(2021){Schuldt}, {Suyu}, {Meinhardt}, {Leal-Taix{\'e}}, {Ca{\~n}ameras}, {Taubenberger}, \& {Halkola}}]{schuldt21}
{Schuldt}, S., {Suyu}, S.~H., {Meinhardt}, T., {et~al.} 2021, \aap, 646, A126

\bibitem[{{Scoville} {et~al.}(2007){Scoville}, {Aussel}, {Brusa}, {Capak}, {Carollo}, {Elvis}, {Giavalisco}, {Guzzo}, {Hasinger}, {Impey}, {Kneib}, {LeFevre}, {Lilly}, {Mobasher}, {Renzini}, {Rich}, {Sanders}, {Schinnerer}, {Schminovich}, {Shopbell}, {Taniguchi}, \& {Tyson}}]{scoville2007}
{Scoville}, N., {Aussel}, H., {Brusa}, M., {et~al.} 2007, \apjs, 172, 1

\bibitem[{{Shu} {et~al.}(2022){Shu}, {Ca{\~n}ameras}, {Schuldt}, {Suyu}, {Taubenberger}, {Inoue}, \& {Jaelani}}]{shu2022}
{Shu}, Y., {Ca{\~n}ameras}, R., {Schuldt}, S., {et~al.} 2022, \aap, 662, A4

\bibitem[{{Sonnenfeld} {et~al.}(2020){Sonnenfeld}, {Verma}, {More}, {Baeten}, {Macmillan}, {Wong}, {Chan}, {Jaelani}, {Lee}, {Oguri}, {Rusu}, {Veldthuis}, {Trouille}, {Marshall}, {Hutchings}, {Allen}, {O'Donnell}, {Cornen}, {Davis}, {McMaster}, {Lintott}, \& {Miller}}]{SpaceWarpsHSC}
{Sonnenfeld}, A., {Verma}, A., {More}, A., {et~al.} 2020, \aap, 642, A148

\bibitem[{{Storfer} {et~al.}(2024){Storfer}, {Huang}, {Gu}, {Sheu}, {Banka}, {Dey}, {Inchausti Reyes}, {Jain}, {Kwon}, {Lang}, {Lee}, {Meisner}, {Moustakas}, {Myers}, {Tabares-Tarquinio}, {Schlafly}, \& {Schlegel}}]{desistorfer2024}
{Storfer}, C., {Huang}, X., {Gu}, A., {et~al.} 2024, \apjs, 274, 16

\bibitem[{{Suyu} \& {Halkola}(2010)}]{suyu10}
{Suyu}, S.~H. \& {Halkola}, A. 2010, \aap, 524, A94

\bibitem[{{Suyu} {et~al.}(2012){Suyu}, {Hensel}, {McKean}, {Fassnacht}, {Treu}, {Halkola}, {Norbury}, {Jackson}, {Schneider}, {Thompson}, {Auger}, {Koopmans}, \& {Matthews}}]{suyu12}
{Suyu}, S.~H., {Hensel}, S.~W., {McKean}, J.~P., {et~al.} 2012, \apj, 750, 10

\bibitem[{Szegedy {et~al.}(2015)Szegedy, Liu, Jia, Sermanet, Reed, Anguelov, Erhan, Vanhoucke, \& Rabinovich}]{2szegedy14}
Szegedy, C., Liu, W., Jia, Y., {et~al.} 2015, in 2015 IEEE Conference on Computer Vision and Pattern Recognition (CVPR) (Los Alamitos, CA, USA: IEEE Computer Society), 1--9

\bibitem[{Szegedy {et~al.}(2016)Szegedy, Vanhoucke, Ioffe, Shlens, \& Wojna}]{szegedy15}
Szegedy, C., Vanhoucke, V., Ioffe, S., Shlens, J., \& Wojna, Z. 2016, in 2016 IEEE Conference on Computer Vision and Pattern Recognition (CVPR) (Los Alamitos, CA, USA: IEEE Computer Society), 2818--2826

\bibitem[{{Tortora} {et~al.}(2019){Tortora}, {Posti}, {Koopmans}, \& {Napolitano}}]{tortora2019}
{Tortora}, C., {Posti}, L., {Koopmans}, L.~V.~E., \& {Napolitano}, N.~R. 2019, \mnras, 489, 5483

\bibitem[{{Treu} {et~al.}(2011){Treu}, {Dutton}, {Auger}, {Marshall}, {Bolton}, {Brewer}, {Koo}, \& {Koopmans}}]{treu11}
{Treu}, T., {Dutton}, A.~A., {Auger}, M.~W., {et~al.} 2011, \mnras, 417, 1601

\bibitem[{Walmsley {et~al.}(2023)Walmsley, Allen, Aussel, Bowles, Gregorowicz, Slijepcevic, Lintott, m.~Scaife, Jabłońska, Karchev, Lanzieri, Mohan, O’Ryan, Saiguhan, Suárez, Guerra-Varas, \& Velu}]{Walmsley2023}
Walmsley, M., Allen, C., Aussel, B., {et~al.} 2023, Journal of Open Source Software, 8, 5312

\bibitem[{{Walmsley} {et~al.}(2024){Walmsley}, {Bowles}, {Scaife}, {Shingirai Makechemu}, {Gordon}, {Ferguson}, {Mann}, {Pearson}, {Popp}, {Bovy}, {Speagle}, {Dickinson}, {Fortson}, {G{\'e}ron}, {Kruk}, {Lintott}, {Mantha}, {Mohan}, {O'Ryan}, \& {Slijepevic}}]{walmsley24}
{Walmsley}, M., {Bowles}, M., {Scaife}, A. M.~M., {et~al.} 2024, arXiv e-prints, arXiv:2404.02973

\bibitem[{{Weaver} {et~al.}(2022){Weaver}, {Kauffmann}, {Ilbert}, {McCracken}, {Moneti}, {Toft}, {Brammer}, {Shuntov}, {Davidzon}, {Hsieh}, {Laigle}, {Anastasiou}, {Jespersen}, {Vinther}, {Capak}, {Casey}, {McPartland}, {Milvang-Jensen}, {Mobasher}, {Sanders}, {Zalesky}, {Arnouts}, {Aussel}, {Dunlop}, {Faisst}, {Franx}, {Furtak}, {Fynbo}, {Gould}, {Greve}, {Gwyn}, {Kartaltepe}, {Kashino}, {Koekemoer}, {Kokorev}, {Le F{\`e}vre}, {Lilly}, {Masters}, {Magdis}, {Mehta}, {Peng}, {Riechers}, {Salvato}, {Sawicki}, {Scarlata}, {Scoville}, {Shirley}, {Silverman}, {Sneppen}, {Smolc̆i{\'c}}, {Steinhardt}, {Stern}, {Tanaka}, {Taniguchi}, {Teplitz}, {Vaccari}, {Wang}, \& {Zamorani}}]{weaver2022}
{Weaver}, J.~R., {Kauffmann}, O.~B., {Ilbert}, O., {et~al.} 2022, \apjs, 258, 11

\bibitem[{{Wilde} {et~al.}(2022){Wilde}, {Serjeant}, {Bromley}, {Dickinson}, {Koopmans}, \& {Metcalf}}]{wilde22}
{Wilde}, J., {Serjeant}, S., {Bromley}, J.~M., {et~al.} 2022, \mnras, 512, 3464

\end{thebibliography}

%

\begin{appendix}
  \onecolumn 

\newpage

\section{{Images of lenses discovered}} 
\label{sec:lenses-found}

\subsection{{Top grade A candidates}} \label{ssec:grade-a}

\begin{figure} [!htp]
    \centering
    \includegraphics[width=1\linewidth]{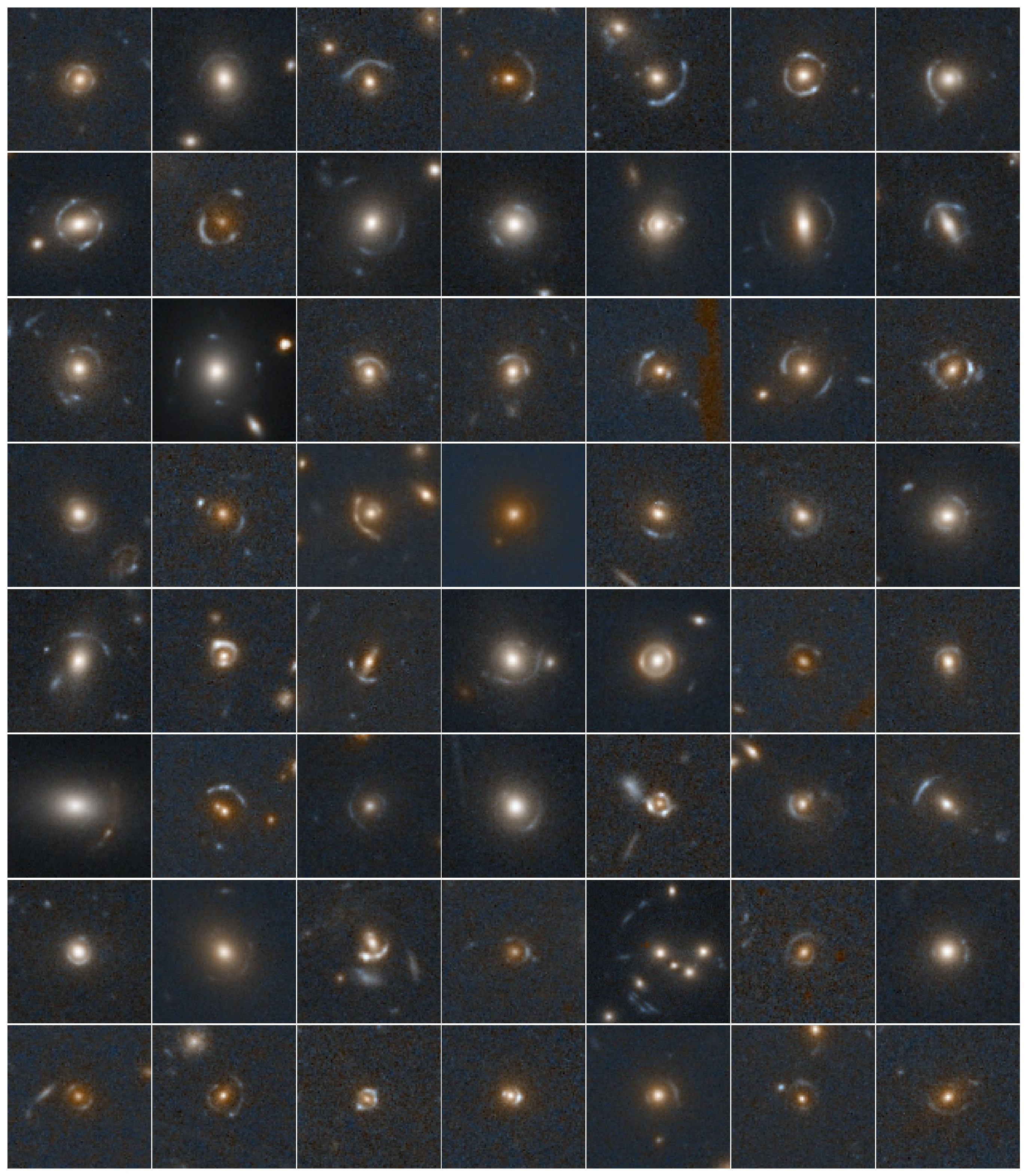}
    \caption{{Top 56/250 grade A lenses discovered in Q1, ranked (left to right, top to bottom) by expert inspection score.}}
    \label{fig:grade-A-fig}
\end{figure}

{In Fig. \ref{fig:grade-A-fig} we show the top 56/250 grade A lenses, ranked according to score assigned during expert visual inspection.}

\subsection{{DSPLs}} \label{ssec:dspl-fig}
\begin{figure} [!htp]
    \centering
    \includegraphics[width=0.8\linewidth]{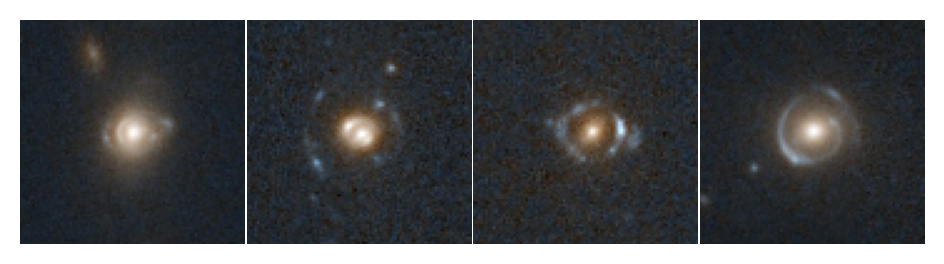}
    \caption{{Four DSPL \textcolor{black}{candidates} discovered in Q1 \textcolor{black}{that have successful DSPL models}. From left to right, they are the `teapot lens', the `cosmic dartboard', `Galileo's lens', and the `cosmic ammonite'.}}
    \label{fig:dspls}
\end{figure}

\noindent{In Fig. \ref{fig:dspls} we show four DSPLs found in Q1 that have been successfully modelled as DSPLs by \citet{Q1-SP054}.}

\subsection{{Edge-on lenses}} \label{ssec:edge-on-fig}
\begin{figure} [!htp]
    \centering
    \includegraphics[width=0.9\linewidth]{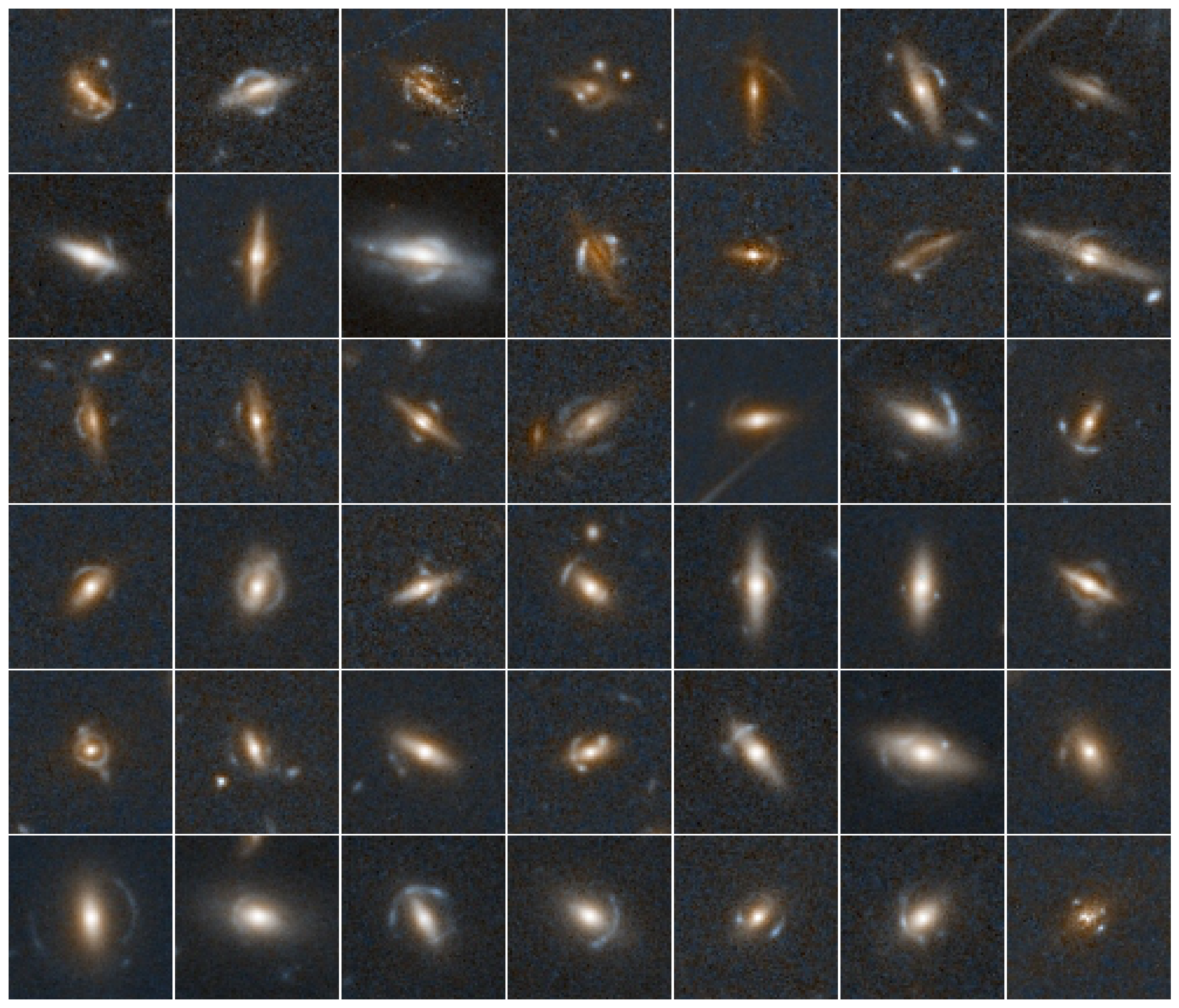}
    \caption{{Selection of 42 of the grade A+B lenses where the lensing galaxy is an edge-on disc or lenticular galaxy. Because these lenses typically have smaller Einstein radii, we used a smaller cutout size of $70 \times 70$ pixels (\ang{;;7}$\times$\ang{;;7}).}}
    \label{fig:edgeons}
\end{figure}

\noindent{In Fig. \ref{fig:edgeons} we show grade A+B lenses discovered in Q1, where the lensing galaxy has a morphology closer to that of a disc or lenticular galaxy that is also at high inclination. Because the boundary between disc, lenticular, and elliptical galaxies is not well defined, the 42 examples presented here should not be considered a complete list, but is included to display the diversity of lenses found in Q1.}

\section{Example grade A, B, and C lenses} \label{sec:grade-abc}
\begin{figure} [!htp]
    \centering
    \includegraphics[width=0.85\linewidth]{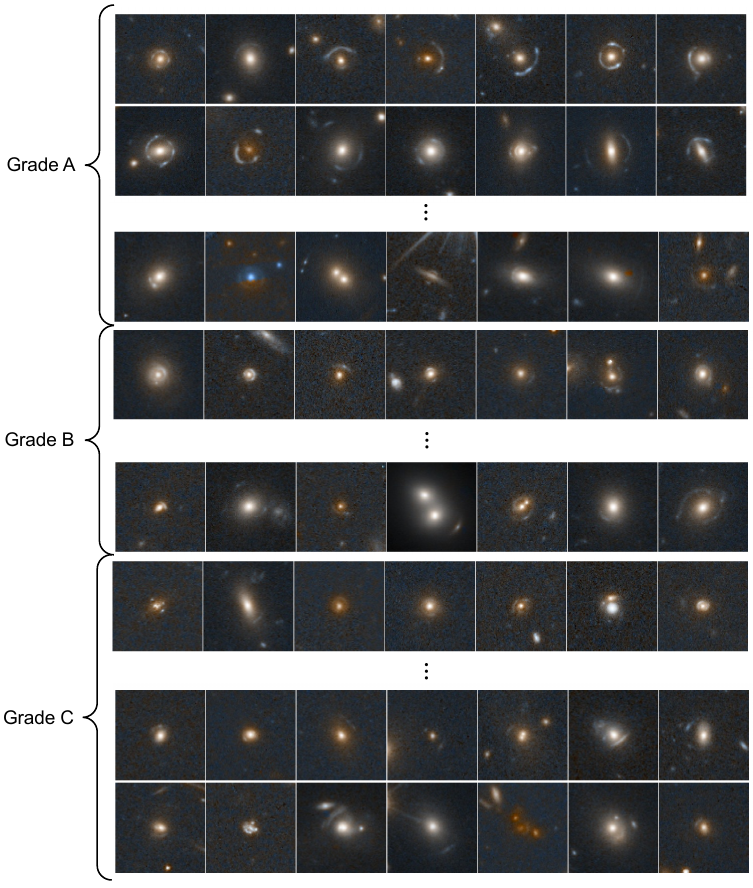}
    \caption{Images ranked by expert classification ordered from more likely to be a lens (top) to less likely to be a lens (bottom) at the upper and lower ends of the grade limits.}
    \label{fig:grade-buckets}
\end{figure}

In Fig. \ref{fig:grade-buckets} we show images order{ed} from most likely to least likely to be a lens according to expert inspection and the boundaries between grades A, B, and C to provide the reader with an understanding of what each grade refers to. For the full catalogue of lenses discovered {we} refer to \citet{Q1-SP048}.

\section{\texttt{Zoobot} false positives} \label{sec:false-positives}
\begin{figure}[!htp]
    \centering
    \includegraphics[width=1\linewidth]{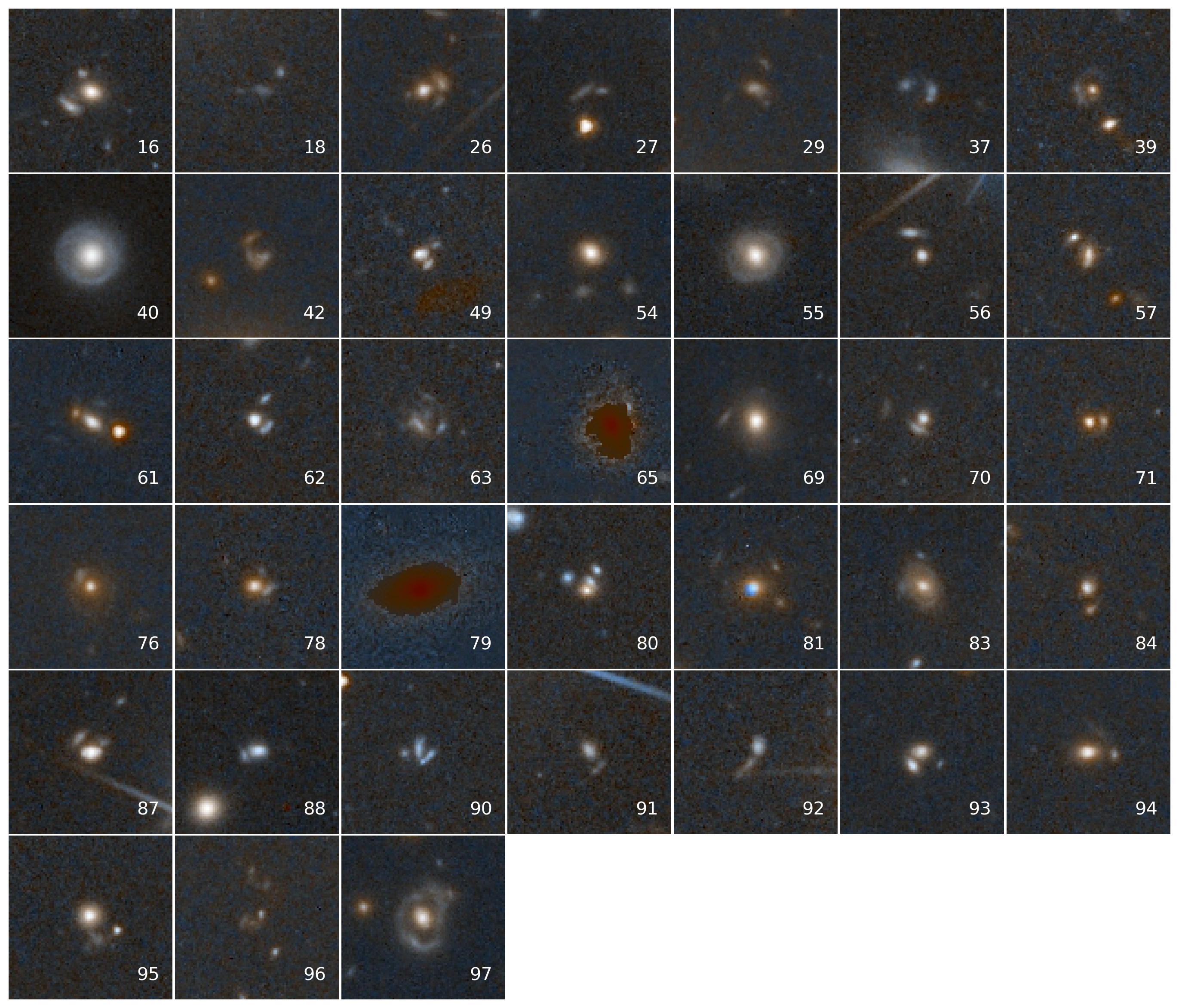}
    \caption{All the non-lenses that were ranked by \texttt{Zoobot} in the top 100 images. The number in each image displays how it was ranked by \texttt{Zoobot}.}
    \label{fig:zoobot_fps}
\end{figure}

In Fig. \ref{fig:zoobot_fps} we show the images that were ranked by \texttt{Zoobot} in the top 100 that were non-lenses according to expert inspection. 
Besides the two artefact images (ranked 65 and 79), the false positives can be loosely be recognised as images containing multiple galaxies in close proximity, where one is typically more elongated. 
Although most of these are very obviously non-lenses, in many cases (e.g. images ranked 27, 39, 55, 70 etc.) it can at least be discerned what is being confused as an arc, and given the typical uninterpretability of CNNs, this is perhaps a good sign.
Furthermore, if what is being confused as lenses are primarily serendipitous alignment{s} of galaxies, these are much rarer than other typical false positives, such as spiral galaxies, and therefore this is less of a concern.
It is notable that in a minority of cases (e.g. image ranked 92) the more elongated galaxy is not centred on the other galaxy, showing that \texttt{Zoobot} is not necessarily learning any physical properties about the relationship between the lens and lensed galaxies. 
Reducing the prevalence of these false positives may be tricky; 
although the negative training set contained a `merger' class that included images of multiple galaxies in different arrangements, the number of ways multiple galaxies can be oriented within one {image} is extremely large, so it is expected that some configurations could be confused with lensing systems.
Additionally, the galaxies in these false positives are generally not very extended or well resolved, meaning they were unlikely to have been included in \texttt{Zoobot}'s pretraining on GalaxyZoo morphologies. However, as the quantity of \Euclid data accumulates, we will be able to use known false positives from previous \Euclid lens searches during the training for the next data releases, meaning we can hope for this to improve over time.

\end{appendix}

\label{LastPage}
\end{document}